\documentclass[
  journal=pasa,
  manuscript=research-paper, 
  year=2024,
  volume=37,
]{cup-journal}
\usepackage{graphicx}
\usepackage{dcolumn}
\usepackage{bm}
\usepackage{multirow}
\usepackage{threeparttable}
\usepackage[figuresright]{rotating}
\usepackage{caption}
\usepackage{subcaption}
\usepackage{hyperref}
\usepackage{float}
\usepackage{footnote}

\usepackage{microtype,siunitx,booktabs}
\usepackage{amsmath}
\usepackage{xcolor}

\hypersetup{colorlinks=true,citecolor=blue,linkcolor=blue,urlcolor=blue}

\newcommand{\pinta}{\texttt{PINTA}}

\newcommand{\rficlean}{\texttt{RFIClean}}

\newcommand{\dspsr}{\texttt{dspsr}}
\newcommand{\pdmp}{\texttt{pdmp}}

\newcommand{\psrchive}{\texttt{PSRCHIVE}}
\newcommand{\psrfits}{\texttt{PSRFITS}}

\newcommand{\tempotwo}{\texttt{TEMPO2}}
\newcommand{\dmcalc}{\texttt{DMCalc}}

\newcommand{\enterprise}{\texttt{ENTERPRISE}}

\newcommand{\astropy}{\texttt{astropy}}

\title[InPTA DR2]{The Indian Pulsar Timing Array Data Release 2: I. Dataset and Timing Analysis}

\author{Prerna Rana}
\affiliation{Department of Astronomy, University of Cape Town, 
Cape Town 7700, South Africa.}
\alsoaffiliation{Department of Astronomy and Astrophysics,  Tata Institute of Fundamental Research, Colaba, Mumbai 400005}

\author{Pratik Tarafdar}
\affiliation{The Institute of Mathematical Sciences, C. I. T. Campus, Taramani, Chennai 600113, India}

\author{Nobleson K}
\affiliation{International Research Organization for Advanced Science and Technology, Kumamoto University, 2-39-1 Kurokami, Kumamoto 860-8555, Japan}

\author{Churchil Dwivedi}
\affiliation{Astronomy and Astrophysics Division, Physical Research Laboratory, Thaltej Campus, Sindhu Bhavan Marg, Ahmedabad 380059, Gujarat, India}
\alsoaffiliation{Department of Earth and Space Sciences, Indian Institute of Space Science and Technology, Valiamala, Thiruvananthapuram 695547, Kerala, India}

\author{Bhal Chandra Joshi}
\affiliation{National Centre for Radio Astrophysics, Pune University Campus, 
Pune 411007, Maharashtra, India}
\alsoaffiliation{Department of Physics, Indian Institute of Technology Roorkee, 
Roorkee 247667, Uttarakhand, India}

\author{Debabrata Deb}
\affiliation{The Institute of Mathematical Sciences, C. I. T. Campus, Taramani, Chennai 600113, India}

\author{Sushovan Mondal}
\affiliation{The Institute of Mathematical Sciences, C. I. T. Campus, Taramani, Chennai 600113, India}
\alsoaffiliation{Homi Bhabha National Institute, Training School Complex, Anushakti Nagar, Mumbai 400094, India}

\author{M. A. Krishnakumar}
\affiliation{National Centre for Radio Astrophysics, Pune University Campus, 
Pune 411007, Maharashtra, India}
\alsoaffiliation{Max-Planck-Institut f{\"u}r Radioastronomie, 
Auf dem H{\"u}gel 69, 53121 Bonn, Germany}

\author{Adya Shukla}
\affiliation{Department of Physics, Indian Institute of Technology Roorkee, 
Roorkee 247667, Uttarakhand, India}

\author{Jaikhomba Singha}
\affiliation{High Energy Physics, Cosmology \& Astrophysics Theory (HEPCAT) Group, Department of Mathematics and Applied Mathematics, University of Cape Town, Cape Town 7700, South Africa.}

\author{Himanshu Grover}
\affiliation{Department of Physics, Indian Institute of Technology Roorkee, 
Roorkee 247667, Uttarakhand, India}

\author{Hemanga Tahbildar}
\affiliation{Department of Physics, IISER Bhopal, Bhauri Bypass Road, Bhopal, 462066, India}

\author{Abhimanyu Susobhanan}
\affiliation{Max-Planck-Institut f{\"u}r Gravitationsphysik (Albert-Einstein-Institut), Leibniz Universit{\"a}t Hannover, Callinstra{\ss}e 38, 30167 Hannover, Germany}

\author{Mayuresh Surnis}
\affiliation{Department of Physics, IISER Bhopal, Bhauri Bypass Road, Bhopal, 462066, India}

\author{Shantanu Desai}
\affiliation{Department of Physics, IIT Hyderabad, Kandi, Telangana 502284, India}

\author{Neelam Dhanda Batra}
\affiliation{Department of Physics \& Astrophysics, University of Delhi, 110007, India}

\author{Aman Srivastava}
\affiliation{Department of Physics, IIT Hyderabad, Kandi, Telangana 502284, India}
\alsoaffiliation{Department of Physics, GLA University, Mathura 281406, India }

\author{Vinay Bharambe}
\affiliation{Department of Physics \& Astrophysics, University of Delhi, 110007, India}

\author{Jibin Jose}
\affiliation{Department of Astronomy, Astrophysics, and Space Engineering, Indian Institute of Technology Indore, Indore 453552, India}

\author{Vaishnavi Vyasraj}
\affiliation{Department of Physics, IIT Hyderabad, Kandi, Telangana 502284, India}

\author{Shebin Jose Jacob}
\affiliation{Department of Physics, Government Brennen College, Thalassery, Kannur University, Kannur 670106, Kerala, India}

\author{Amarnath}
\affiliation{Raman Research Institute, Bengaluru 560080, Karnataka, India}
\alsoaffiliation{Department of Physics, Cochin University of Science and Technology, Kochi, Kerala, 682022, India}

\author{Manpreet Singh}
\affiliation{Department of Physical Sciences, Indian Institute of Science Education and Research (IISER) Mohali, Sector 81, SAS Nagar, Mohali, Punjab, 140306, India}

\author{Zenia Zuraiq}
\affiliation{Department of Physics, Indian Institute of Science, Bengaluru, Karnataka, 560012, India}

\author{Sarbartha Sengupta}
\affiliation{The Institute of Mathematical Sciences, C. I. T. Campus, Taramani, Chennai 600113, India}
\alsoaffiliation{Homi Bhabha National Institute, Training School Complex, Anushakti Nagar, Mumbai 400094, India}

\author{Toki Ogi}
\affiliation{Kumamoto University, Graduate School of Science and Technology, Kumamoto,860-8555,Japan}

\author{Dhruv Kumar}
\affiliation{Department of Physics, National Institute of Technology Agartala, Tripura 799046, India}

\author{S Jagadeesh}
\affiliation{Department of Engineering Science, IIT Hyderabad, Kandi, Telangana 502284, India}

\author{Fazal Kareem}
\affiliation{Max-Planck-Institut f{\"u}r Radioastronomie, 
Auf dem H{\"u}gel 69, 53121 Bonn, Germany}

\author{Deep Maity}
\affiliation{The Institute of Mathematical Sciences, C. I. T. Campus, Taramani, Chennai 600113, India}
\alsoaffiliation{Homi Bhabha National Institute, Training School Complex, Anushakti Nagar, Mumbai 400094, India}

\author{Kaustubh Rai}
\affiliation{Department of Physics, IISER Bhopal, Bhauri Bypass Road, Bhopal, 462066, India}

\author{Kunjal Vara}
\affiliation{Department of Physics, IISER Bhopal, Bhauri Bypass Road, Bhopal, 462066, India}

\author{Shaswata Chowdhury}
\affiliation{The Institute of Mathematical Sciences, C. I. T. Campus, Taramani, Chennai 600113, India}

\author{Ryo Kato}
\affiliation{Mizusawa VLBI Observatory, National Astronomical Observatory of Japan, 2-21-1 Osawa, Mitaka, Tokyo 181-8588, Japan}

\author{Swetha Arumugam}
\affiliation{Department of Physics and Astronomy, Clemson University, Clemson, SC 29634, USA}

\author{Pragna Mamidipaka}
\affiliation{Department of Electrical and Computer Engineering, Carnegie Mellon University, Pittsburgh, PA 15213, USA}

\author{Arul Pandian B}
\affiliation{Raman Research Institute, Bengaluru 560080, Karnataka, India}
\alsoaffiliation{Department of Physics and Electronics, CHRIST (Deemed to be University), Bengaluru - 560029, India}

\author{Kavya Shaji}
\affiliation{School of Physics, University of Sydney, Australia }
\alsoaffiliation{National Centre for Radio Astrophysics, Pune University Campus, 
Pune 411007, Maharashtra, India}

\author{Prabu Thiagaraj}
\affiliation{Raman Research Institute, Bengaluru 560080, Karnataka, India}

\author{P. Arumugam}
\affiliation{Department of Physics, Indian Institute of Technology Roorkee, 
Roorkee 247667, Uttarakhand, India}

\author{Manjari Bagchi}
\affiliation{The Institute of Mathematical Sciences, C. I. T. Campus, Taramani, Chennai 600113, India}
\alsoaffiliation{Homi Bhabha National Institute, Training School Complex, Anushakti Nagar, Mumbai 400094, India}

\author{Manoneeta Chakraborty}
\affiliation{Department of Astronomy, Astrophysics, and Space Engineering, Indian Institute of Technology Indore, Indore 453552, India}

\author{A. Gopakumar}
\affiliation{Department of Astronomy and Astrophysics,  Tata Institute of Fundamental Research, Colaba, Mumbai 400005}

\author{Yashwant Gupta}
\affiliation{National Centre for Radio Astrophysics, Pune University Campus, 
Pune 411007, Maharashtra, India}

\author{Yogesh Maan}
\affiliation{National Centre for Radio Astrophysics, Pune University Campus, 
Pune 411007, Maharashtra, India}

\author{Avinash Kumar Paladi}
\affiliation{Department of Physics, Indian Institute of Science, Bengaluru, Karnataka, 560012, India}

\author{Keitaro Takahashi}
\affiliation{Faculty of Advanced Science and Technology, Kumamoto University, 2-39-1 Kurokami, Kumamoto 860-8555, Japan}
\alsoaffiliation{International Research Organization for Advanced Science and Technology, Kumamoto University, 2-39-1 Kurokami, Kumamoto 860-8555, Japan}

\doi{xx.yyyy/pasa.zzzz.tt}

\received {dd Mmm YYYY}
\revised  {dd Mmm YYYY}
\accepted {dd Mmm YYYY}
\published{dd Mmm YYYY}


\begin{document}








\begin{abstract}
The Indian Pulsar Timing Array (InPTA) employs unique features of the upgraded Giant Metrewave Radio Telescope (uGMRT) to monitor dozens of  the International Pulsar Timing Array (IPTA) millisecond pulsars (MSPs), simultaneously in the 300-500 MHz and the 1260-1460 MHz bands. This dual-band approach ensures that any frequency-dependent delays are accurately characterized, significantly improving the timing precision for pulsar observations, which is crucial for pulsar timing arrays. We present details of InPTA's second data release that involves 7 yrs of data on 27 IPTA MSPs. This includes sub-banded Times of Arrival (ToAs), Dispersion Measures (DM), and initial timing ephemerides for our MSPs. A part of this dataset, originally released in InPTA's first data release, is being incorporated into IPTA's third data release which is expected to detect and characterize nanohertz gravitational waves in the coming years. The entire dataset is reprocessed in this second data release providing some of the highest precision DM estimates so far and interesting solar wind related DM variations in some pulsars. This is likely to characterize the noise introduced by the dynamic inter-stellar ionised medium much better than the previous release thereby increasing sensitivity to any future gravitational wave search.
 



\end{abstract}




\section{Introduction} 

Pulsars are rapidly rotating, highly magnetised neutron stars. They emit beams of coherent radiation from their magnetic poles. When these beams intersect our line of sight, we see periodic pulses that arrive very regularly \citep{Gold1968}.
Pulsars are classified as slow pulsars (typically with $P > 30$ ms and $B_s > 10^{10}$ G) and millisecond pulsars (MSPs; typically with $P < 30$ ms and $B_s < 10^{10}$ G) based on their rotation periods ($P$) and the surface magnetic fields ($B_s$). MSPs, which were discovered by \citet{bkh+1982},
are often part of binary systems and their faster spins are 
due to the transfer of (orbital) angular momentum from their companions \citep{rs1982}. 
In other words, MSPs are old pulsars 
that have been spun up by accretion thereby burying their earlier magnetic fields producing low magnetic field pulsars, which spin down very slowly \citep{acr+1982}.
These conditions ensure that MSPs behave as extremely stable clocks and their stability is comparable in accuracy to the best atomic clocks on the earth on time scales of years, thanks to the technique of pulsar timing \citep{lk2004}.
 The availability of such precise celestial clocks allowed astronomers to test general relativity \citep{tw1982}, create a new time scale \citep{hgc+2020}, discover exoplanets \citep{wf1992}, provide constraints on nuclear equations of state, etc., by timing MSPs \citep{lpm+1990}. 

Persistent long-term monitoring of the first binary pulsar PSR B1913+16 and developments in various aspects of pulsar timing provided the first indirect astrophysical proof for the existence of gravitational waves  \citep[GWs;][]{tw1982,dt1991,dt1992,taylor1994}. 
GWs are ripples in the curvature of space-time that are caused by violent events in the universe like the inspiral and subsequent merger of compact objects like Neutron Stars (NSs) or Black Holes (BHs) \citep{ss2009}.
The subsequent direct detection of GWs from a merging stellar mass BH binary, namely GW150914, by the two multi-kilometer-scale 
Laser Interferometer Gravitational-Wave Observatory (LIGO), inaugurated the era of GW astronomy \citep{aaa+2016}.
Further, LIGO observations of GWs from an inspiraling NS binary and its electromagnetic counterparts opened the era of multi-messenger GW astronomy \citep{aaa+2017}.
Similar to the laser-mirror system of the LIGO, an array of accurately timed MSPs can create a galaxy-sized GW detector with individual MSPs acting as arms of the interferometer with respect to the Earth. Such observatories are typically sensitive to GWs in the nanohertz frequency regime  \citep{Sazhin1978,detweiler1979,hd1983,fb1990}. 
This nanohertz detector, called the Pulsar Timing Array (PTA), relies on measuring minute changes in the times of arrival (ToA) of the radio pulses from the individual MSPs at the earth, as the passing GWs perturb the space-time between the pulsar and the earth \citep{EW1975}. These perturbations induce modulations in the measured ToAs that are correlated spatially across pulsars. The way to measure such a correlated signal is through a joint analysis of the timing residuals\footnote{Timing residual is generated after subtracting the expected ToA estimated through a timing model from the observed ToA.} for all the pulsars in the array. A PTA experiment operates by routinely monitoring an ensemble of MSPs and catalog their pulse ToAs over decades. It is customary for PTAs to publish such catalogs regularly, usually referred to as data releases. For example, the International Pulsar Timing Array is currently working on its third Data Release (IPTA DR3) \citep{ipta2024}.
The most-up-to-date IPTA DR3 is expected to include recent data releases from its constituents, namely the European PTA \cite[EPTA:][]{dcl+2016, epta2023}, the Indian PTA \cite[InPTA:][]{jab+2018,tnr+2022}, the North American Nanohertz Observatory for Gravitational waves \cite[NANOGrav:][]{2009arXiv0909.1058J,aaa+2023a}, the Australia-based Parkes PTA \cite[PPTA:][]{mhb+2013,zrk+2023},  and MeerKAT PTA \citep{msb+2023}.


It turns out that PTA data releases are critical to advance the nascent field of nanohertz (nHz) GW astronomy and the IPTA DR3 is expected to significantly further the progress made so far in the nHz GW astronomy \citep{ipta2024}.
This is mainly because of the compelling evidences, reported by NANOGrav \citep{aaa+2023b}, EPTA+InPTA \citep{epta+inpta2023a}, PPTA \citep{rzs+2023} and the Chinese PTA \citep{xcg+2023} in 2023, for the presence of a stochastic GW background (GWB) in nHz frequencies in their respective data sets.
The most likely source for such a GWB is expected to be the superposition of nHz GW emissions from an ensemble of inspiralling Supermassive Black Hole binaries (SMBHBs) though there are other possible exotic explanations \citep{BTC+2019}.
The evidence for the GWB reported by various PTAs in 2023 was the result of a highly coordinated and demanding effort by EPTA, InPTA, NANOGrav, and PPTA, where the first data release of the InPTA experiment \citep[InPTA DR1:][]{tnr+2022} was included in the EPTA+InPTA dataset. This effort built upon earlier investigations that had identified a common red noise process in previous data releases from NANOGrav, EPTA, PPTA, and IPTA  \citep{abb+2020,gsr+2021,ccg+2021,aab+2022}. 
Unfortunately, these earlier data sets and associated  nHz GWB searches did not reveal the required quadrupolar Hellings-Downs (HD) angular correlation, critical to clarify 
the general relativistic nature of the reported common red signals \citep{hd1983}.
Interestingly, recent NANOGrav and EPTA+InPTA investigations 
also report tentative evidence for GWs from individual SMBH binaries in their latest data sets  \citep{aaa+2023d,epta+inpta2023b}.
Additionally, PTAs are sensitive to certain Burst With Memory events in their data sets \citep{dsd+2024,aab+2024}.
Critical ingredients to all these exciting possibilities are regularly updated PTA data releases. In this paper, we report the second data release from the InPTA collaboration that we refer to as the InPTA DR2  and detail its various attributes. This dataset is proposed to be combined with the latest datasets from other PTAs for the upcoming IPTA data releases.



The InPTA experiment adapts the unique strengths of the Giant Metrewave Radio Telescope \cite[GMRT:][]{swarup1991} and  its recent major upgrade \cite[uGMRT:][]{gak+2017} for the ongoing  IPTA efforts by providing a unique low-radio frequency view of many IPTA pulsars, see figure \ref{fig:skymap}. 
It turns out that we are the only IPTA constituent who simultaneously record MSP data in both 300–500 MHz and 1260–1460 MHz ranges, which enable us to make some of the most precise dispersion measure (DM) estimates
for many IPTA pulsars \citep{tnr+2022}. 
Interestingly, uGMRT’s status as an SKA pathfinder ensures that such simultaneous low/high radio frequency MSP observations will also be replicated during Square Kilometer Array-based PTA efforts \citep{jgp+2022}.

In what follows, we provide an overview of the InPTA observations and dataset included in InPTA DR2 in section \ref{sec:obs}. In section \ref{sec:datareduction}, we detail the procedure for pre-processing and data reduction. We then describe the techniques adopted for generating noise-free templates in section \ref{sec4.1}, optimizing the selection of frequency sub-bands used for each pulsar in section \ref{sec4.2}, an iterative procedure to calculate the fiducial DM for each pulsar in section \ref{sec4.3}, and estimating ToAs and DMs from our observations in section \ref{sec4.4}. In section \ref{sec:timing}, we discuss our procedure for the deterministic timing analysis adopted in this work. Finally, we discuss our results, conclusions and future directions in section \ref{sec:discussion}.

\section{Observations} \label{sec:obs}

The InPTA observations are carried out using the uGMRT, which is an interferometer with thirty antennas, each with a diameter of  45-m  \citep{gak+2017}. Fourteen of these antennas are located in a central square, while the remaining antennas are distributed along three arms in a  `Y' shape.
The uGMRT provides four observing bands: band 2 (120-250 MHz), band 3 (250-500 MHz), band 4 (550-850), and band 5 (1050-1450) \citep{gak+2017}. The InPTA observations were carried out by splitting the uGMRT antennas into multiple subarrays, observing the same source in different frequency bands simultaneously, with 100 MHz or 200 MHz bandwidth in each band depending on the observing epoch. 

Typically, the nearest arm antennas with a subset of central square antennas were used in each subarray. The outer arm antennas were not used as phase of the voltage outputs from these antennas drift significantly during the observations of a target source. The geometrical and instrumental phase delays associated with each antenna were calibrated by observing a point source before each target pulsar. Then, the voltage signal from each hand of polarization were compensated by the estimated phase delays 
and added into a phased array sum using a separate beam-former for each subarray.

Digital processing of the phased array sum was employed to subdivide the observing band-pass into 1024 to 2048 channels. This was necessary to remove the 
dispersive effects of the ionized inter-stellar medium (IISM). While these channelized data were helpful in reducing the pulse smear well below the sampling time for band 5 and to a large extent for band 4 data, a real-time coherent dedispersion pipeline \citep{dg2016} was employed to completely remove the dispersive smear for band 3 data, thereby enabling a fine sampling of data. The former data stream is referred to as phased array (PA) beam, while the latter is called coherent dedispersion pipeline (CDP) beam in this paper. The data for both the streams were sampled based on a clock derived from an active Hydrogen maser and recorded to disk for offline data reduction 
(described in section \ref{sec:datareduction}). A time-stamp for the first sample of each data stream was derived from a GPS-disciplined Rubidium clock. Hence, the pulse arrival times for the InPTA data, measured at the uGMRT site, are based on GPS time-scale and no clock corrections were applied.

The InPTA DR1 included observations of 14 MSPs conducted during 2018-2021 (observation cycles 34-35 and 37-40 of the uGMRT) typically with $\sim$ 15 days cadence. 
In the present work, in addition to the observations included in the InPTA-DR1, we include observations of 13 more pulsars (in addition to the earlier 14 MSPs) and extend our observing time span to approximately 7.5 years covering the period between 2016$-$2024 (observation cycles 31-35 and 37-45). 

A pilot campaign for the InPTA was started in 2016$-$2017 (observation cycle 31) with observations of 16 pulsars. The main purpose of these observations was to characterize the timing precision achievable with the newly commissioned broadband systems in the uGMRT upgrade and quantify the advantages of broadband observations with the upgraded instrument. These early observations used two subarrays in band 3 (300–500 MHz) and band 5 (1260–1460 MHz) with 200 MHz bandwidth. The campaign continued during 2017-2018 (observation cycles 32-33) with three subarrays in band 3 (300–500 MHz), band 4 (550–750 MHz or 650-850 MHz), and band 5 (1260–1460 MHz) with 200 MHz bandwidth where CDP pipeline \citep{dg2016} was used in band 5. A different observing strategy was used during 2018-2019 (observation cycles 34-35) where a lower bandwidth was adopted, i.e., band 3 (400–500 MHz), band 4 (650–750 MHz), and band 5 (1360–1460 MHz), but with CDP in both band 3 and band 5. The observations during 2019-2022 (observation cycles 37-42) were carried out using two subarrays in band 3 (300–500 MHz) and band 5 (1260–1460 MHz) with CDP in band 3. Since 2022 (observation cycle 43), InPTA has followed a hybrid observing strategy where alternate observing sessions are held in band 3-only (only one subarray) or band 3 + band 5 (two subarrays) configurations with 200 MHz bandwidth in each band and CDP in band 3. This adjustment allowed us to include more pulsars in the band 3-only observation mode in this period (2022-2024) by utilizing more uGMRT antennas in a single subarray, and hence significantly increasing the sensitivity and enabling pulsar detection within shorter observation periods.

\begin{figure*}[h!]
    \centering
    \includegraphics[width=1.0\textwidth]{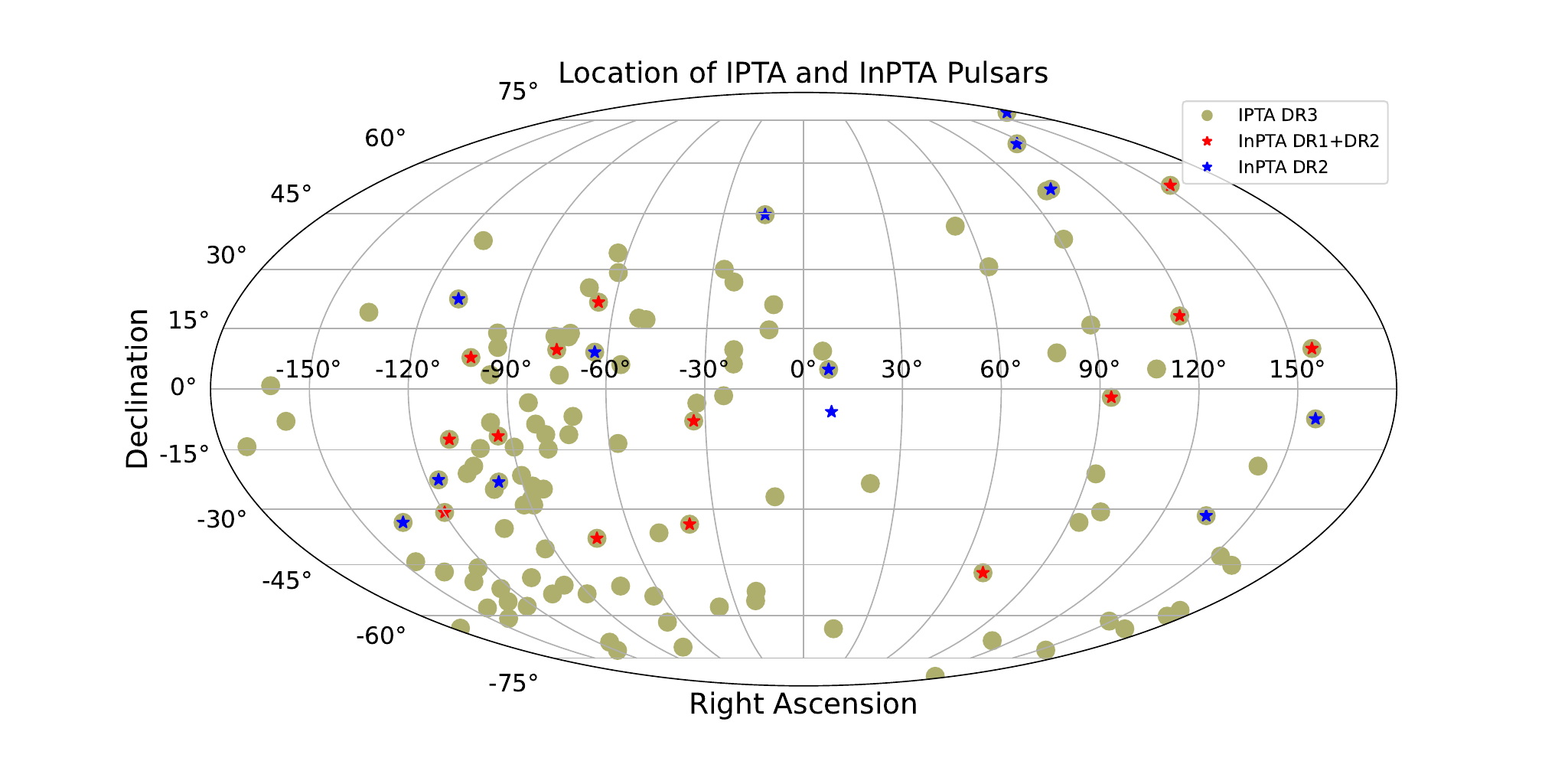}
    \caption{The sky distribution for 27 pulsars included in this data release is shown, marked by red and blue stars, representing observations made with the InPTA experiment between November 2016 and March 2024. 14 pulsars indicated by red stars were part of the InPTA DR1, whereas pulsars marked by blue stars are added in the present data release along with 14 InPTA DR1 pulsars. Green circles indicate pulsars that are planned to be included in the upcoming third data release of IPTA.}
    \label{fig:skymap}
\end{figure*}

\begin{figure*}[h!]
    \centering
    \includegraphics[width=1.08\textwidth]{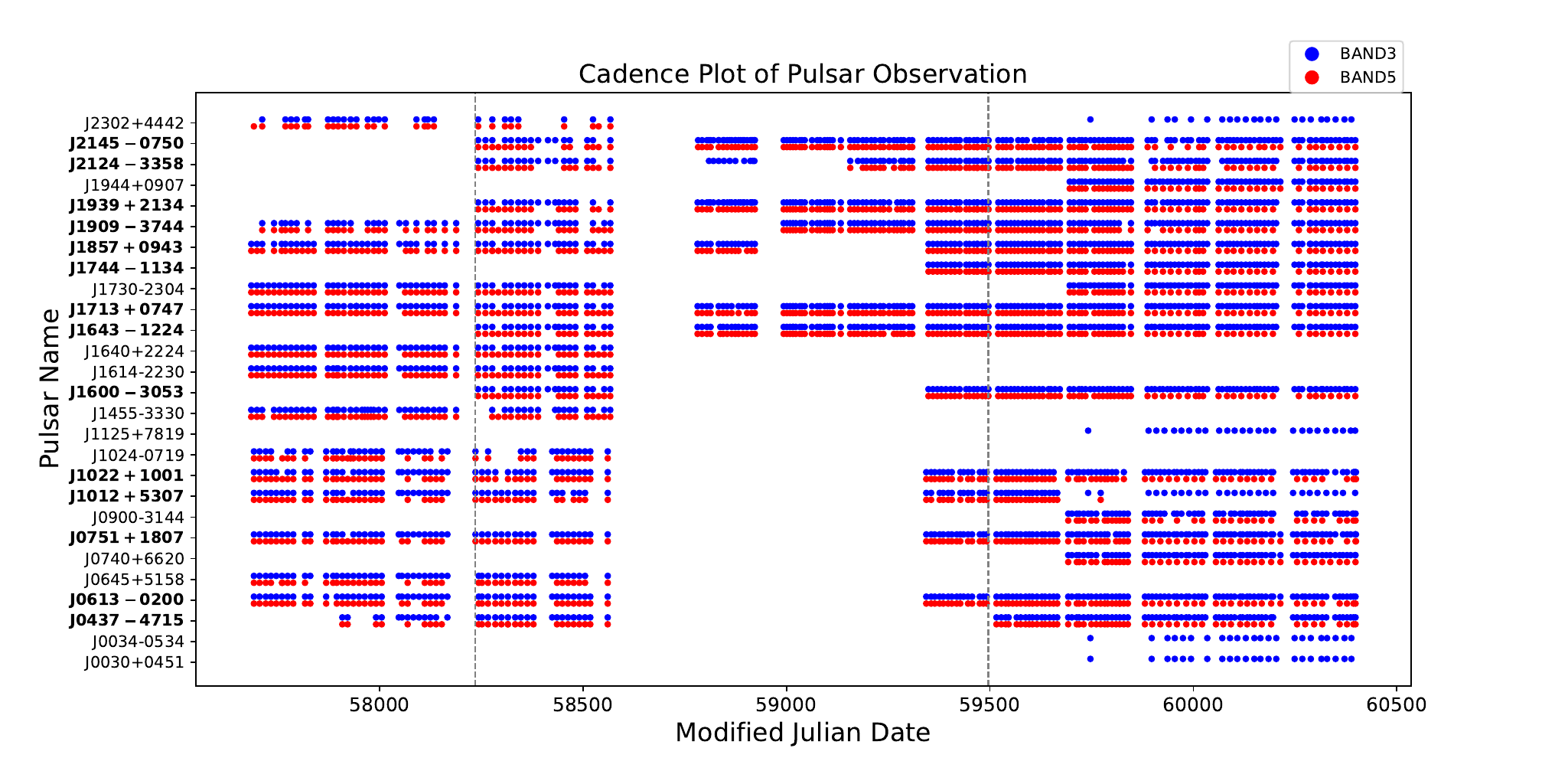}
    \caption{The observation cadence for 27 pulsars included in the present data release is shown across a range of Modified Julian Dates (MJD). Most of these pulsars were observed concurrently in band 3 (blue circles) and band 5 (red circles) of the uGMRT as part of the InPTA experiment. 14 pulsars highlighted in bold were also part of the InPTA DR1 \citep{tnr+2022}, and the vertical dashed lines indicate the time span of data that was included in InPTA DR1 for these pulsars. Starting from MJD 59881 (observation cycle 43 of the uGMRT), the InPTA experiment adopted a new observation strategy, alternating between dual-band (band 3 and band 5) and single-band (band 3) observation modes, effectively reducing the band 5 observation cadence by half. PSRs J0030+0451, J0034$-$0534, J1125+7819, J1012+5307, and J2302+4442 were observed only in the single-band (band 3) configuration during this phase. Additionally, an experimental single-band (band 3) observation for all pulsars was conducted for the first time on MJD 59746 (observation cycle 42). PSRs J0740+6620, J0900-3144, and J1944+0907 were also recently added in the sample list of the InPTA experiment in band 3 + band 5 observation mode. For PSR J2302+4442, band 5 data products from observation cycles 31-35 were excluded from the present data release, see section \ref{sec:discussion} for details. }
    \label{fig:cadence}
\end{figure*}

Table \ref{tab:observations} provides a summary of the typical observation settings, such as the number of frequency channels and sampling time, used across various observation cycles of the uGMRT included in InPTA DR2, with details of non-standard epochs noted in the Table caption. Note that we exclude the band 4 observations taken during observation cycles 31-35 in this data release, as will be justified in section \ref{sec4.1}. A sky distribution of pulsars included in this work is shown in Figure \ref{fig:skymap} along with the pulsars that are planned to be included in the upcoming IPTA DR3. The cadence of InPTA observations for 27 pulsars included in InPTA DR2 is shown in Figure \ref{fig:cadence} where 14 pulsars which were part of InPTA DR1 are highlighted and the time span of InPTA DR1 dataset is marked by vertical dashed lines.

\begin{table*}[!ht]
\begin{tabular}{|c|c|c|c|c|c|c|c|c|}
\hline
\hline
\textbf{\begin{tabular}[c]{@{}c@{}}Observation\\Cycle\end{tabular}} &
\textbf{\begin{tabular}[c]{@{}c@{}}No. of\\PSRs\end{tabular}} &
\textbf{\begin{tabular}[c]{@{}c@{}}MJD \\ Start\end{tabular}} & \textbf{\begin{tabular}[c]{@{}c@{}}MJD\\ End\end{tabular}} & \textbf{\begin{tabular}[c]{@{}c@{}}Band\\ No\end{tabular}} & \textbf{\begin{tabular}[c]{@{}c@{}}Frequency\\ Band (MHz)\end{tabular}} & \textbf{\begin{tabular}[c]{@{}c@{}}Number of \\ channels\end{tabular}} & \textbf{\begin{tabular}[c]{@{}c@{}}Sampling\\ time ($\mu$s)\end{tabular}} & \textbf{\begin{tabular}[c]{@{}c@{}}Coherent\\ Dedispersion\end{tabular}} \\
\hline
                             & 16 &     57684             &     57838                     &  3   &  300$-$500    &  2048  &  81.92  &  No   \\ \cline{5-9} 
\multirow{-2}{*}{ 31}      & \multirow{-2}{*}{}    & \multirow{-2}{*}{}       & &  5  &  1260$-$1460   &  2048   & 81.92  &  No   \\
\hline
                             & 17 &      57868               &    58012                      & 3   & 300$-$500     & 2048   & 81.92  & No   \\ \cline{5-9} 
                            &                       & &                          & 4   & 550$-$750     & 2048   & 81.92  & No    \\ \cline{5-9} 
\multirow{-3}{*}{32}        & \multirow{-3}{*}{}    & \multirow{-3}{*}{}       & & 5   & 1260$-$1460   & 2048   & 81.92  & Yes   \\
\hline
                             & 19 &            58047         &    58188                      & 3   & 300$-$500     & 2048   & 81.92  & No   \\ \cline{5-9} 
                            &                       & &                          & 4   & 550$-$750     & 2048   & 81.92  & No    \\ \cline{5-9} 
\multirow{-3}{*}{33}        & \multirow{-3}{*}{}    & \multirow{-3}{*}{}       & & 5   & 1260$-$1460   & 2048   & 81.92  & Yes   \\
\hline
                             & 22 &    58235                   &    58389                      & 3   & 400$-$500     & 1024   & 81.92  & Yes   \\ \cline{5-9} 
                            &                       & &                          & 4   & 650$-$750     & 1024   & 81.92  & No    \\ \cline{5-9} 
\multirow{-3}{*}{34}        & \multirow{-3}{*}{}    & \multirow{-3}{*}{}       & & 5   & 1360$-$1460   & 1024   & 81.92  & Yes   \\ \hline
                             & 22 &  58413                     &      58524                    & 3   & 400$-$500     & 1024   & 81.92  & Yes   \\ \cline{5-9} 
                            &                       & &                          & 4   & 650$-$750     & 1024   & 81.92  & No    \\ \cline{5-9} 
\multirow{-3}{*}{35}        & \multirow{-3}{*}{}    & \multirow{-3}{*}{}       & & 5   & 1360$-$1460   & 1024   & 81.92  & Yes   \\ \hline
                             & 6 &  58781                     &     58922                     & 3   & 300$-$500     & 512    & 20.48  & Yes   \\ \cline{5-9} 
\multirow{-2}{*}{37}        & \multirow{-2}{*}{}    & \multirow{-2}{*}{}       & & 5   & 1260$-$1460   & 1024   & 40.96  & No    \\ \hline
                             & 5 &    58990                   &    59133                      & 3   & 300$-$500     &   512     & 20.48       & Yes   \\ \cline{5-9} 
\multirow{-2}{*}{38}        & \multirow{-2}{*}{}    & \multirow{-2}{*}{}       & & 5   & 1260$-$1460   &    1024    &   40.96     & No    \\ \hline
                             & 6 &  59156                     &      59309                    & 3   & 300$-$500     &    256    &  10.24      & Yes   \\ \cline{5-9} 
\multirow{-2}{*}{39}        & \multirow{-2}{*}{}    & \multirow{-2}{*}{}       & & 5   & 1260$-$1460   &    1024    &   40.96     & No    \\ 
\hline
                             & 13 &   59343                    &     59496                     & 3   & 300$-$500     &    128    &    5.12    & Yes   \\ \cline{5-9} 
\multirow{-2}{*}{40}     & \multirow{-2}{*}{}    & \multirow{-2}{*}{}       & & 5   & 1260$-$1460   &    1024    &   40.96     & No    \\ 
\hline
                             & 14 &        59516             &      59672                  & 3   & 300$-$500     &    128    &    5.12    & Yes   \\ \cline{5-9} 
\multirow{-2}{*}{41}     & \multirow{-2}{*}{}    & \multirow{-2}{*}{}       & & 5   & 1260$-$1460   &    1024    &   40.96     & No    \\ 
\hline
                             & 22 &     59692, 59881, 60055            &       59846, 60034, 60203                & 3   & 300$-$500     &    128    &    5.12    & Yes   \\ \cline{5-9} 
\multirow{-2}{*}{42, 43, 44}     & \multirow{-2}{*}{}    & \multirow{-2}{*}{}       & & 5   & 1260$-$1460   &    1024    &   40.96     & No    \\ 
\hline
                             & 28 &     60245             &       60399                 & 3   & 300$-$500     &    128    &    5.12    & Yes   \\ \cline{5-9} 
\multirow{-2}{*}{45}     & \multirow{-2}{*}{}    & \multirow{-2}{*}{}       & & 5   & 1260$-$1460   &    1024    &   40.96     & No    \\ 
\hline

\end{tabular}
\caption{The observation settings for simultaneous multi-band InPTA observations using multiple sub-arrays, with data recorded through the GWB backend. A correction of +10 kHz was applied for MJDs between 59217 and 59424 (cycles 39–40) to account for an offset in local oscillator frequencies at the observatory during this period. Observations from cycles 31 to 33 were conducted during the early phase of the upgraded GMRT, when various combinations of observational settings were still being tested. PSRs observed on MJDs 58413 and 58431 (cycle 35) used a bandwidth of 200 MHz, though the standard bandwidth for cycles 34 and 35 was 100 MHz. Band 5 data between MJDs 58411 and 58436 (cycle 35) were recorded without coherent dedispersion. Non-standard observations were conducted on MJDs 59376 and 59380 (cycle 40) with the Polyphase Filterbank (PFB) setting enabled, although PFB is generally turned off for our observations. Band 4 data set from Cycles 32 to 35 is not included in the present data release. }
\label{tab:observations}
\end{table*}


\section{Data reduction} \label{sec:datareduction} 
The first step in processing uGMRT data for InPTA is converting it into a standard pulsar archive format~\citep{hsm+2004} which is widely used across the pulsar community. To ensure consistency in PTA analysis and prevent systematic errors, a uniform data reduction process is critical. In our earlier work~\citep{smj+2021}, we introduced the ``Pipeline for the Indian Pulsar Timing Array" (\pinta{}), designed to streamline and standardize data reduction, facilitating its use directly at the observatory to minimize data transfer and enhance efficiency.

The InPTA observations are conducted at the uGMRT in pulsar total intensity mode. These observations are recorded by the GMRT Wideband Backend \citep[GWB:][]{rkg+2017} as a frequency-resolved time series in a raw data format described in \citet{smj+2021}. 
These data are further processed offline using the \pinta{} pipeline which performs RFI mitigation and partial folding using \rficlean{} \citep{mlv+2021} and \dspsr{} \citep{sb2011} packages respectively, producing \psrfits{} archives \citep{hsm+2004}. The data were folded in the early days using the timing solutions published in the first IPTA data release \citep{vlh+2016}. Subsequently, pulsar ephemerides from either our own updated solutions or the subsequent IPTA data release 2 \citep{pdd+2019} were used. Since 2022, we have been using our own updated solutions from the InPTA DR1 or from the EPTA+InPTA data release \citep{tnr+2022,epta2023,epta+inpta2023c}\textcolor{black}.
\pinta{} also performs the backend delay corrections as listed in \citet{rks+2021} by updating the \texttt{be:delay} field in the \psrfits{} header \citep{tnr+2022} as well as by populating the metadata such as the time-stamp in the header. Typically, data were partially 
folded with full frequency resolution (indicated in Table \ref{tab:observations}) to several 10 s sub-integrations. We currently utilize a fully automated pipeline integrated with \pinta{} for data reduction, which standardizes the process with a single command, reduces the potential for human error and simplifies the workflow. Additionally, another automated pipeline integrated with \dmcalc{}\footnote{https://github.com/inpta/DMCalc-upgrade-22Oct2024} \citep{kmj+2021} provides the initial estimates of ToAs and DM after each epoch’s data reduction. This enables rapid detection of any significant variations, such as those caused by solar activity \citep{yhc+2007,tsb+2021,sct+2024}  or astrophysical events (for example, DM events in PSR J1713+0747 \citep{jml+2017,leg+2018}), ensuring timely insights.

At early stage of InPTA observations (observation cycles 31-33), the rawdata was reduced using the \texttt{PRESTO} pipeline \citep{rem2002,rce2003}. Later, the data reduction process was transitioned to our indigenously developed \pinta{} pipeline. As a part of this work, we reprocessed the data set of observation cycles 31 to 33 (MJD 57684 to 58188) with \pinta{} to maintain conformity across the reduced archive files. To ensure consistency, a comparison was undertaken between the results produced by \texttt{PRESTO} and \pinta{}. This involved cross-checking the signal-to-noise ratios (S/N) and examining the pulse profiles obtained from both the pipelines to verify pulsar detection. We found consistency between the two pipelines without any significant discrepancies.

 These \psrfits{} archives are further processed by a pre-processing script\footnote{\texttt{dr2\_preprocess.sh} - this is included with the data release} to convert the data to a standard form for the entire analysis of the InPTA DR2. This script performs the following tasks: (i) reversing the frequency order from the lowest frequency to the highest frequency\footnote{The uGMRT data can have different sideband, upper sideband (USB) or lower sideband (LSB), based on different observing settings. We have chosen a standard frequency order for the InPTA data to keep consistency. }  and (ii) time-collapsing all the sub-integrations. (iii) The pre-processing script also corrects the frequency labels by adding 10 kHz to the central frequencies for MJDs between 59217 and 59424. This is to account for a systematic offset of 10 kHz in the local oscillator system at uGMRT that was identified in the observations between MJD 59217 to 59424 \citep{tnr+2022}. These steps help ensure a consistent data formatting and accuracy.

\section{Data Analysis and Processing Methodology} \label{sec4}

In this section, we discuss the procedures adopted for template generation, selection of optimum number of sub-bands used in each band for each pulsar, estimating the fiducial DM, and the estimation of ToAs and DMs from our observations. 

\subsection{Template generation} \label{sec4.1}
The precise estimation of the ToAs and DMs of a pulsar at low radio frequencies is challenging for a few reasons. 
The frequency-dependent profile shape evolution, which is most prominent at low frequencies, could be strongly covariant with the DM, and this makes the estimation of the true DM of a pulsar through timing practically impossible. 
The DM of a pulsar varies over time, typically in the range 10$^{-3}$-10$^{-5}$ pc cm$^{-3}$, due to the pulsar's relative motion to the observer, the influence of the variable solar wind, and the dynamic nature of IISM \citep{dvt+2020, tnr+2022}.
If uncorrected, this DM variability will cause smearing of the pulse shape while frequency-collapsing a profile, apart from introducing unmodeled epoch-dependent delays varying with 
the observing frequency. 
Another limitation arises from the inhomogeneities in electron density along the line of sight, which leads to multi-path propagation of radio waves, resulting in pulse broadening which is stronger at low frequencies \citep{rickett1977,sjk+2024}.
To minimize the effects due to these limitations, we follow a careful approach for the generation of noise-free templates in each frequency band for each pulsar based on \citet{tnr+2022}
with certain modifications as described below:
\begin{itemize}
    \item \textit{Selection of the template epoch}: We first identify a high-S/N observation  (template epoch) without any artifacts in both band 3 and band 5 for each pulsar preferably from the recent observations done between April 2022 to September 2024, MJD 59692 to 60582 (observation cycles 42-46). We ensure that the same epoch is used to make templates in both bands to take advantage of our concurrent observations. The following five pulsars, J0645+5158, J1024$-$0719, J1455$-$3330, J1614$-$2230, J1640+2224, were not observed after MJD 58524 (observation cycle 35) as a part of the InPTA experiment. Hence the template epochs for these pulsars were chosen from the earlier days of uGMRT observations with 200 MHz bandwidth (observation cycles 31-33, see Table \ref{tab:observations}). We construct the templates utilising the full frequency resolution of the band 3 and 5 profiles from the selected high-S/N template epochs. The current data release does not include the band 4 dataset for two main reasons: (i) identifying a high-S/N template epoch concurrent across all three bands is challenging, and (ii) only two years of band 4 data (from April 2017 to February 2019) is available. Therefore, prioritizing the identification of a high-S/N template epoch concurrent in bands 3 and 5 was considered more beneficial than including the limited band 4 dataset.

    \item \textit{Preliminary alignment}: As described in \citet{kmj+2021} and \citet{tnr+2022}, the DM values estimated using a template can show a constant offset from the pulsar's true DM, and this depends on the fiducial DM used to dedisperse the template itself. To avoid such offsets between different bands in our dataset, we align the frequency-resolved template profiles across both bands using the same fiducial DM (choosing the same epoch for template generation across both bands 3 and 5 also ensures the validity of this procedure). This alignment is initially performed using the method described by \citet{kmj+2021}, where the initial fiducial DM is estimated from the band 3 profile using the \pdmp{} command of \psrchive{}\footnote{Version $-$ 2021-06-03+ }.
    \item \textit{Band equalising}: After the preliminary alignment using the \pdmp{} DM obtained from band 3 archive of the template epoch, we correct for the shape of the bandpass used to record the data, thus equalizing the signal power across all frequency channels in each band. To do this, we equalise the off-pulse root-mean-square (RMS) signal in all channels and this is done by giving weights to the signal in each sub-band based on the signal amplitude therein, see Figure \ref{fig:bandequalising} for instance. This step was not part of the template generation method used in the InPTA DR1 \citep{tnr+2022}.

    \begin{figure*}
    \centering
    \includegraphics[width=0.86\linewidth]{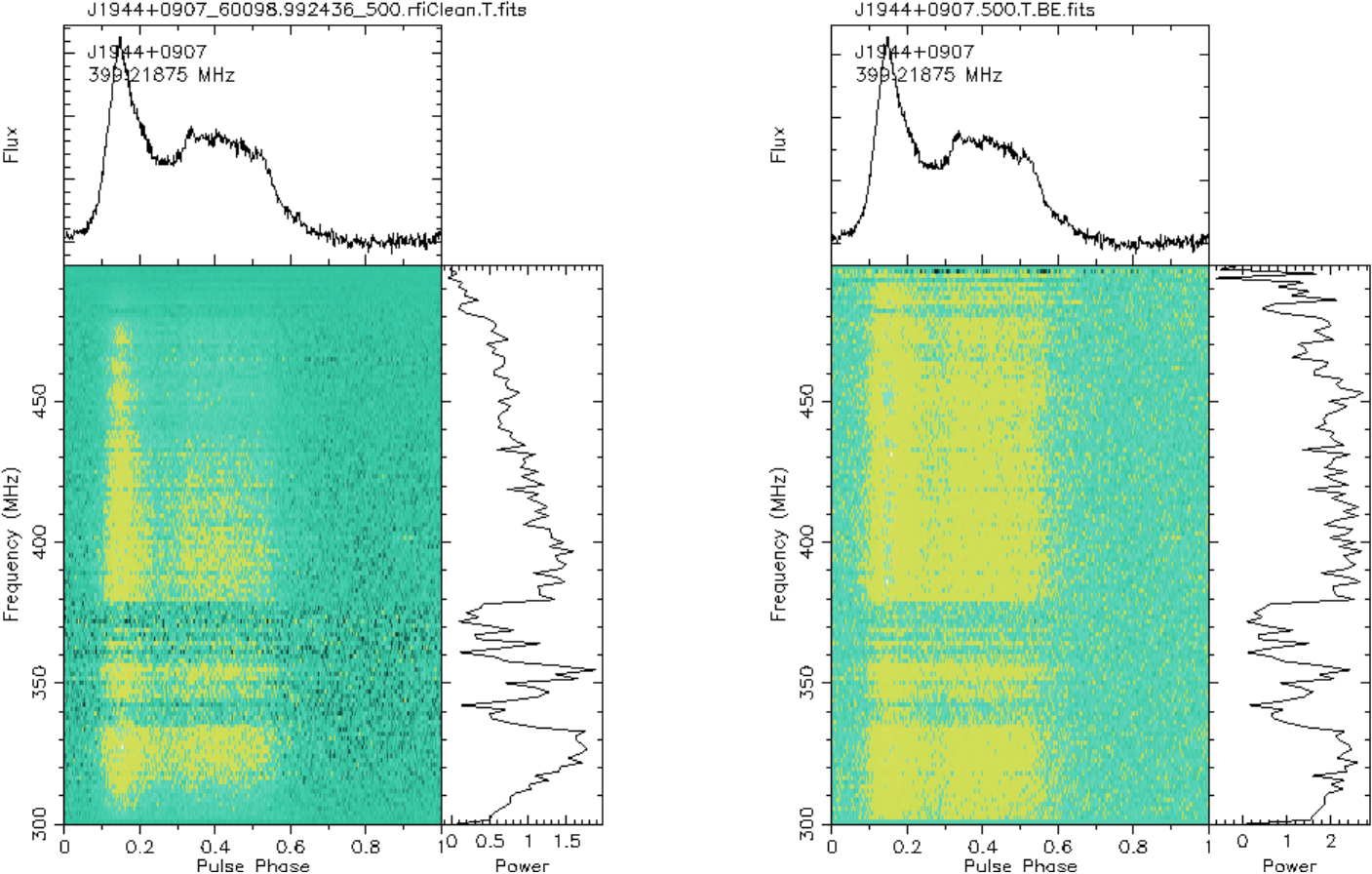}
    \caption{ The intensity in arbitrary units is shown as a function of observing frequency and pulse phase for PSR J1944+0907 using band 3 of the uGMRT. The right plot displays the band-equalized pulse profile from the template epoch, which was used to generate the template for this pulsar. In contrast, the left plot shows the same pulse profile without band equalization, where the effect of the bandpass shape is clearly visible, causing the signal to appear brighter at lower frequencies and fainter at higher frequencies. We equalize the off-pulse RMS signal  across all channels to correct for the effects of the bandshape, while preserving the pulsar's intrinsic spectral features. }
    \label{fig:bandequalising}
\end{figure*}
    
    \item \textit{Wavelet Smoothing}: The noise-reduced and frequency-resolved templates for each band are generated using an optimal wavelet smoothing algorithm, implemented in \psrchive{} through the \texttt{psrsmooth} command \citep{CD1995}, from the band-equalized profiles. We tested various wavelets from the “undecimate” version, namely  the UB and UD series, of wavelets provided in the smoothing algorithm of the psrchive/psrsmooth routine. The “undecimate” version is translation-invariant, with the factor indicating the number of coefficients used. For each pulsar and band, we selected a wavelet that (i) preserves the pulse profile's features and shape (verified visually) and (ii) provides a reasonably good S/N for the smoothed template (though not necessarily the highest). In our analysis, we find that the chosen wavelet is unique for each pulsar in each band.

\end{itemize}

Finally, after making high-S/N templates with reduced noise, in comparison to noise observed in the corresponding template epoch archive, using the aforementioned method, we dedisperse frequency-resolved templates of both band 3 and band 5 using the fiducial DM that is estimated using an iterative procedure described in section \ref{sec4.3}.

\subsection{Selecting optimum number of sub-bands} \label{sec4.2}
Since pulsar radio data is recorded with a finite frequency resolution given by the number of sub-bands in the profile (as discussed in section \ref{sec:obs}), the original high-frequency-resolution profiles may not have good enough S/N in each sub-band to estimate ToAs of sufficiently high precision. Hence, we have devised a method to estimate the optimal number of sub-bands to partially frequency-scrunch the data in each band of each pulsar ensuring enough S/N per sub-band and also the reduction of radio frequency-dependent profile shape evolution effects. This optimal selection is guided by two key constraints: (i) the signal in each sub-band must be strong enough, (ii) there should be minimal frequency-dependent profile shape evolution across adjacent sub-bands in both band 3 and band 5. Since each pulsar has its unique intrinsic brightness and profile shape evolution with frequency, this approach must be tailored to the specific characteristics of each pulsar, requiring close attention to the individual pulsar behavior.

We briefly explain our procedure of optimal selection of sub-bands below and further details are provided in \ref{sec:appendix-subbands}. 

\begin{itemize}
  \item   \textit{Selecting three different epochs:} We begin by selecting data archives from three kinds of epochs for each band (band 3 and band 5) and each pulsar: (a) the epoch that was used to create the templates (template epoch), (b) the epoch having the highest recorded S/N, and (c) the epoch with S/N approximately equal to the median S/N of the pulsar in that band. This approach ensures that the estimated number of sub-bands is not biased exclusively towards high-S/N epochs. The following steps are then applied to all three epochs. 

  \item   \textit{Initial guess of sub-bands:}  To estimate the initial number of sub-bands for each band, we aim to ensure that the S/N per sub-band is similar for both band 3 and band 5. This is done by dividing the integrated S/N by the square root of the number of sub-bands (see equation \ref{eqAppendix1} in \ref{sec:appendix-subbands}). A typical S/N cutoff of 20 per sub-band is applied, although for very faint pulsars, a lower cutoff of 15 may be acceptable. 

  \item  \textit{Equivalent ToA precision in both bands:} In addition to the criteria explained in the previous step, we also ensure that the median ToA precision is optimal and consistent across both bands for the selected number of sub-bands. To achieve this, we first time-scrunch the archive for each selected epoch into three sub-integrations in time, then frequency-scrunch it into the initial number of sub-bands (determined in the previous step) using the \texttt{pam} command in \psrchive. Next, we generate frequency-resolved ToAs for each sub-integration by cross-correlating with the template (that was generated using the procedure described in section \ref{sec4.1}) using the \texttt{pat} command in \psrchive. Finally, the number of sub-bands are tuned through this iterative procedure by ensuring that the median ToA precision is both equivalent and optimized in each band, while maintaining sufficient S/N per sub-band (as verified in the previous iteration). 

  \item  \textit{Accounting for radio frequency-dependent profile shape evolution:}  In the final iteration, we focus on accounting for radio frequency-dependent profile shape evolution, with the primary objective of minimizing profile shape variation between adjacent sub-bands. The number of sub-bands determined in the previous iteration, denoted as $N_0$, serves as the starting point in the current iteration. To avoid any bias from profile smearing caused by interstellar dispersion, we first dedisperse the pulse profiles of each epoch and each band using a DM that is estimated by fitting the band 3 ToAs of the corresponding epoch (referred to as the band 3 DM of that epoch, this DM is obtained using the method and pipeline described in section \ref{sec4.4}). Additionally, the pulse profiles from all three selected epochs are subjected to bandshape equalization, as outlined in section \ref{sec4.1}. We then calculate the difference in pulse profiles between any two adjacent sub-bands, referred to as the profile residuals, $p_i$, where $i$ ranges from 1 to ($N-1$) with $N$ being the number of sub-bands used. The profile residuals are normalized by the peak signal level, and sub-bands potentially contaminated by RFI are eliminated based on the criteria outlined in \ref{sec:appendix-subbands}. These profile residuals are obtained by varying the number of sub-bands between $N=2$ and $N_0$. The profile residuals are then subjected to the Anderson-Darling test \citep[][]{ad1952,ad1954} for Gaussianity and also visually checked, with the goal of achieving Gaussian residuals that indicate negligible profile evolution across adjacent sub-bands.  Finally, we chose the optimal number of subbands within the range $\mathbf{N_0 \geq N \geq 2}$ such that the profile residuals are all Gaussian based on an analysis of all three epochs selected in the first step for each band. This is to make sure that the optimum number of sub-bands estimate is robust.  
\end{itemize}

 The above procedure was applied separately to datasets recorded with 100 MHz and 200 MHz observing bandwidths. The final and optimum number of sub-bands of each pulsar for each band and observing bandwidth are listed in Table \ref{tab:subband}.

\begin{table*}[!ht]
\centering
\begin{tabular}{|c|cc|cc|c|cc|cc|}
\hline
\textbf{Pulsar} & \multicolumn{2}{c}{\textbf{Band 3  }} & \multicolumn{2}{c|}{\textbf{Band 5  }} & \textbf{Pulsar} & \multicolumn{2}{c}{\textbf{Band 3  }} & \multicolumn{2}{c|}{\textbf{Band 5 }}\\
\cline{2-5} \cline{7-10}
 & \textbf{100 MHz} & \textbf{200 MHz} & \textbf{100 MHz} & \textbf{200 MHz} & & \textbf{100 MHz} & \textbf{200 MHz} & \textbf{100 MHz} & \textbf{200 MHz}\\
\hline
\hline
J0030+0451 &  --  &  32  & -- &  -- & J1614$-$2230  &  2  &  4  &  2  &   4  \\
J0034$-$0534 & --  & 32  & -- & -- & J1640+2224 &  2   & 4  &   4  &  4  \\
J0437$-$4715 &  64  &  128  & 16  & 32  & J1643$-$1224  &  8  & 32   &  4  & 8  \\
J0613$-$0200 &  32  &  64 &  2 &  2  & J1713+0747  &  8   &  16   &  2  &  4 \\
J0645+5158 &  8  &  16 &  4 & 8  & J1730$-$2304  & 8  & 64  &  2 & 8  \\
J0740+6620 & --  &  8 &  --  &  4  &  J1744$-$1134 &  --  & 32  & --  & 8  \\
J0751+1807 &  8  & 16  &  2   & 2  &  J1857+0943 &  8  &  16  &  2   &   2  \\
J0900$-$3144$^{*}$ &  --  & --    & --  &  4  &  J1909$-$3744 & 8   &  32  &  2  &  2 \\
J1012+5307 &  16  &  128  &  2  &  4  &  J1939+2134  &  64  & 128  &  2  &  4 \\
J1022+1001 & 16  & 32  &  4  &  8  &  J1944+0907 &  --  &  16  & --  &  4  \\
J1024$-$0719 & 4   &  4  & 2   &   4  &  J2124$-$3358 &  32  &  64  &  2 &  4  \\
J1125+7819 & -- &  8   & -- & -- &  J2145$-$0750 &  4  &  32   &   4  &  8 \\
J1455$-$3330 & 4  &  8   &  2  &  4 & J2302+4442  &  4  &  16  & -- & -- \\
J1600$-$3053 & 2  &  8 & 2  &  4  &   & & & & \\

\hline
\end{tabular}
\footnotetext{$^{*}$ We have included only band 5 data for PSR J0900$-$3144 in the present data release.}
\caption{Pulsar names along with their optimum number of selected sub-bands are provided for both band 3 and band 5 using the method described in Section \ref{sec4.2}. The sub-bands were chosen separately for datasets recorded with 100 MHz (MJD 58235 to 58524) and 200 MHz observing bandwidths to account for the dependence of SNR of pulsar detection on observing bandwidth and the pulse-profile shape evolution across the band. PSRs J0030+0451, J0034$-$0534, J1125+7819, and J2302+4442 were recently added in the InPTA's pulsar sample in the band3-only observation mode (see Section \ref{sec:obs}). PSRs J0030+0451, J0034$-$0534, J0740+6620, J0900$-$3144, J1125+7819, J1744$-$1134, and J1944+0907 were not observed between MJD range 58235 to 58524 (observing cycles 34 and 35 of the uGMRT) when pulsar data was recorded with 100 MHz observing bandwidth.}
\label{tab:subband}
\end{table*}

\subsection{Estimating the Fiducial DM} \label{sec4.3} 
We use an iterative procedure, as described in \cite{tnr+2022}
,  to estimate the fiducial DM for each pulsar. First, the band 3 template epoch archive is dedispersed with an initial value of the \pdmp{} DM. It should be noted that we are using the archive of the template epoch, which was used for making the template (refer to section \ref{sec4.1}), and not the template itself. This distinction ensures that our analysis is based on the original observational data associated with that epoch. The band 3 template epoch archive (obtained using the procedure described in section \ref{sec4.1}) is time-scrunched into three sub-integrations and frequency-scrunched to the optimum number of sub-bands, selected using the procedure described in section \ref{sec4.2}, using the \texttt{pam} command of \psrchive. Finally, the frequency-resolved ToAs are estimated for all three sub-integrations by cross-correlating the band 3 sub-integrations with the frequency-resolved template, which is generated using the method described in section \ref{sec4.1}, using the \texttt{pat} command of \psrchive. We then fit a DM to these band 3 ToA residuals to align the residuals in frequency using \tempotwo{} \citep{hem2006, ehm2006}. The DM fitted from aligning the band 3 residuals for the template epoch is then used to dedisperse both the band 3 and band 5 templates. 

In the next step, we leverage the unique capability of simultaneous dual-band observations with the uGMRT. We have concurrent band 3 and band 5 template epochs, from which we construct the templates as detailed in section \ref{sec4.1}. Using these dedispersed, frequency-resolved templates, we generate concurrent sub-banded ToAs for both bands by cross-correlating them with the corresponding sub-integrations from the template epoch. A DM is then fitted to align the multi-band ToAs, providing tighter constraints on the DM value. This iteratively refined DM is subsequently used as the fiducial DM for the corresponding pulsar.

\subsection{Generation of ToAs and DM} \label{sec4.4}

Following template generation, sub-band optimization, and fiducial DM estimation, we proceed to run \dmcalc{} \citep{kmj+2021} on each pulsar for all the available epochs. \dmcalc{} requires the frequency-aligned templates, the optimum number of sub-bands for partially collapsing the templates and archives, and the processed archives themselves, in each band as inputs. It also needs an ephemeris file of the pulsar in order to run \tempotwo{} in order to fit epoch-wise DMs. We use ephemeris files from the most recent releases of EPTA+InPTA, NANOGrav, and PPTA as the starting points of our analysis \citep{epta+inpta2023c,aaa+2023a,zrk+2023}.
The parameter values of the ephemeris files are first rotated to a reference epoch at MJD 59000 using \tempotwo{}, aligning them with the midpoint of the InPTA DR2 time baseline. Initially, \dmcalc{} applies the fiducial DM from these ephemeris file to the header of the archives. This ensures that the estimations are relative to a uniform reference. Then, it cross-correlates each sub-band (estimated from sub-band optimization as explained earlier) of the archive with the corresponding sub-band of the template in the Fourier domain to obtain frequency-resolved or sub-banded ToAs for each epoch. These ToAs are estimated using cross-correlation with the standard template using Fourier-domain with Markov-chain Monte Carlo (FDM) method as implemented in PSRCHIVE \citep{hsm+2004} and briefly described in \cite{vlh+2016}. For this purpose, \dmcalc{} utilizes the Python interface of \psrchive{}.

\begin{figure*}[!ht]
    \centering
    \includegraphics[width=0.9\linewidth]{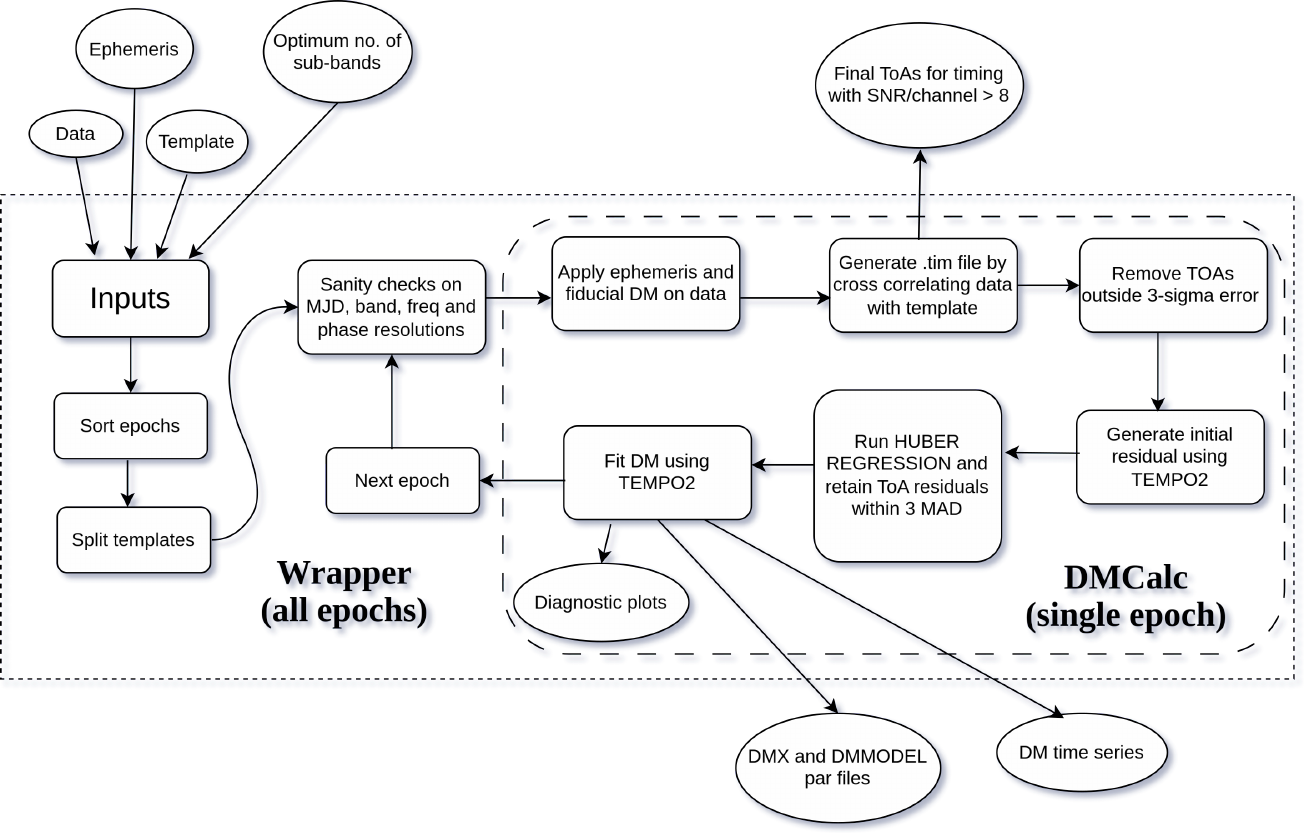}
     \caption{This workflow diagram illustrates the comprehensive process used to estimate ToA and DM using \dmcalc{} and the wrapper script. The wrapper applies the \dmcalc{} processing steps across all epochs, performing essential sanity checks. For each epoch, \dmcalc{} fits DM values, refines ToA residuals, and generates diagnostic plots. The final outputs include DMX and DMMODEL ephemeris files (DMX and DMOFF values are not piece-wise linear fits, but obtained from measured simultaneous epoch-wise DMs as explained in section \ref{sec4.4}), DM time series, and ToAs of high-S/N} epochs. 
    \label{fig:flowchart}
\end{figure*}

The next step involves using these ToAs for DM estimation. \dmcalc{} achieves an initial iterative flagging of the outlier ToAs for a given epoch which is done as follows. Firstly, ToAs with error bars greater than thrice the median ToA error bar are flagged. With the remaining ToAs and the input ephemeris, \dmcalc{} calculates the ToA residuals by invoking \tempotwo{}. Secondly, Huber regression method is employed \citep{hr2011} to fit a quadratic trend to the ToA residuals and remove any ToAs beyond $3\sigma$ of the median absolute deviation of the residuals. The remaining frequency-resolved filtered ToAs are again passed on to \tempotwo{} for fitting an average DM to the sub-banded ToAs for a given epoch with the following combinations -- (a) ToAs from the 300--500 MHz uGMRT band (band 3 or B3) alone, (b) ToAs generated from both the 300--500 MHz and 1260--1460 MHz (band 5 or B5) uGMRT bands. The resulting DM values -- B3 DMs from combination (a), and B3+5 DMs from combination (b) -- as elaborated above, their corresponding precisions, reduced chi-squared estimates, and the pre- and post-fit weighted RMS values are stored in a file along with appropriate internal flags to facilitate the separate identification of B3 and B3+5 DMs for further analysis. The minimum and median values of estimated B3 and B3+5 DM uncertainties for each pulsar are provided in Table \ref{tab:DMprecision}. A copy of the sub-banded ToAs generated for each epoch along with relevant flags, including IPTA-specific flags, necessary for timing and combination is appended to a final \texttt{{tim}} file. This final copy of ToAs is not flagged with the $3\sigma$ cutoffs on ToA error bars or Huber regression fits as mentioned above. Such outliers are carefully examined in the timing phase later. However, the final set of ToAs preserved for timing have an adhoc cutoff of signal-to-noise ratio per channel less than $8$, as also proposed in Appendix B of \cite{abb+2015}, as a preliminary filter for sub-bands with very poor or no signal.

\begin{figure*}
  \centering
    \includegraphics[width=0.97\linewidth]{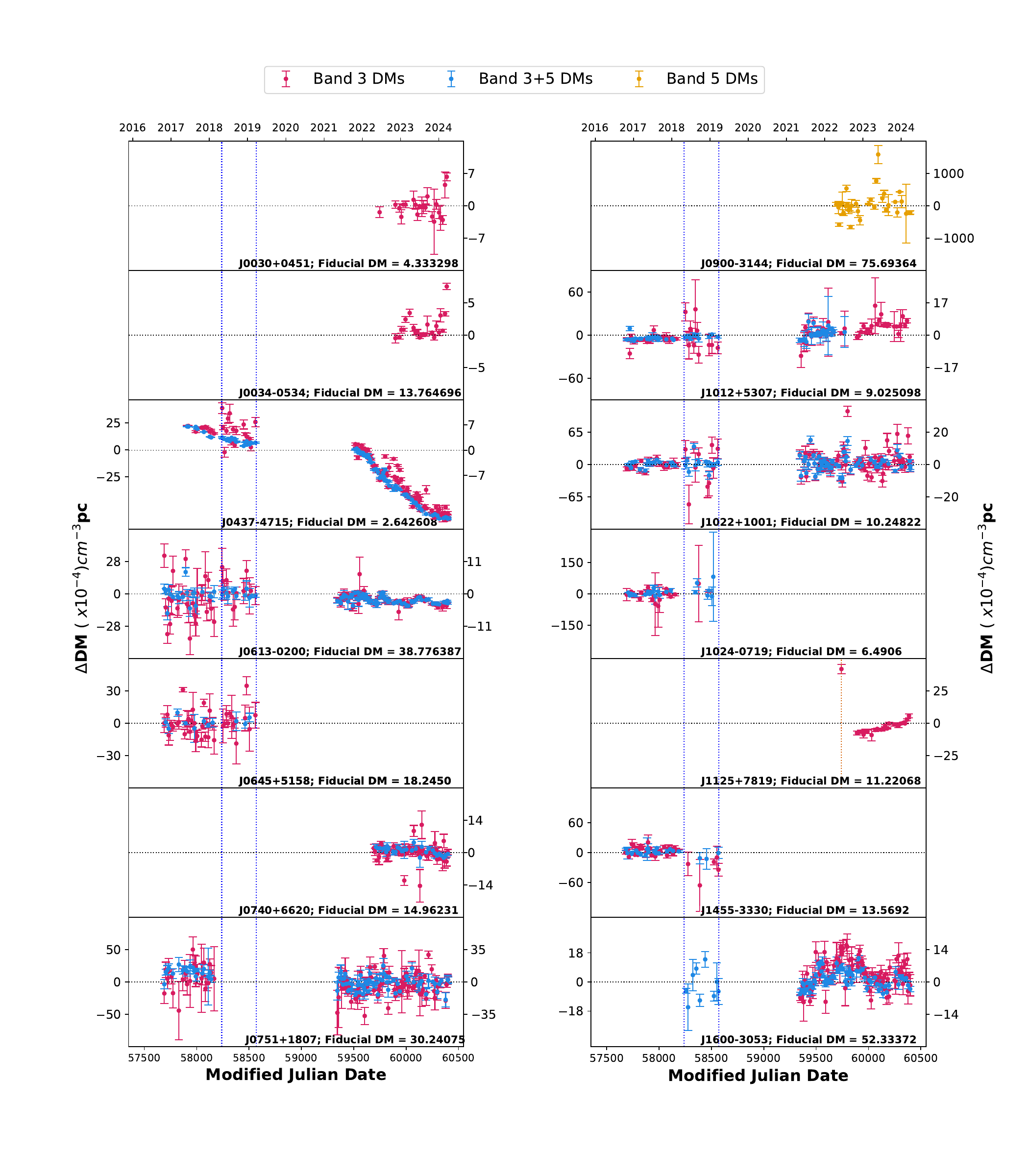}
     \caption{This Figure presents the DM time-series for 14 pulsars, showing the differences ($\Delta$DM, in units of $10^{-4}$ cm$^{-3}$ pc) between the fiducial DM and the DMs estimated using two approaches: (i) fitting ToAs from band 3 only (red points) and (ii) fitting ToAs from both band 3 and band 5 simultaneously (blue points). The method for DM estimation is described in detail in section \ref{sec4.4}. The two vertical lines at MJD 58239 and 58569 in each panel divide the DM time-series into three distinct sections: (i) the left section displays DM values derived from early uGMRT observations, recorded with a 200 MHz bandwidth, (ii) the middle section shows DM values estimated from data recorded with a reduced 100 MHz bandwidth during observation cycles 34 and 35 of uGMRT observations and (iii) the right section represents the DM estimates after the InPTA experiment optimized its observation strategy using 200 MHz bandwidth to produce higher-precision data. Consequently, DM estimates in this section show significantly improved precision. Hence, the vertical axes are scaled differently for epochs before and after MJD 58569 to reflect the improved DM precision achieved from cycle 37 onward. These outliers may represent significant scientific phenomena worthy of further investigations. In the special case of PSR J0900-3144, band 3 dataset is not included in the present data release, hence DMs were obtained by fitting band 5 ToAs. For PSRs J0030+0451, J0034$-$0534, and J1125+7819, only band 3 DMs are available, as these pulsars were added later when a new observation strategy of alternating between dual-band (band 3 and band 5) and single-band (band 3) configurations was adopted. We see a clear signature of solar wind adding excess DM with annual variation in the DM time series of PSR J0034-0534, as also seen in the DM time series shown in \cite{dvt+2020,tsb+2021}, and J0613$-$0200. We also see a sudden jump in DM value for PSR J1125+7819 at MJD 59741 which will be investigated in a separate work. }
    \label{fig: DMtimeseriesplot1}
\end{figure*}

\begin{figure*}
  \centering
    \includegraphics[width=1.0\linewidth]{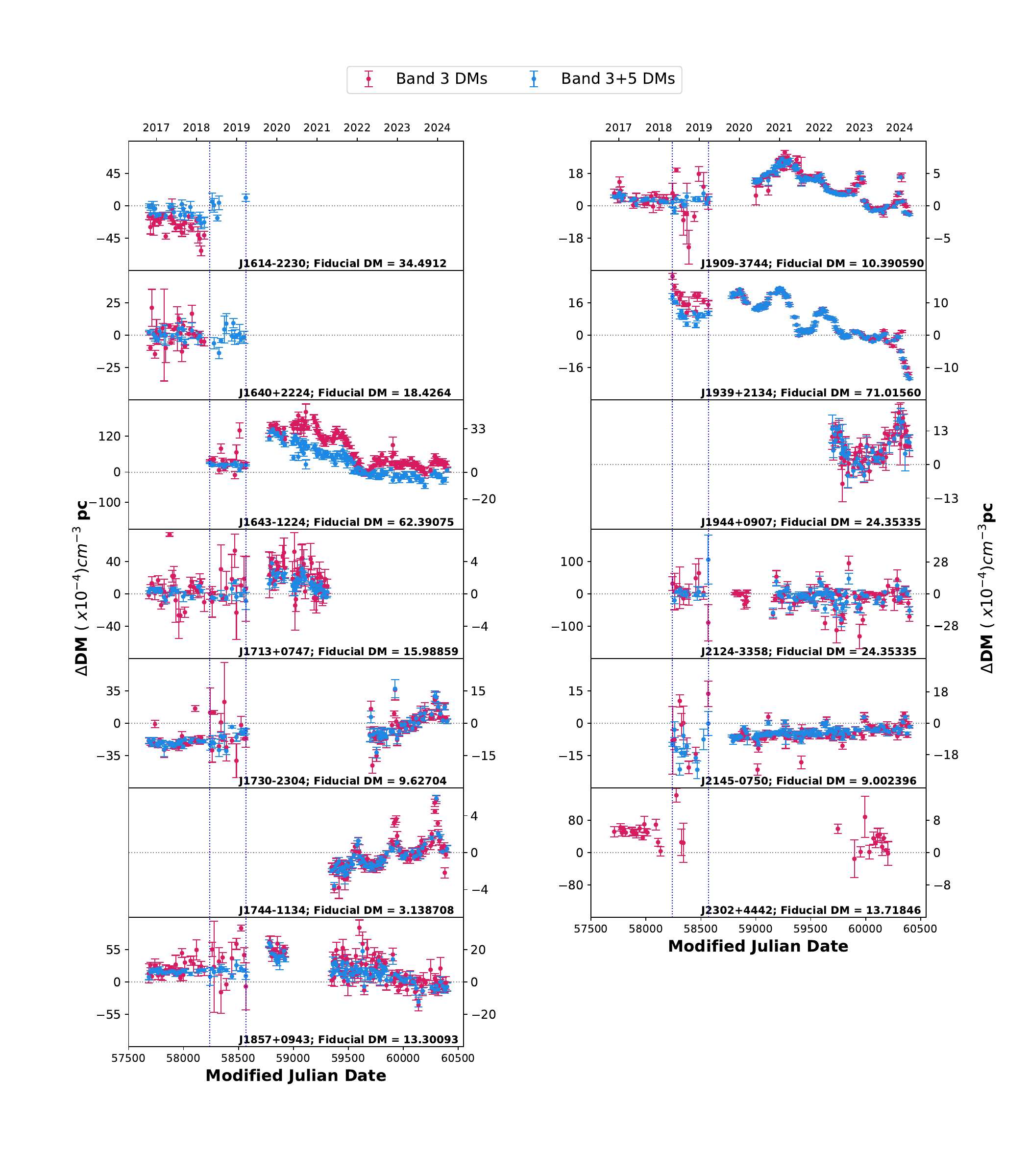}
     \caption{Same as in Fig. \ref{fig: DMtimeseriesplot1}, the DM time-series is shown for 13 pulsars. These outliers may represent significant scientific phenomena worthy of further investigations. PSR J2302+4442 has only band 3 DMs because we did not include band 5 data of this pulsar in this data release, see section \ref{sec:discussion} for details. We see a clear signature of solar wind adding excess DM with annual variation in the DM time series of PSRs J1744-1134 and J2145-0750, as also seen in the DM time series shown in \cite{dvt+2020,tsb+2021}, and also in the DM time series of PSR J1909-3744. The offset between band 3 and band 3+5 DMs for PSR J1643-1224 arises from a bias in the DM estimation caused by unmodeled scatter-broadening of the pulse profile dominant in band3 \citep{sjk+2024}. For PSR J1713+0747, we have included dataset (ToAs and DMs) before pulse-profile shape event, that is up to MJD 59309, in the present data release.}
    \label{fig: DMtimeseriesplot2}
\end{figure*}

This process is iterated over all epochs using a wrapper script around \dmcalc{} which prepares \dmcalc{} for each individual run by accomplishing the following preliminary processing of inputs -- (a) sorting the archives in the ascending order of the MJDs, (b) splitting the templates in each band with an original bandwidth of 200 MHz, into 100 MHz components, (c) selecting the relevant 100 MHz component for cross-correlation with 100 MHz bandwidth archives from some of the earlier uGMRT observation cycles (cycles 34-35), and (d) performing basic sanity checks related to the MJDs, central frequencies, fiducial DMs, frequency and phase resolutions, etc. of archives and templates before cross-correlation of the respective bands. After a complete run of \dmcalc{}, we obtain frequency-resolved ToAs, a DM time series, two ephemeris files -- a DMX par file containing DMX values, and a DMMODEL par file containing DMOFF values. It is important to mention here that the DMX or DMOFF values we use are \textit{not piece-wise linear fits over time bins covering several days}. Our DMX values represent the DM measured from simultaneous multi-band observations relative to a fiducial DM. \dmcalc{} calculates the difference between measured DM and fiducial DM, and includes them in a set of two new par files -- in one as DMX value over a single epoch, and in another as epoch-wise DMOFF parameters. Thus, the DMX and DMOFF values in our case incorporate actually measured simultaneous epoch-wise DMs and not piece-wise linear \tempotwo{} fits. This is conceptually different from the usual method of incorporating DMXs through \tempotwo{} fits, and hence should be distinguished carefully. Between these two methods (DMX and DMOFF) of DM time series incorporation in the par files, we have used the estimated DMX values to calculate timing residuals. The DMX values are estimated using more precise B3+5 DMs, unless band 5 ToAs are not available (for example, due to non-detection in band 5) in which case band 3 DM is used. These files are subsequently used for timing analysis, noise analysis, and further studies. A flowchart depicting the workflow of ToA and DM estimations using \dmcalc{} is provided in Figure \ref{fig:flowchart}. We obtain high-precision DMs using the procedure outlined above. Table \ref{tab:DMprecision} provides a summary of the median and minimum DM precisions achieved for each pulsar. The DM time series for all pulsars are illustrated in Figures \ref{fig: DMtimeseriesplot1} and \ref{fig: DMtimeseriesplot2}, and the related details are discussed in section \ref{sec:discussion}.

\begin{table*}[!ht]
\centering
\begin{tabular}{|c|cc|cc|c|cc|cc|}
\hline
\textbf{Pulsar} & \multicolumn{2}{c|}{\textbf{Band 3 }} & \multicolumn{2}{c|}{\textbf{Band 3+5 }} & \textbf{Pulsar} & \multicolumn{2}{c|}{\textbf{Band 3 }} & \multicolumn{2}{c|}{\textbf{Band 3+5 }} \\
\cline{2-5} \cline{7-10} 
 & \textbf{Median} & \textbf{Min} & \textbf{Median} & \textbf{Min} & & \textbf{Median} & \textbf{Min} & \textbf{Median} & \textbf{Min} \\
  & \multicolumn{2}{c|}{\textbf{$\times$ 10$^{-5}$ (pc cm$^{-3}$)}} & \multicolumn{2}{c|}{\textbf{$\times$ 10$^{-5}$ (pc cm$^{-3}$)}} &  & \multicolumn{2}{c|}{\textbf{$\times$ 10$^{-5}$ (pc cm$^{-3}$)}} & \multicolumn{2}{c|}{\textbf{$\times$ 10$^{-5}$ (pc cm$^{-3}$)}} \\
\hline
\hline
J0030+0451 & 11.3   & 1.1  & -- &  -- & J1614$-$2230  & 51.9   &  5.1  & 56.4  &   28.2 \\
J0034$-$0534 & 3.6 & 1.7 & -- & -- & J1640+2224 &  26.9  &  1.1  &  37.0  &  10.4  \\
J0437$-$4715 & 4.2  & 0.6  & 2.5  & 0.5 & J1643$-$1224  &  18.2  &  1.5  &  18.2  &  1.3 \\
J0613$-$0200 & 6.1  & 2.1   & 6.4   & 2.2   & J1713+0747  &  21.8  &  1.6  &  7.5  & 1.0 \\
J0645+5158 & 68.0  & 20.4 & 37.9  & 12.9  & J1730$-$2304  &  19.5  & 0.8  &  19.0 &  0.8  \\
J0740+6620 & 14.7  & 3.9  & 12.3  & 4.4  &  J1744$-$1134 &  4.1  &  0.8 &  3.1  & 0.5  \\
J0751+1807 & 75.7  & 3.0 &  51.3  & 2.8  &  J1857+0943 &  36.5 &  1.4  &  24.6  &  0.8  \\
J0900$-$3144$^{*}$ &  -- &  -- & 986   & 8.5    &  J1909$-$3744 &  2.3  &  0.4  &  1.9  & 0.3  \\
J1012+5307 & 21.2    & 0.3   & 16.5   & 0.3   &  J1939+2134  &  2.9  &  0.3  &  2.8 &  0.3 \\
J1022+1001 &  28.2   & 1.5   & 22.5   & 1.7   &  J1944+0907 &  28.8  &  3.5  &  27.1 &  3.9  \\
J1024$-$0719 & 92.4   &  3.7   &  90.5   &  12.2   &  J2124$-$3358 &  34.8  &  4.4  &  31.2   &  7.3  \\
J1125+7819 & 9.8    &  1.1    & -- & -- &  J2145$-$0750 &  13.0   &  0.8  &  9.9  & 0.7  \\
J1455$-$3330 & 36.4  &  10.2   &  29.8  & 14.8   & J2302+4442  & 74.6  &  2.1  & -- & -- \\
J1600$-$3053 & 27.4  & 4.7  & 14.2  &  1.4  &   & & & & \\

\hline
\end{tabular}
\footnotetext{$^{*}$ The reported DM precision for PSR J0900$-$3144 is obtained using only band 5 data because  band 3 dataset is excluded  for this pulsar from the present data release.}
\caption{Pulsar names with their estimated band 3 and band 3+5 DM uncertainties. The information of band 3+5 DMs is not available for PSRs J0030+0451, J0034$-$0534, J1125+7819, and J2302+4442 because these were observed in band3-only observation mode, recently adopted by InPTA (see Section \ref{sec:obs}).  }
\label{tab:DMprecision}
\end{table*}

\section{Deterministic timing of our MSPs } \label{sec:timing}

The DM estimation analysis using \dmcalc{}, described in the previous section, produces a parameter file (\texttt{par file}) with one DMX parameter for each epoch. These DMXs are estimated as DM corrections over the fixed fiducial DM (instantaneous DM $ -$ fiducial DM) and represent the contribution of IISM propagation delays and delays due to solar wind for each epoch. \dmcalc{} also provides a file with ToAs (\texttt{tim file}). 

In the timing analysis, the \texttt{tim} and \texttt{par} files are 
analysed using the latest version of \tempotwo{}\footnote{Version $-$  2024.04.1 \url{https://bitbucket.org/psrsoft/tempo2/}} to obtain the timing residuals. In brief, first bad ToAs were removed as explained in the subsequent paragraph. Then, DMX for epochs with ToA residuals showing parabolic trends across frequency (if any, due to wrong DM estimate) were refitted to improve DMX estimates for such specific epochs. This was followed by obtaining a timing solution using the remaining good ToAs. Finally, appropriate EFACs were used to scale the ToA error bars to obtain the final parameter estimates and their uncertainties in the timing solution.

The bad ToAs in the data are usually due to poor S/N in some of the subbands. Often, these are ToAs corresponding to timing residuals with very large error bars. In the first step, such bad ToAs were removed from the data (\texttt{tim} file). First, ToAs with timing error bars larger than 100 $\mu$s were flagged/removed from the analysis. Then, the median ToA error bar was computed and all ToAs with typically 10 times the median ToA error bar were also flagged. The channels at the edge of the 300$-$500  and 1260$-$1460 MHz band have poor signal due to out of band signal aliasing within the band in the edge channels and/or radio-frequency interference (RFI).  Often, this leads to erroneous cross-correlation producing outlier residuals. Additionally, the narrowband RFI\footnote{Narrowband RFI is defined as RFI affecting adjacent frequencies over a small fraction of the overall band (typically 10 MHz or less over 200 MHz band)} from satellites and other sources leads to a loss of signal in the template epoch leading to a very poor template in corresponding sub-bands. All ToAs in such frequency channels, irrespective of S/N in the relevant channel of individual data epochs, are associated with residuals which appear as outliers. Hence, such outlier ToAs are also flagged.

In rare cases, few outliers are also seen, which may be due to discrete events, such as a jump in DM or a rare profile change. An example is PSR J1125+7819, where we see ToA outliers associated with a possible event of jump in DM value at MJD 59741 (see Figure \ref{fig: DMtimeseriesplot1}). The difference between fiducial DM and the epoch DM is $\sim$0.0042 pc cm$^{-3}$. Such ToAs are noted for future investigations, but these are flagged for subsequent timing analysis.

Analysis of the cleaned \texttt{tim} file sometimes shows a few epochs with a parabolic trend across frequency. Residuals with such epochs correspond to DMXs with large DM error bars or poorly estimated DMXs. Typically, 1 \% to 12 \% 
of such epochs were noticed in the analysis of most of the pulsars in our sample, whereas 8 pulsars in our sample did not require any DMX to be fitted. Often, the large DM error bar is caused by absence of band 3 data or due to poor band 3 S/N in such epochs. ToAs of such epochs were also flagged in the analysis. In the remaining epochs with poor DMXs, a fresh DMX fit was carried out and the DMX in the original par file was replaced by the resulting higher precision DMX estimate. The DM measurements of these epochs 
were also replaced in the DM time series generated by \dmcalc{}, as shown in Figures \ref{fig: DMtimeseriesplot1} and \ref{fig: DMtimeseriesplot2}.

The improved \texttt{tim} and \texttt{par} files were then used to obtain a timing solution for each pulsar. Note that due to the unique nature of the InPTA experiment, the propagation effects of IISM are already accounted for in this analysis as we use concurrent high precision spot DM measurements. This is a significant difference in our analysis compared to other PTA experiments.

The timing analysis used DE440 ephmeris and Bureau International des Poids et Mesures (BIPM) 2023 realisation of Terrestrial Time (TT) time-scale to translate each site-arrival-times (SAT) to barycentric-arrival-times (BAT) at the solar system barycentre (SSB) into the Barycentric Coordinate Time (TCB) frame. The period and position epochs were fixed at MJD 59000 for all pulsars, whereas the DM epoch was chosen as the epoch of the fiducial DM (template epoch). No solar wind model was fitted in the first part of the analysis as DMXs will take any solar system propagation effect as well, although a spherically symmetric solar wind model was fitted at the later step of analysis. Starting from rotational frequency (F0) and its first derivative (F1), the rest of the parameters were fitted in the order of their variation with the available time baseline for each pulsar avoiding any highly covariant parameter. We scale ToA error bars in the analysis to constrain the reduced $\chi^2$ to unity. Based on various observation settings (such as band 3 or 5, 100 MHz or 200 MHz bandwidth, and CDP or PA beam pipeline), as outlined in Table \ref{tab:observations}, and whether the observations were taken before or after observing cycle 36\footnote{Since the uGMRT was being upgraded during earlier observation cycles, we treat the ToAs recorded before and after cycle 36 separately. }, we categorize the ToAs into seven distinct groups using the group flag in the parameter files. As a result, we fit a maximum of seven different scaling factors, specifically the T2EFAC parameters defined in \tempotwo. These scaling factors typically range from 0.3 to 7 for band 3 ToAs, except for PSR J1939$+$2134, where the scaling factor is approximately 14 for band 3 ToAs recorded with a 200 MHz bandwidth and the CDP pipeline.  Whereas, the scaling factors typically range from 0.4 to 2.3 for band 5 ToAs. In the case of PSRs J0613$-$0200, the scaling factor decreases to approximately 0.2 for band 5 ToAs recorded with a 200 MHz bandwidth and the CDP pipeline (pre-cycle 36). For PSR J0645+5158, the scaling factor is between 0.2 to 0.3 for band 5 ToAs recorded with either a 100 MHz or 200 MHz bandwidth using the CDP pipeline (pre-cycle 36). We plan to have better constraints on such scaling factors by modeling them as white noise parameters in subsequent noise analysis (planned as a part of subsequent paper). We find that the optimal selection of sub-bands, as described in section \ref{sec4.2}, along with frequency resolved templates takes care of profile shape evolution across a band, hence we do not require to fit for Frequency-Dependent (FD) parameters, which are  polynomial coefficients in log-frequency space \citep{abb+2015}, even with low-frequency data set where such effects are dominant.

\begin{figure*}
   \centering
    \includegraphics[width=1.0\linewidth]{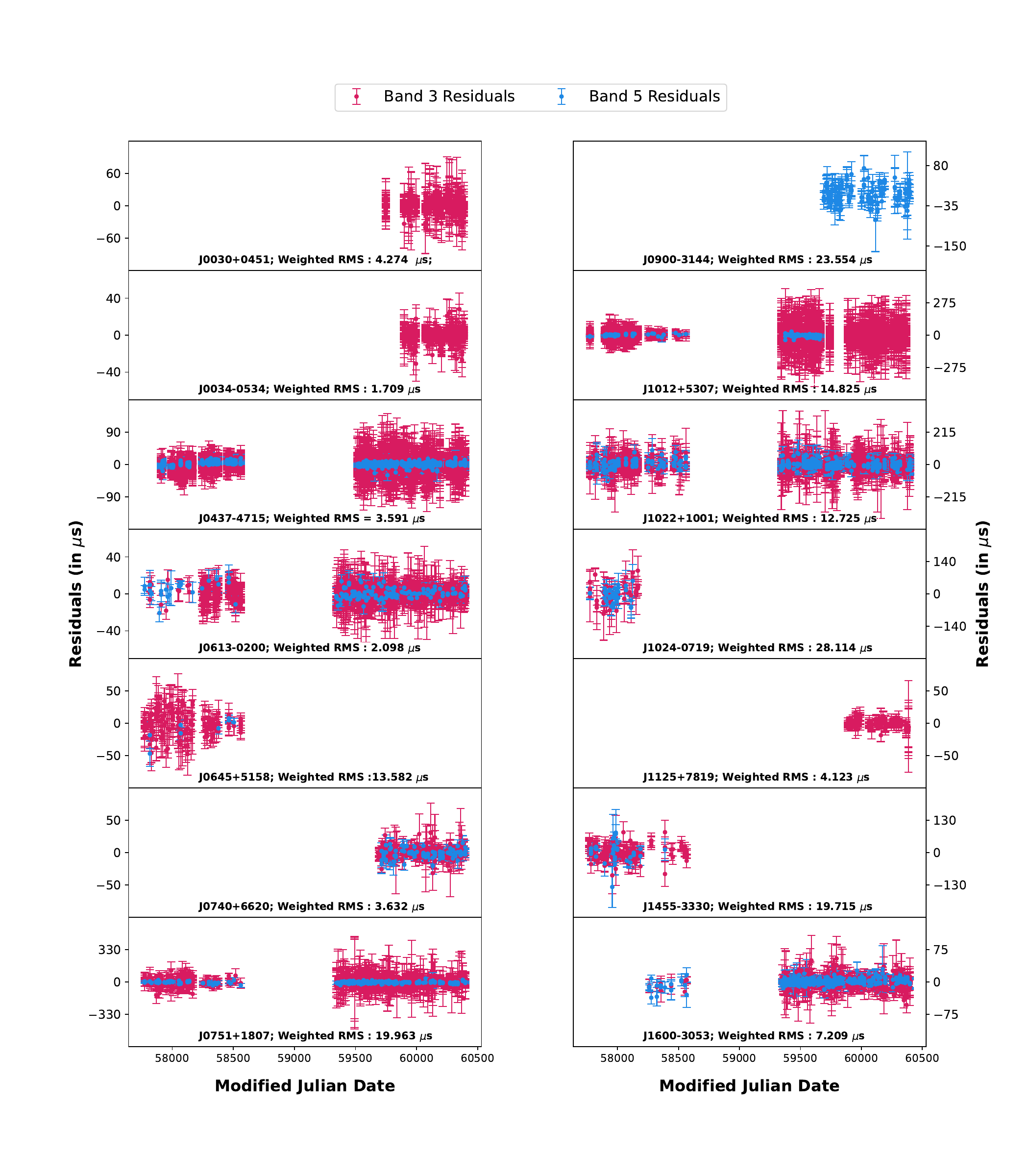}
    \caption{The timing residuals obtained from band 3 and band 5 data
for 14 pulsars are plotted against corresponding epochs. The IISM trends are modeled using the epoch-wise DMXs obtained from DMs estimated using \dmcalc{}. Red points represent band 3 and blue points represent band 5 residuals. Pulsar names and their respective post-fit weighted RMS of residuals are mentioned at the bottom of the respective panels. Epochs in terms of Modified Julian Date are depicted on the consolidated horizontal axes at the bottom. Band 3 dataset is not included in this data release for PSR J0900-3144, hence only band 5 ToAs are shown. For PSRs J0030+0451, J0034$-$0534, and J1125+7819, only band 3 timing residuals are shown, because these pulsars were added later in the experiment when a new observation strategy was introduced, which alternated between dual-band (band 3 and band 5) and single-band (band 3) observation modes. }
    \label{fig:timing-residuals1}
\end{figure*}

\begin{figure*}
    \centering
    \includegraphics[width=1.0\linewidth]{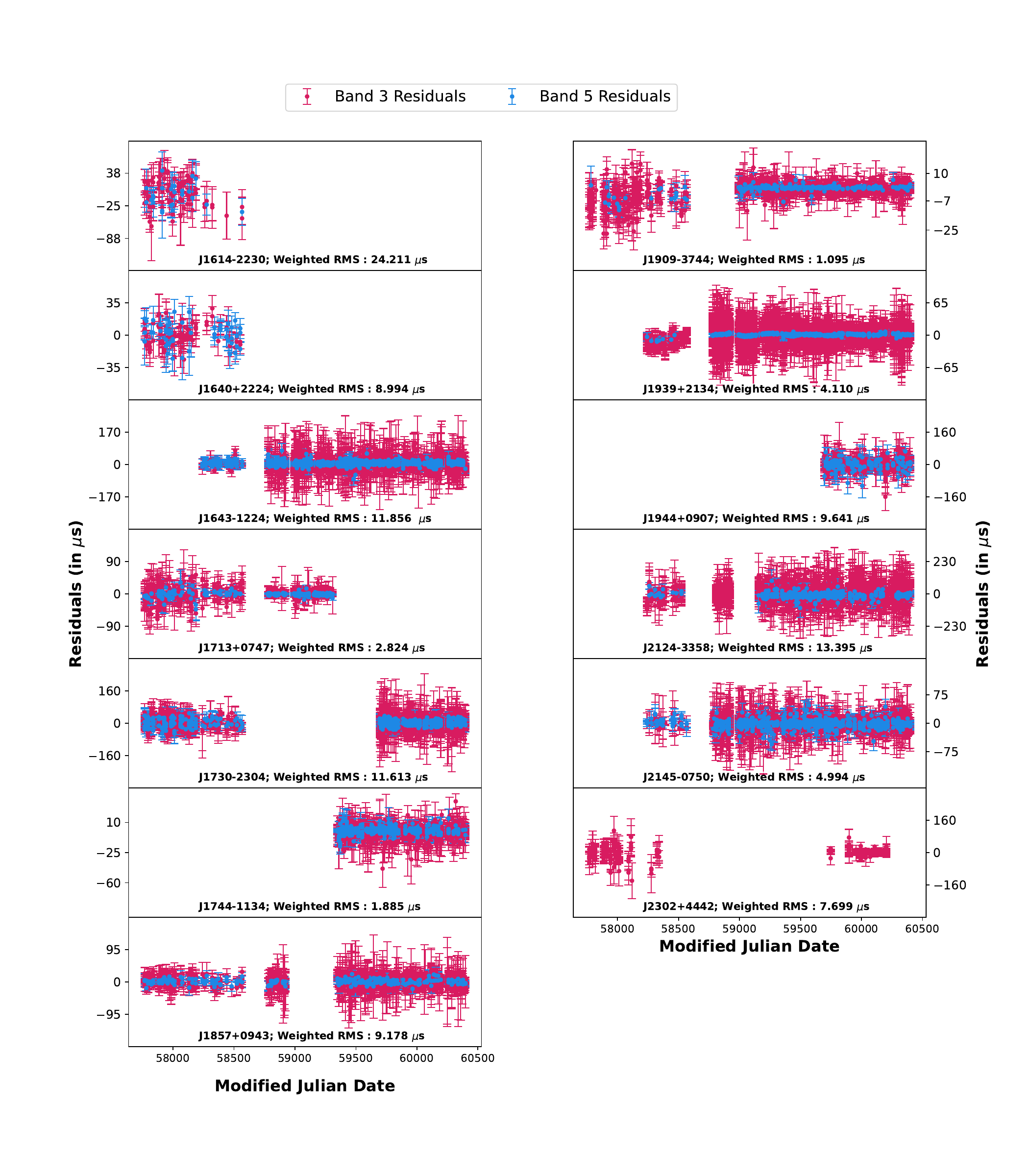}
    \caption{Same as Figure \ref{fig:timing-residuals1}. The timing residuals obtained from band 3 and band 5 data for 13 pulsars are plotted against corresponding epochs. The current data release includes ToAs up to MJD 59309 for PSR J1713+0747, prior to the pulse-profile shape event.}
    \label{fig:timing-residuals2}
\end{figure*}

Once a reasonable fit was obtained, the corresponding solution was perturbed in each parameter by about 10 $times$ the formal error bar followed by a refit to check the stability of the solution. The final solution was written to an output \texttt{par} file. The weighted RMS of residuals for the pulsars in our sample ranged from 1.095 to 28.114 $\mu$s. The timing residuals obtained from this procedure are shown in Figures \ref{fig:timing-residuals1} and \ref{fig:timing-residuals2}.

In the second step of the analysis, the \texttt{par} file fitted above was stripped of all DMXs keeping the other parameters of the solution same. The ToAs were analysed again by fitting DM and its first and second derivative to model effects due to DM variations (as realised by a Taylor series) and a solar wind model to remove IISM trends. For pulsars J0437-4715, J0645+5158, J0740+6620, J0900-3144, J1024-0719, J1125+7819, J1455-3330, J1640+2224, J1857+0943, J1939+2134, J1944+0907, J2302+4442 the solar wind model produced unrealistic results, such as negative values.  In these instances, the solar wind parameter was not fitted. For PSR J1713+0747, we have included data (ToAs and DMs) up to MJD 59309 in the present data release, as we are investigating the profile shape event reported around MJDs 59320-59321 \citep{ssj+21,jcc+2024} using post-event InPTA dataset in a separate study. Therefore, the final data product of this data release consists of a list of good ToAs in the cleaned \texttt{tim} file alongwith two timing solutions, one with DMXs modelling the IISM noise and the other with a Taylor series model for DM variations.

The resulting \texttt{par} file obtained by fitting DM and its derivatives is planned to be utilized in the subsequent noise analysis to model the white as well as achromatic and chromatic Gaussian noise processes. The estimates of the amplitude and spectral slope of these noise processes allow scaling the ToA error bars appropriately and subtraction of the red-noise fluctuations in the residuals. These will be incorporated in the subsequent paper. Note that the \texttt{tim} file, used in the analysis described above, consists of multiple frequency resolved ToAs across a frequency band for each epoch. In this sense, we have carried out a narrow-band timing as opposed to wide-band timing where a single ToA across each band for a given epoch is used. Such wide-band timing is planned as a part of subsequent paper.

\section{Summary and Future Directions} \label{sec:discussion}

In this work, we present the second data release of the InPTA experiment, which provides precise ToAs and DMs for 27 MSPs. These pulsars were observed simultaneously in three frequency bands: 300–500 MHz (band 3), 550-850 MHz (band 4) and 1260–1460 MHz (band 5), using the uGMRT. Table \ref{tab:observations} details the observation settings used for InPTA observations across different uGMRT observation cycles included in this data release. Figure \ref{fig:skymap} shows the sky distribution of InPTA DR1 and DR2 pulsars alongside pulsars that are planned to be included in the third data release of IPTA. Figure \ref{fig:cadence} depicts the cadence of the InPTA observations for all 27 pulsars included in the present work. We have included band 3 and band 5 dataset in the present study. Due to the limited band 4 dataset, with band 4 observations of pulsars available only from April 2017 to February 2019, it was challenging to identify high-S/N concurrent observations across all three bands for creating noise-free templates, hence we did not include band 4 dataset in the present data release. The time baseline for the sample of pulsars in this data release varies from $\sim$ 1.2 years to 7.3 years. Since 2022 (observation cycle 43), the InPTA experiment has adopted a hybrid observing strategy with alternating observations modes: band 3-only (using one subarray) or band 3 + band 5 (using two subarrays), each with 200 MHz bandwidth. This change allowed us to observe more pulsars in the band 3-only mode during 2022–2024, by utilizing more uGMRT antennas in a single subarray. Hence, we added PSRs J0030+0451, J0034$-$0534, J1125+7819, and J2302+4442 to our sample in band 3-only observation mode. PSRs J0740+6620, J0900-3144, and J1944+0907 were also recently added, starting from MJD 59692, to the sample of InPTA experiment in the band 3 + band 5 observation mode. We have excluded band 5 data of PSR J2302+4442 from this data release due to insufficient S/N in both bands during concurrent observations, which is required to create templates (as detailed in section \ref{sec4.1}). PSR J2302+4442 is the only pulsar in our sample which was observed in both band 3 and band 5 during observation cycles 31-35, but later observed only in band 3 during observation cycles 43-45, see Figure \ref{fig:cadence}. In the band 3-only observation mode (cycles 43-45), the observed S/N in band 3 for this pulsar ranges from 54 to 245, while it ranges from 8 to 57 in the band 3 + band 5 observation mode (cycles 31-35). Whereas, the median observed S/N in band 5 is $\sim$12 where only a few epochs have S/N > 15. Consequently, we decided to use the highest S/N epoch from the band 3-only observations (cycles 43-45) to create the band 3 template rather than including band 5 data. All other pulsars in our sample were observed in either band 3 + band 5 observation mode across their  individual timing baselines or in band 3-only observation mode starting in year 2022.

We start with creating noise-free frequency-resolved templates made using concurrent observations of band 3 and band 5. We also correct for the bandpass shape used in recording the data for the template epoch by equalizing the off-pulse RMS signal across each band. An example of this process is illustrated in Figure \ref{fig:bandequalising}. 
We then do a careful selection of number of sub-bands used to estimate frequency-resolved ToAs in each band for each pulsar. Such optimization of number of sub-bands accounts for pulse profile-shape evolution within a band, eliminating the need for frequency-dependent (FD) parameters in timing analysis. Table \ref{tab:subband} presents the optimal number of sub-bands chosen for each pulsar in each band, separately listed for datasets recorded with 100 MHz and 200 MHz observing bandwidths. We utilize the concurrent and dual-band observations of the high-S/N template epoch to estimate a precise DM, called the fiducial DM, to dedisperse and align the noise-free templates of both bands. We employed \dmcalc{} to generate precise DMs and ToAs (see Figure \ref{fig:flowchart} for a workflow of this pipeline). The precise DM estimation using the low-frequency uGMRT data effectively mitigates IISM effects, improving timing precision by leveraging concurrent dual-band observations. This methodology sets our approach apart from other PTA experiments.

The ability to estimate high-precision DMs with uncertainties ranging from 10$^{-4}$ to $10^{-6}$ pc cm$^{-3}$ is the key advantage of the InPTA experiment, a low radio frequency PTA experiment. Such measurements are unprecedented and these  represent the highest precision DM measurements ever reported for some of these pulsars. Table \ref{tab:DMprecision} lists the median and minimum DM uncertainties achieved for each pulsar in our sample. The DM time series for all pulsars is shown in Figures \ref{fig: DMtimeseriesplot1} and \ref{fig: DMtimeseriesplot2}. These high-precision DMs also show clear effects due to solar wind, that is adding an excess in the DM value with an annual trend in the DM time series of PSRs J0034$-$0534, J0613$-$0200, J1744$-$1134, J1909$-$3744, and J2145-0750. Among these pulsars, similar annual trends were also seen for PSRs J0034$-$0534, J1744$-$1134 and J2145-0750 in the DM time series reported in \cite{dvt+2020,tsb+2021}. We also compared and found that the DM variation trends seen in our data are consistent with those seen by LOw Frequency ARray (LOFAR) telescope \citep{dvt+2020} or seen by NANOGrav \citep{aaa+2023a} for pulsars common with the InPTA, specially PSRs J1909$-$3744 and J1939$+$2134 where the DM variation is large. A careful investigation of DM time series also highlighted potential DM outliers associated with discrete events of scientific significance, for example a possible DM jump event in PSR J1125$+$7819 near MJD 59741. Such cases will be investigated in a future work. We have utilized our high-precision DMs to estimate epoch-wise DMXs to model IISM propagation delays (see section \ref{sec4.4} for details), which enables us to obtain accurate timing solutions for each pulsar. The timing residuals obtained using our timing procedure, described in section \ref{sec:timing}, are shown in Figures \ref{fig:timing-residuals1} and \ref{fig:timing-residuals2}. 

PSR J0900-3144 has the highest DM of $\sim$ 75 pc cm$^{-3}$ among all pulsars in our sample. The recorded band 3 pulse-profile of this pulsar does not show enough signal close to 300 MHz for most of the epochs. This may be possible either due to high scatter-broadening of the pulse at low radio frequencies or due to imprecise DM used by the CDP pipeline during observations. We plan to investigate this in a separate study. Hence, we included only band 5 data of PSR J0900-3144 in this data release. The DM time series of PSR J1643$-$1224 (Figure \ref{fig: DMtimeseriesplot2}) reveals an offset between band 3 and band 3+5 DMs, which varies over time. This offset arises from a bias in DM estimation caused by unmodelled scatter-broadening of the pulse profile dominant at low radio frequencies and specifically in band 3 for InPTA data. \cite{sjk+2024} provides a detailed study and presents a method to mitigate these biases using both simulated data and InPTA observations of PSR J1643-1224. As detailed in section \ref{sec:obs}, we began conducting alternating band 3 + band 5 and band 3-only observations since 2022. As a result, the band 3 + band 5 and band 3 DMs were used to estimate DMX parameters (calculated within the DMCalc pipeline, see Figure \ref{fig:flowchart}) for alternating epochs. Note that these DMX parameters are defined differently within our scheme (see section \ref{sec4.4} for details) compared to the standard practice, for example used in \cite{aaa+2023a}.  For PSR J1643$-$1224, this leads to a nearly constant offset between the ToAs from band 3 + band 5 and band 3-only observations, caused by the bias in DM estimation due to unmodelled scatter-broadening, as explained above. To mitigate this offset, we shifted the reference DM for band 3-only observations using the FDJUMPDM parameter, defined in \tempotwo, by applying a group flag to the ToAs belonging to the band 3-only observations.

 We now list ongoing and preliminary efforts that employ the current data set. The customized noise modeling of our MSPs is being pursued to characterize various noise contributions and to probe the implications of our inherent simultaneous L and P band MSP observations, with improvements over previous such analyses \citep[e.g.,][]{cbp+2022,sdk+2023,khb+2024}.
Efforts are ongoing to search for signatures of GWB present in our data sets, prompted by the recently concluded EPTA+InPTA 3P+ efforts \citep{epta+inpta2023c,epta+inpta2023a,epta+inpta2024,epta+inpta2023b}.
Additionally, wideband technique is being updated to treat our data set by extending earlier InPTA efforts \citep{nag+2022,pdr+2024}. We plan to conduct wide-band timing for concurrent band 3 and band 5 dataset as part of a subsequent paper. 
Efforts are initiated to adapt our existing pipeline to search for linear memory events in our data set \citep{dsd+2024}.
Further, we are initiating efforts to pursue a number of auxiliary science projects that deal with DM, scattering and solar winds in the coming days using InPTA DR2. This data set, as noted earlier, is proposed to  be combined with the latest data sets from EPTA, MPTA, NANOGrav and PPTA to form the upcoming IPTA data releases, which are expected to be critical to detect and characterize nHz GW sources \citep{ipta2024}.
In closing, we plan to continue uGMRT monitoring of 
an interesting subset of IPTA pulsars to provide the consortium with our simultaneous L and P band observations of these MSPs till the Square Kilometer Array comes online.


\section*{Data Availability}
The InPTA DR2 dataset, detailed in this paper, includes ToA measurements, pulsar ephemerides, DM measurements, and associated codes. The data will be made publicly available at \\ \href{https://github.com/inpta/InPTA.DR2}{https://github.com/inpta/InPTA.DR2}. 

\section*{Software}
\tempotwo{} \citep{hem2006, ehm2006}, \psrchive{} \citep{hsm+2004}, \dspsr{} \citep{sb2011}, \rficlean{} \citep{mlv+2021}, \pinta{} \citep{smj+2021}, \dmcalc{} \citep{kmj+2021}, \enterprise{} \citep{evt+2019}, \texttt{matplotlib} \citep{hunter2007}, \astropy{} \citep[][]{astropy+2022}

\section*{Facilities}
Upgraded Giant Metrewave Radio Telescope (uGMRT)

\section*{Acknowledgements}
InPTA acknowledges the support of the GMRT staff in resolving technical difficulties and providing technical solutions for high-precision work. We acknowledge the GMRT telescope operators for the observations. The GMRT is run by the National Centre for Radio Astrophysics of
the Tata Institute of Fundamental Research, India. Authors acknowledge the National Supercomputing Mission (NSM) for providing computing resources of 'PARAM Ganga' at the Indian Institute of Technology Roorkee, 'PARAM Smriti' at National Agri-Food Biotechnology Institute Mohali and 'PARAM Seva' at Indian Institute of Technology, Hyderabad.  PARAM Ganga,   PARAM Smriti and PARAM Seva are implemented by C-DAC and supported by the Ministry of Electronics and Information Technology (MeitY) and Department of Science and Technology (DST), Government of India. \\
AG acknowledges support of the Department of Atomic Energy, Government of India, under Project Identification No. RTI 4002. \\
AKP is supported by CSIR fellowship Grant number 09/0079 \\
(15784)/2022-EMR-I. \\
AmS is supported by CSIR fellowship Grant number 09/1001 \\
(12656)/2021-EMR-I. \\
BCJ acknowledges the support from Raja Ramanna Chair fellowship of the Department of Atomic Energy, Government of India (RRC - Track I Grant 3/3401 Atomic Energy Research 00 004 Research and Development 27 02 31 1002//2/2023/RRC/R\&D-II/13886 and 1002/2/2023/RRC/R\&D-II/14369). BCJ also acknowledges support from the Department of Atomic Energy Government of India, under project number 12-R\&D-TFR-5.02-0700.  \\
CD acknowledges the Param Vikram-1000 High Performance Computing Cluster of the Physical Research Laboratory (PRL) on which computations were performed. \\
DD acknowledges the support from the Department of Atomic Energy, Government of India through ‘Apex-I Project - Advance Research and Education in Mathematical Sciences’ at The Institute of Mathematical Sciences. \\
HT is supported by DST INSPIRE Fellowship, INSPIRE code IF210656. \\
JS acknowledges funding from the South African Research 
Chairs Initiative of the Department of Science and Technology and 
the National Research Foundation of South Africa. \\
KT was supported by JSPS KAKENHI Grant Numbers 20H00180, 21H01130 and 21H04467, and Japan-India Science Cooperative Program between JSPS and DST JPJSBP120237710. \\
KV acknowledges CSIR fellowship , grant number 09/1020(20166) \\
/2024-EMR-I. \\
MB acknowledges the support from the Department of Atomic Energy,
Government of India through Apex-I Project - Advance Research and Education in Mathematical Sciences at IMSc. \\
PR acknowledges the financial assistance of the South African Radio Astronomy  Observatory (SARAO) towards this research (www.sarao.ac.za). \\
RK was supported by JSPS KAKENHI Grant Numbers 24K17051. \\
YG acknowledges support from the Department of Atomic Energy Government of India, under project number 12-R\&D-TFR-5.02-0700. \\
YM acknowledges support from the Department of Atomic Energy Government of India, under project number 12-R\&D-TFR-5.02-0700. \\
ZZ acknowledges the Prime Minister’s Research Fellows (PMRF) scheme, with Ref. No. TF/ PMRF-22-7307, for providing fellowship. 

\bibliography{InPTA-dr2}

\onecolumn
\appendix

\section{Empirical method to optimize frequency sub-bands}
\label{sec:appendix-subbands}
We perform a customized selection of the number of sub-bands used in band 3 and band 5 for each pulsar. The empirical procedure used to optimize the number of sub-bands, outlined in section \ref{sec4.2}, addresses following three key factors:  \\
(i) The primary requirement is to ensure that there is sufficient and comparable S/N per sub-band in band 3 and band 5 so that the estimated ToAs are reliable. We initially make a first-order guess for the number of sub-bands to use in each band. In order to do that, we make use of the dependence of integrated S/N on observation bandwidth as given by \citet{lk2004}. 
\begin{equation} 
    S/N = \sqrt{n_{p} t \Delta \nu} \left( \frac{T_{peak}}{T_{sys}} \right) \frac{\sqrt{W(P-W)}}{P} \propto \sqrt{N \delta \nu}   \label{eqAppendix1}
\end{equation}
where $n_{p}=1$ for single-polarisation observations and $n_{p}=2$ if two orthogonal polarisations are summed (for InPTA observations), $t$ is observation duration, $P$ is pulse period, $W$ is pulse width, $T_{peak}$ is peak amplitude and $T_{sys}$ is system noise temperature. Therefore, for an observing bandwidth of $\Delta\nu$ that is divided into $N$ sub-bands each with a bandwidth of $\delta\nu$, S/N per sub-band scales as integrated S/N/$\sqrt{N}$.  \\
(ii) Next, we ensure to achieve optimal and equivalent median ToA precision across both bands for the selected number of sub-bands. \\
(iii) Finally, we optimize the number of sub-bands in each band such that the radio frequency-dependent profile shape evolution is accounted for. This is done by ensuring that difference in pulse profiles between any two adjacent sub-bands, called the profile residuals ($p_i$), follows a Gaussian distribution. As a result, we find that frequency-dependent parameters are not required for timing of pulsars in our sample (see section \ref{sec:timing}). 
While visually examining the profile residuals, it is crucial to exclude sub-bands which are contaminated by RFI. To do so, we exclude sub-bands which do not obey the following criteria
\begin{equation}
    \left|\mathcal{F}(p_i)-\mathcal{F}(\overline{p_i})\right| \leq \epsilon\left[\text{median}\left|\mathcal{F}(p_i)-\mathcal{F}(\widetilde{p})\right|\right]\times 1.4826
\end{equation}
where $p_i$ is the $i$th difference of pulse profiles of two adjacent subbands or $i$th profile residuals, $\overline{p_i}$ denotes the mean signal, $\epsilon$ is a threshold value with $\epsilon\in(0, 1]$, $\tilde{p}$ is the median signal in the entire band and the function $\mathcal{F}$ corresponds to the root mean square estimate (the multiplicative constant corresponds to conversion to Median Absolute Deviation). The threshold $\epsilon$ signifies the \textit{tightness} of filtering where smaller values give more stringent filtering and vice versa.

\newpage

\section{DM time-series and timing residuals}
\label{sec:appendix}
The DM time series and the timing residuals for each of the 27 pulsars are presented here.

\begin{figure*}[h!]
\includegraphics[width=0.9\linewidth]{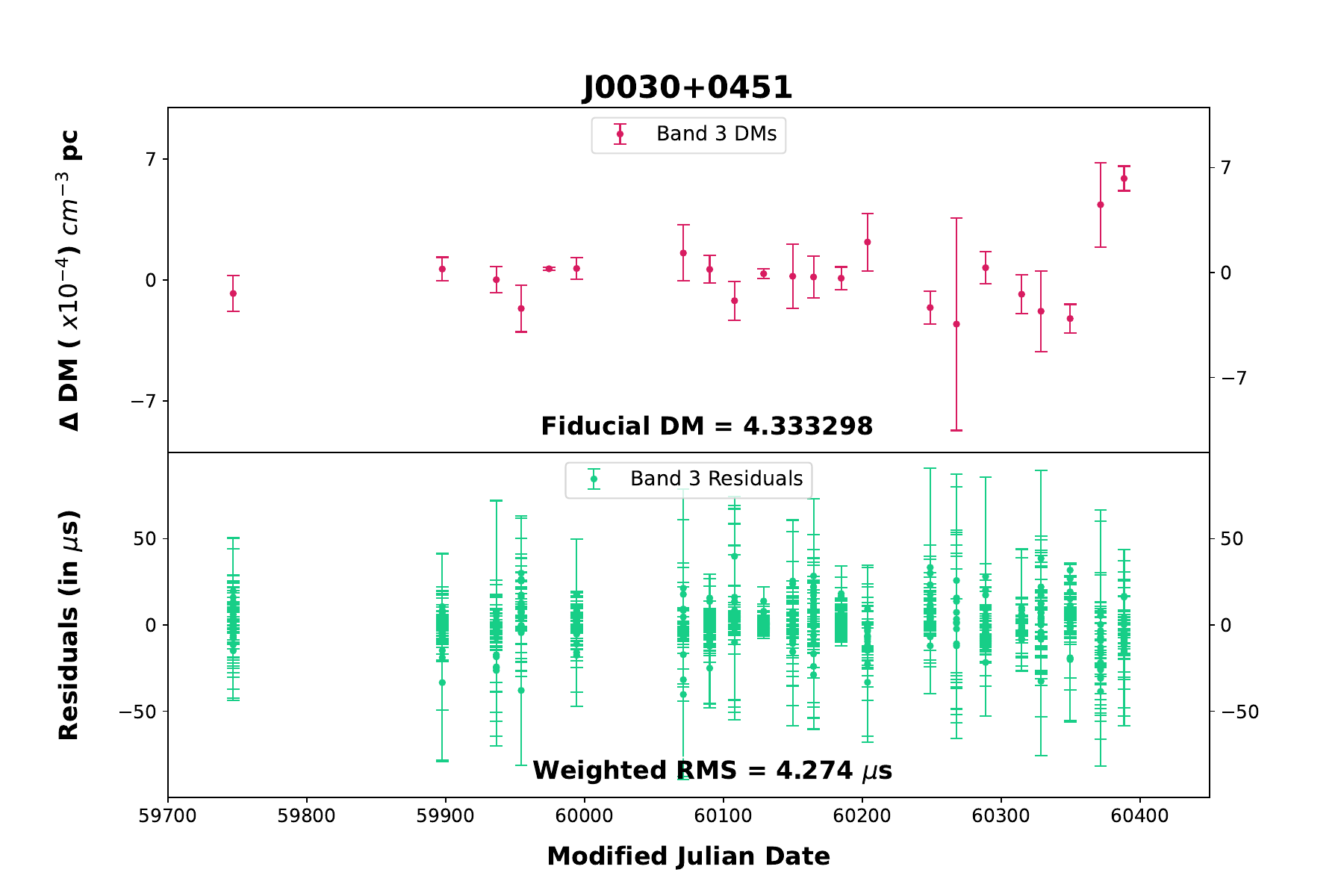}
\caption{Dispersion measure variations and timing residuals for J0030$+$0451. $\Delta$DMs (band 3) represent the difference between estimated DMs and the fiducial DM (mentioned at the bottom of the corresponding panels). Narrowband timing residuals are shown in the bottom panel (post-fit weighted RMS at the bottom of the respective panels).}
\label{fig:J0030}
\end{figure*}

\begin{figure*}[h!]
\includegraphics[width=0.88\linewidth]{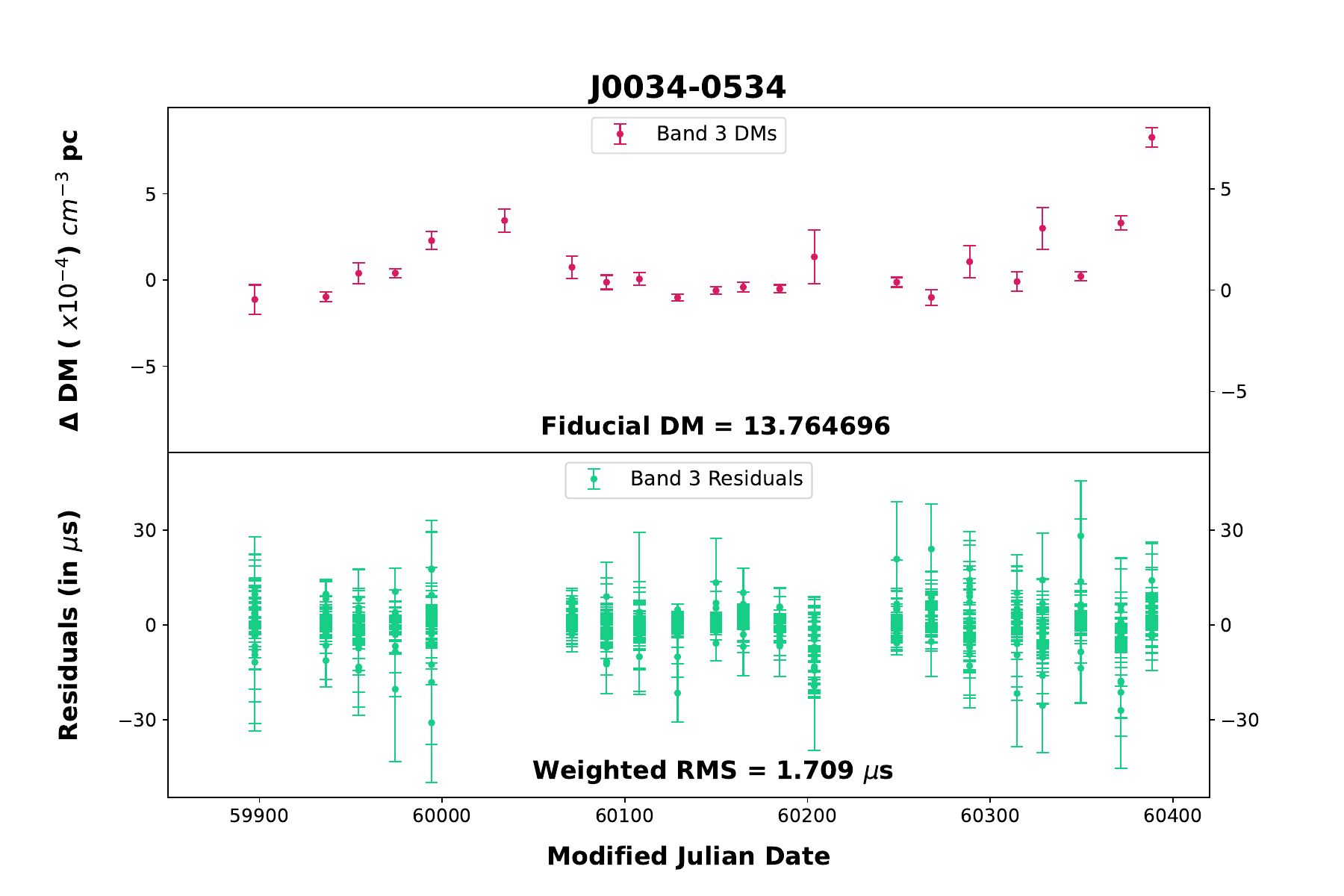}
\caption{Same as Figure \ref{fig:J0030}. Dispersion measure variations and timing residuals for J0034$-$0534. }
\label{fig:J0034}
\end{figure*}

\begin{figure*}[h!]
\includegraphics[width=0.88\linewidth]{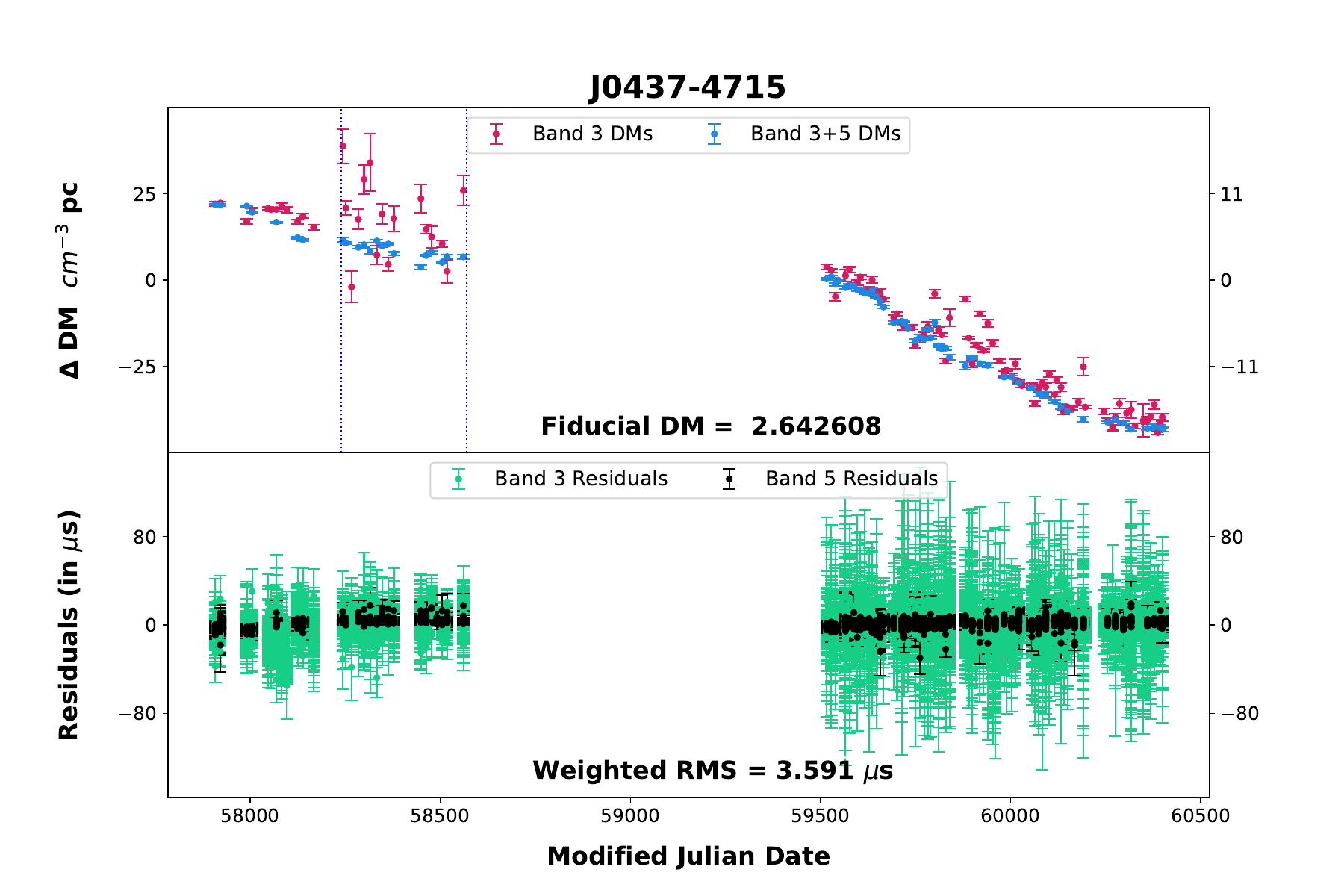}
\caption{Same as Figure \ref{fig:J0030}. Dispersion measure variations and timing residuals for J0437$-$4715 (B3 and B3+5). The two vertical lines at MJD 58239 and 58569 divide the DM time-series into three distinct sections as described in Figure \ref{fig: DMtimeseriesplot1}. The vertical axes of DM time-series plot are scaled differently for epochs before and after MJD 58569 to reflect the improved DM precision achieved from cycle 37 onward. }
\label{fig:J0437}
\end{figure*}

\begin{figure*}[h!]
\includegraphics[width=0.88\linewidth]{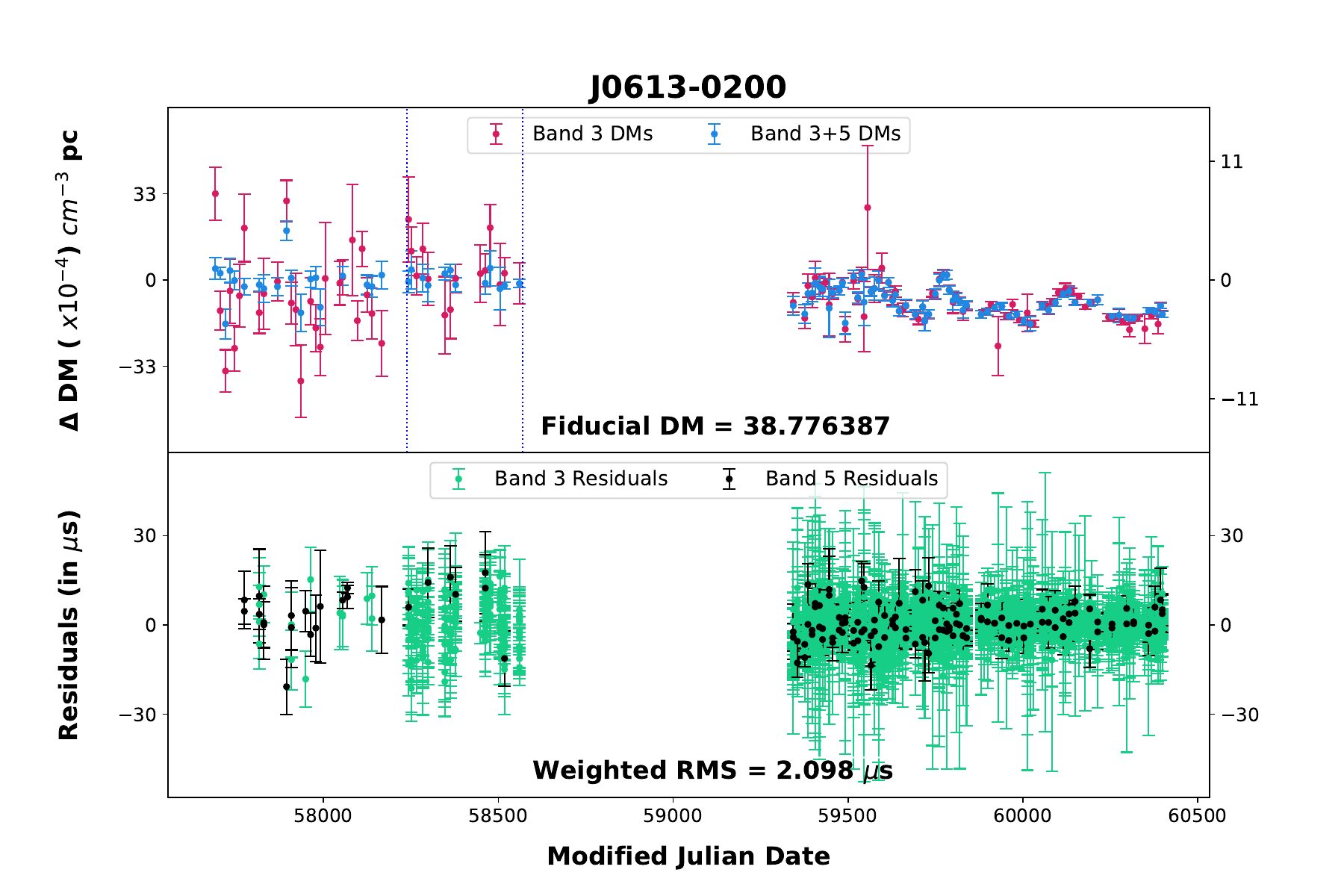}
\caption{Same as Figure \ref{fig:J0030}. Dispersion measure variations and timing residuals for J0613$-$0200 (B3 and B3+5). The two vertical lines at MJD 58239 and 58569 divide the DM time-series into three distinct sections as described in Figure \ref{fig: DMtimeseriesplot1}. The vertical axes of DM time-series plot are scaled differently for epochs before and after MJD 58569 to reflect the improved DM precision achieved from cycle 37 onward. }
\label{fig:J0613}
\end{figure*}

\begin{figure*}[h!]
\includegraphics[width=0.88\linewidth]{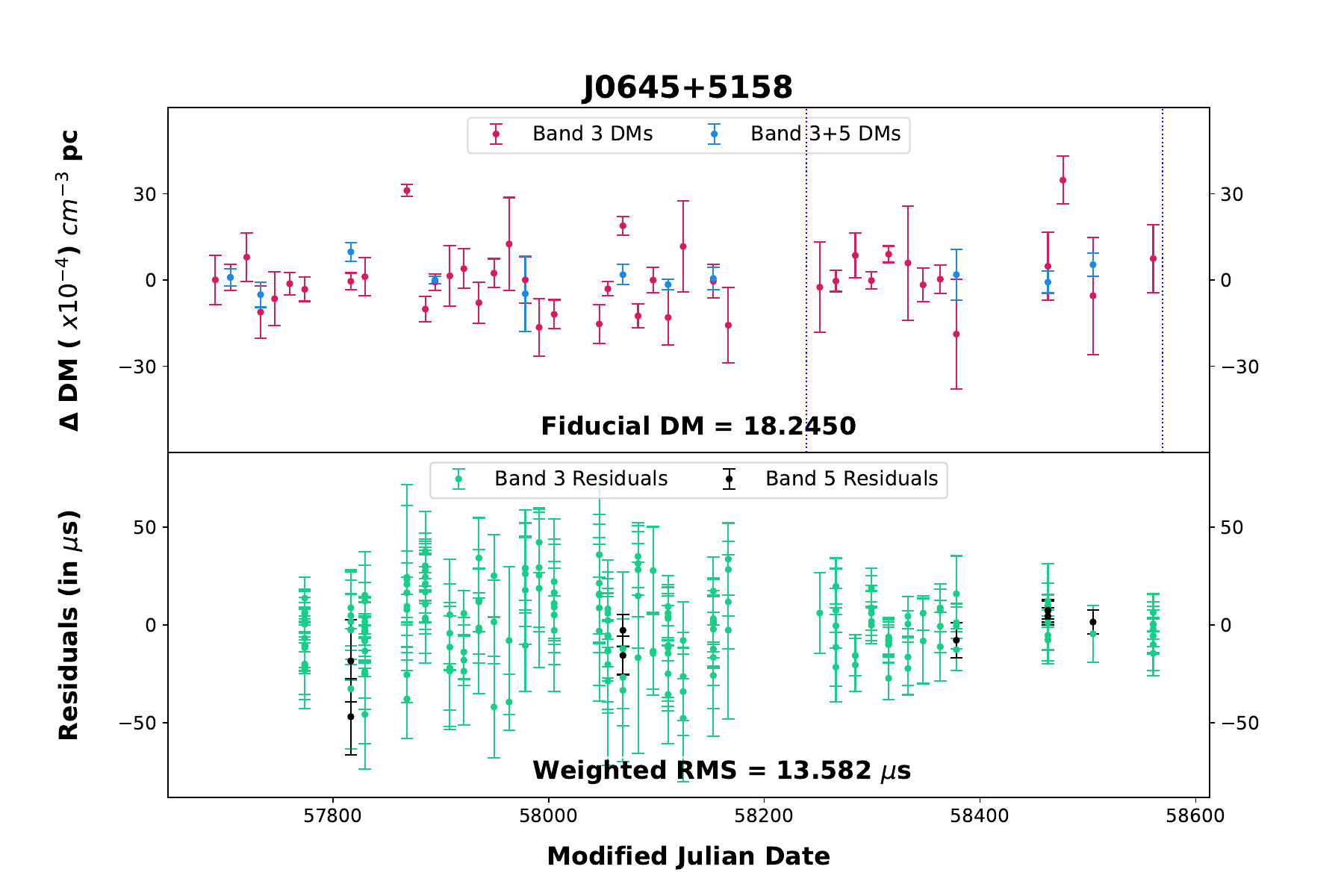}
\caption{Same as Figure \ref{fig:J0030}. Dispersion measure variations and timing residuals for J0645$+$5158 (B3 and B3+5). The two vertical lines at MJD 58239 and 58569 divide the DM time-series into two distinct sections as described in Figure \ref{fig: DMtimeseriesplot1}. }
\label{fig:J0645}
\end{figure*}

\begin{figure*}[h!]
\includegraphics[width=0.88\linewidth]{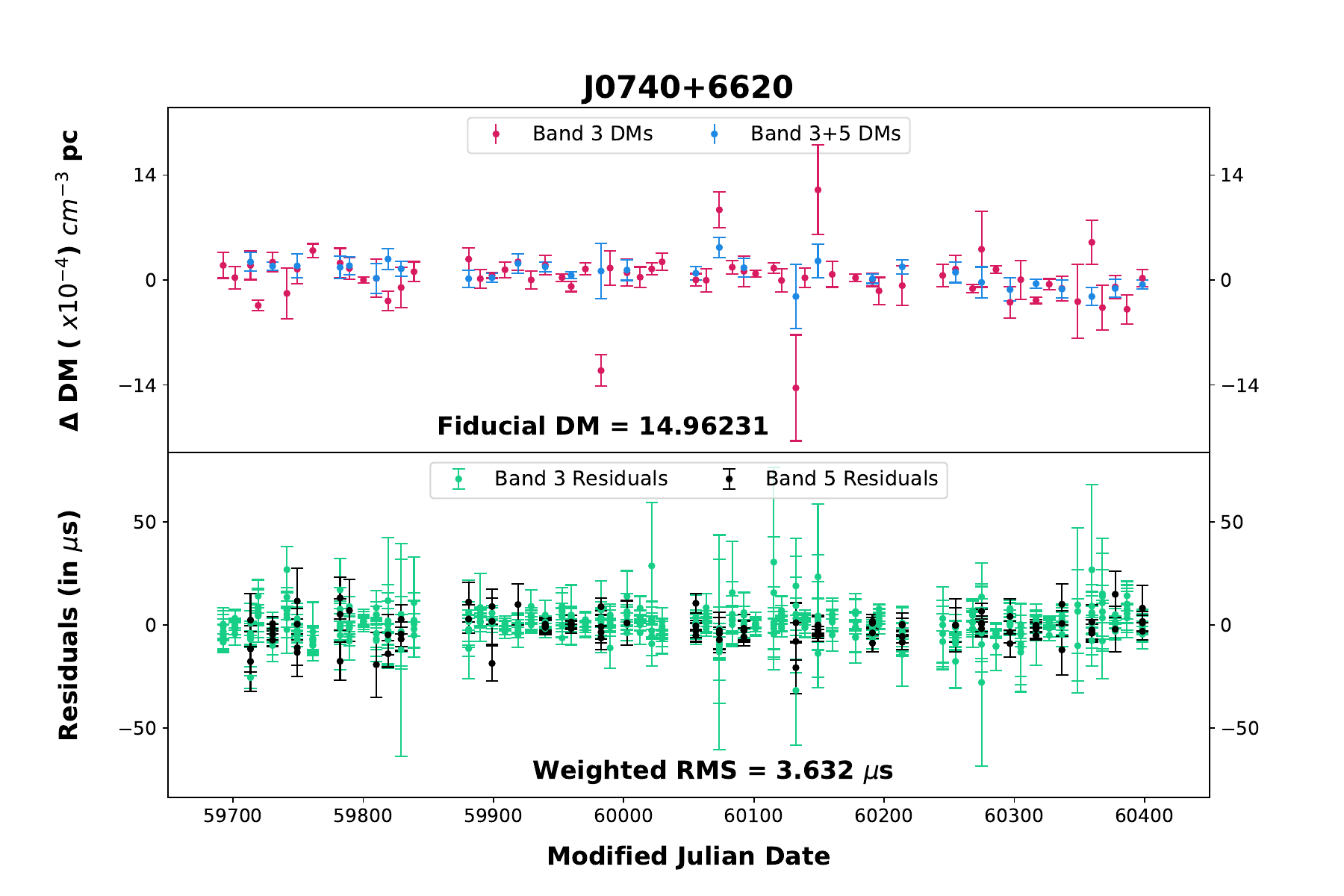}
\caption{Same as Figure \ref{fig:J0030}. Dispersion measure variations and timing residuals for J0740$+$6620 (B3 and B3+5).}
\label{fig:J0740}
\end{figure*}

\begin{figure*}[h!]
\includegraphics[width=0.88\linewidth]{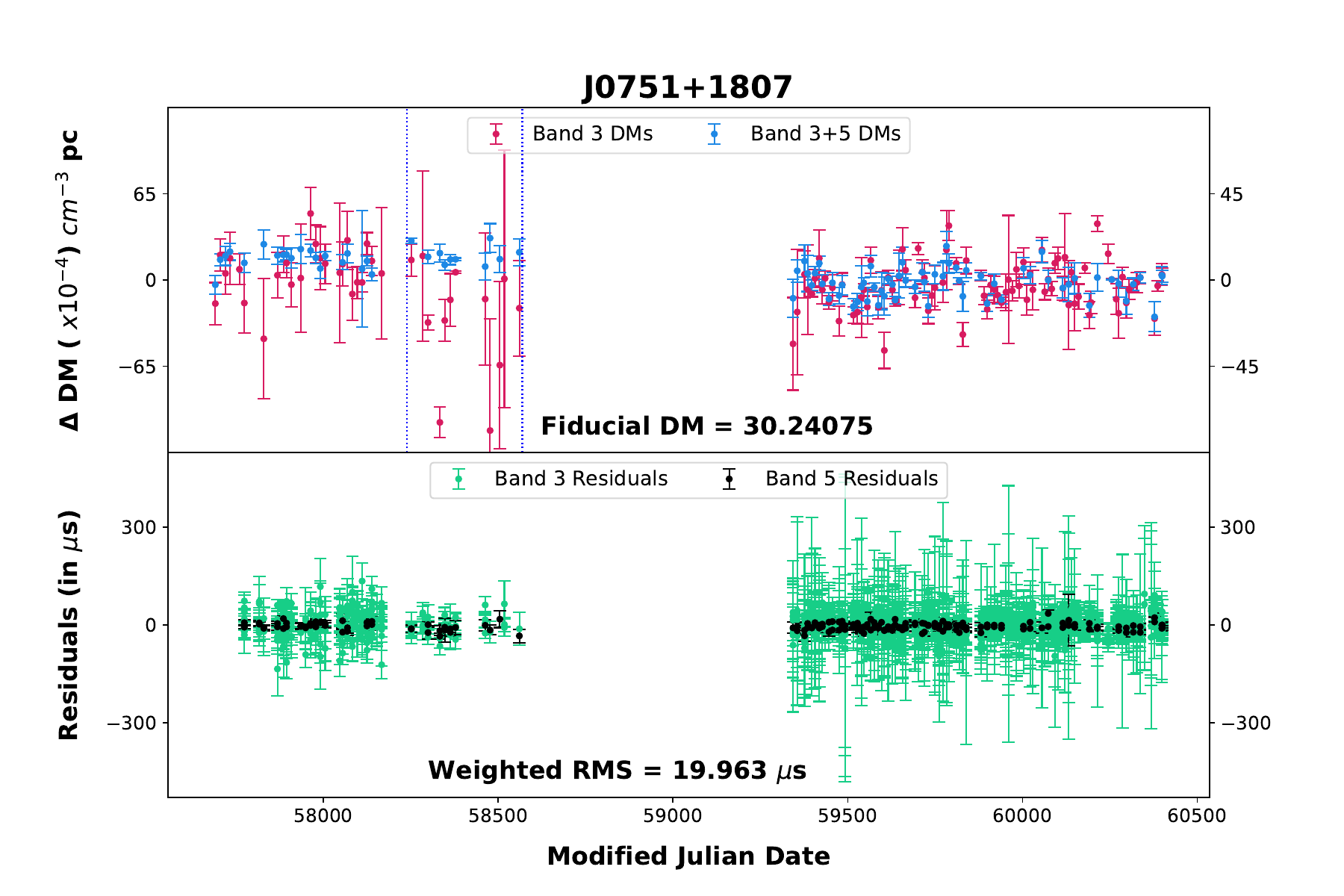}
\caption{Same as Figure \ref{fig:J0030}. Dispersion measure variations and timing residuals for  J0751$+$1807 (B3 and B3+5). The two vertical lines at MJD 58239 and 58569 divide the DM time-series into three distinct sections as described in Figure \ref{fig: DMtimeseriesplot1}. The vertical axes of DM time-series plot are scaled differently for epochs before and after MJD 58569 to reflect the improved DM precision achieved from cycle 37 onward.  }
\label{fig:J0751}
\end{figure*}

\begin{figure*}[h!]
\includegraphics[width=0.88\linewidth]{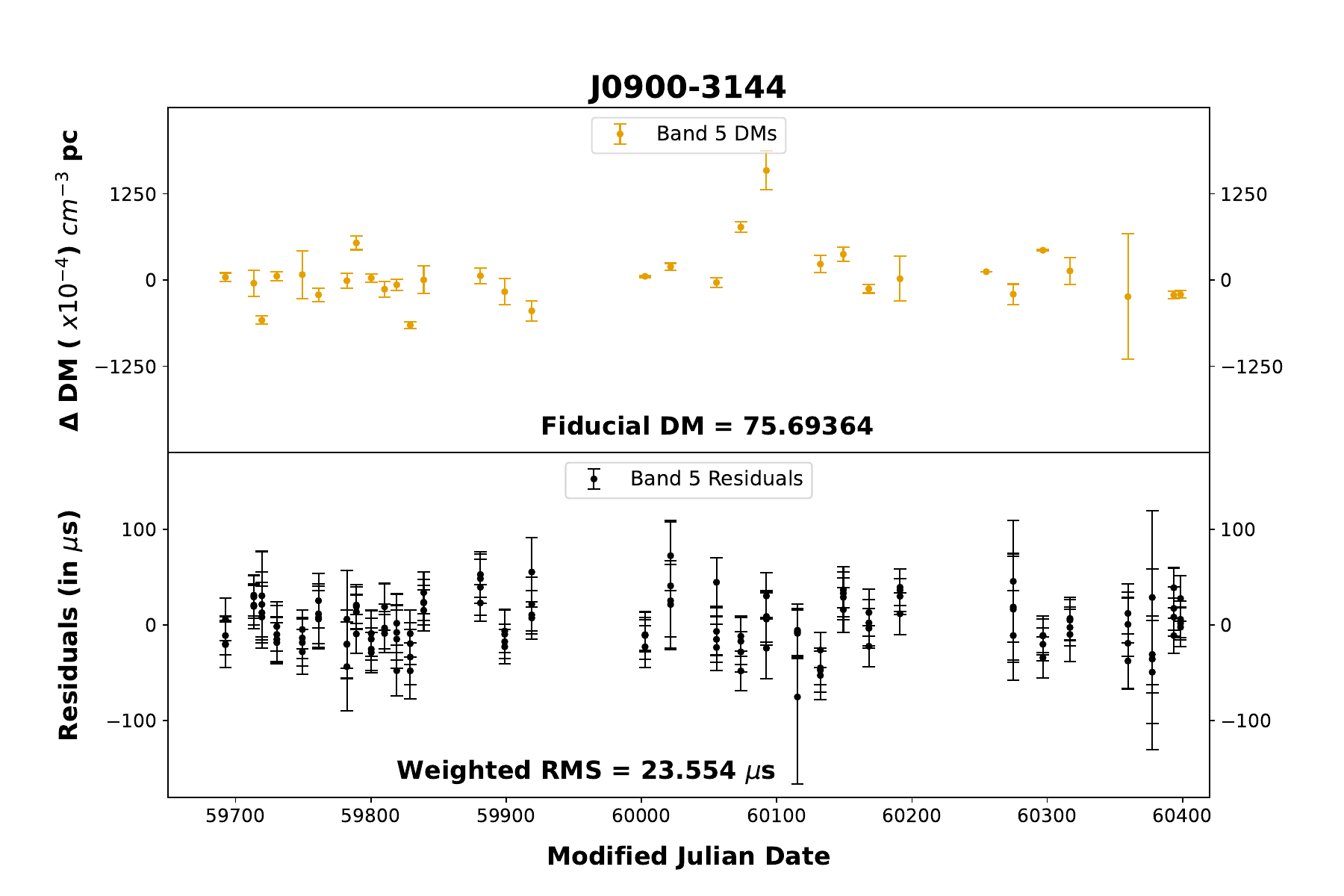}
\caption{Same as Figure \ref{fig:J0030}. Dispersion measure variations and timing residuals for J0900$-$3144 (B5).}
\label{fig:J0900}
\end{figure*}

\begin{figure*}[h!]
\includegraphics[width=0.88\linewidth]{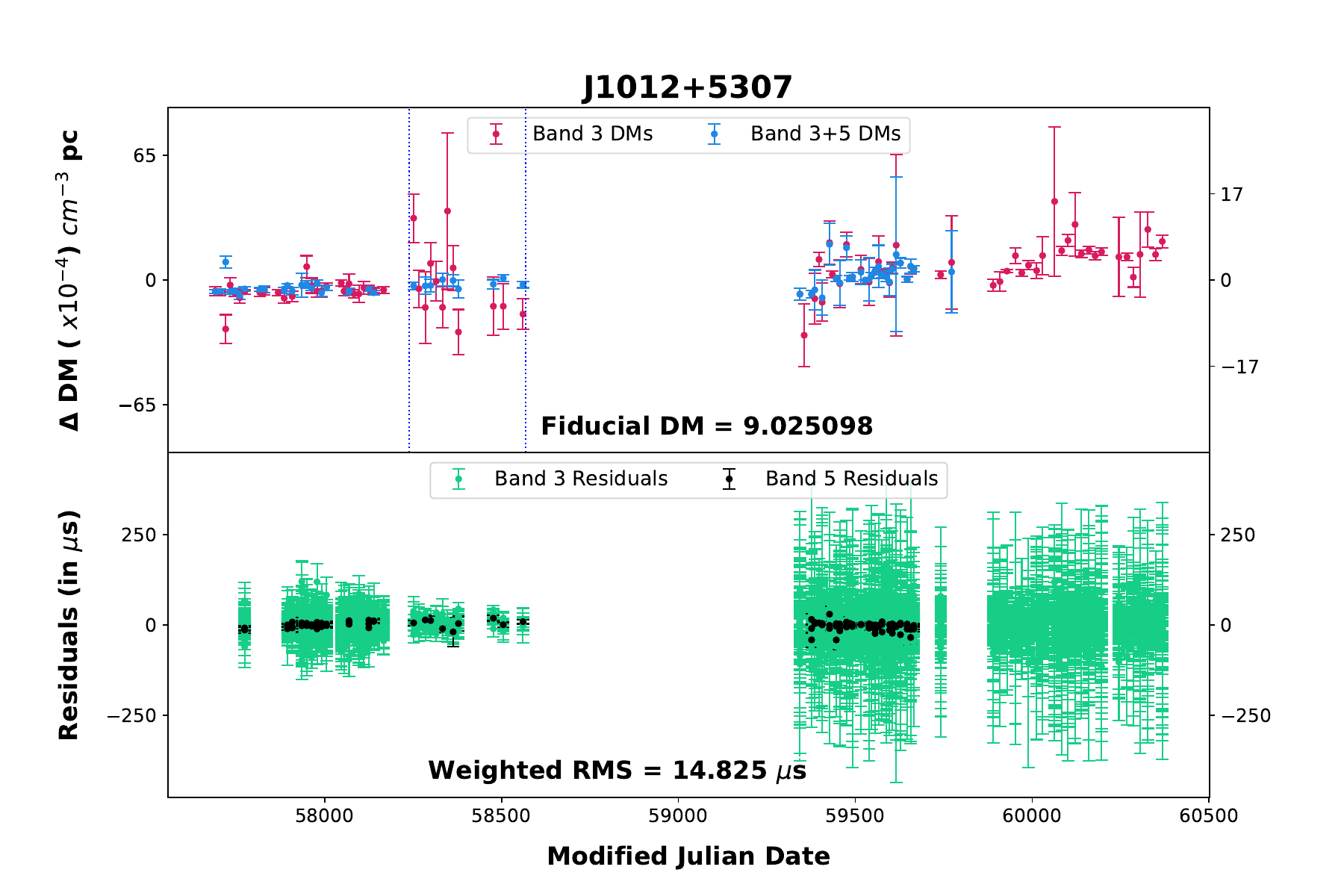}
\caption{Same as Figure \ref{fig:J0030}. Dispersion measure variations and timing residuals for J1012$+$5307 (B3 and B3+5, B3 only in the latest uGMRT cycles). The two vertical lines at MJD 58239 and 58569 divide the DM time-series into three distinct sections as described in Figure \ref{fig: DMtimeseriesplot1}. The vertical axes of DM time-series plot are scaled differently for epochs before and after MJD 58569 to reflect the improved DM precision achieved from cycle 37 onward.  }
\label{fig:J1012}
\end{figure*}

\begin{figure*}[h!]
\includegraphics[width=0.88\linewidth]{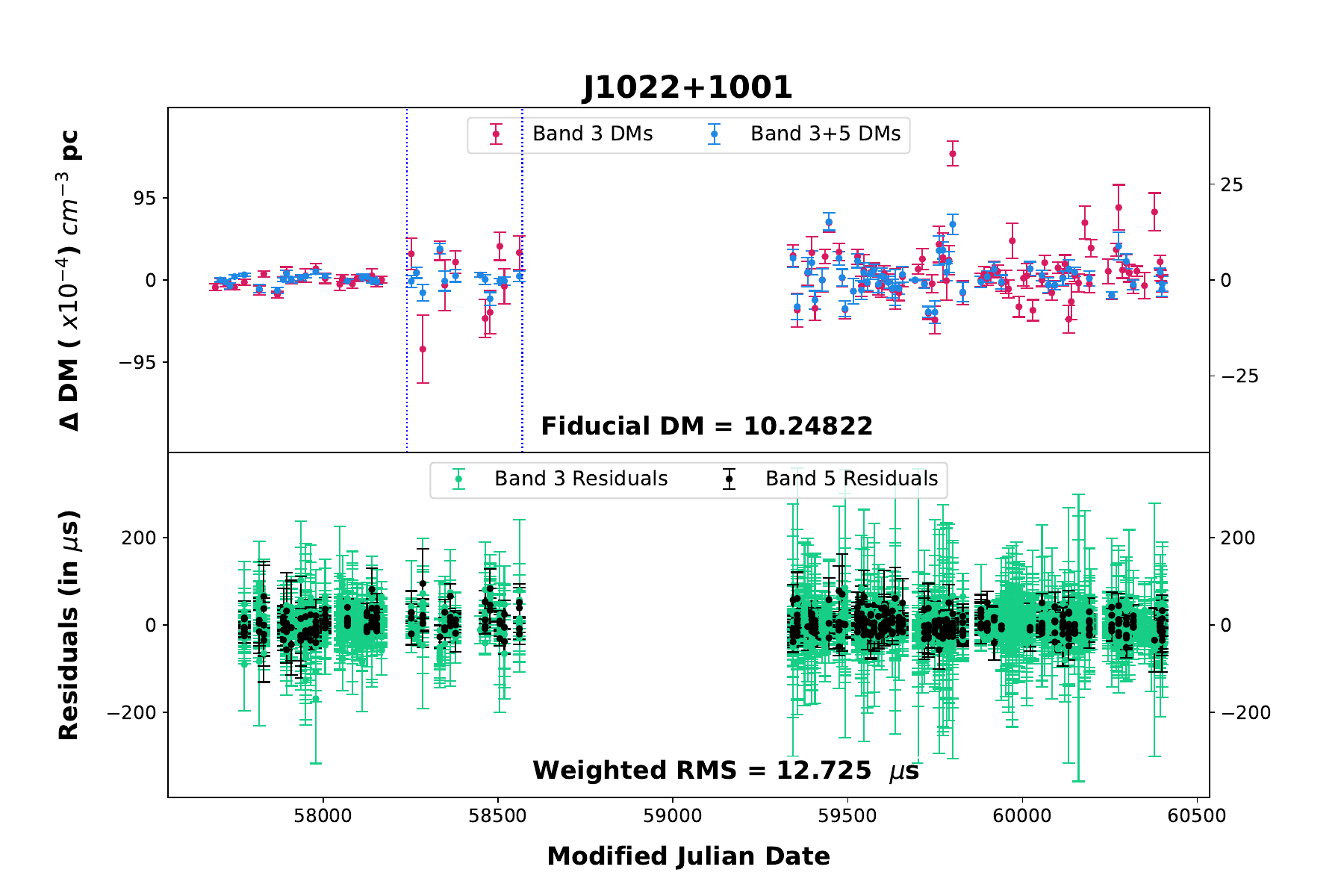}
\caption{Same as Figure \ref{fig:J0030}. Dispersion measure variations and timing residuals for  J1022$+$1001 (B3 and B3+5). The two vertical lines at MJD 58239 and 58569 divide the DM time-series into three distinct sections as described in Figure \ref{fig: DMtimeseriesplot1}. The vertical axes of DM time-series plot are scaled differently for epochs before and after MJD 58569 to reflect the improved DM precision achieved from cycle 37 onward. }
\label{fig:J1022}
\end{figure*}

\begin{figure*}[h!]
\includegraphics[width=0.88\linewidth]{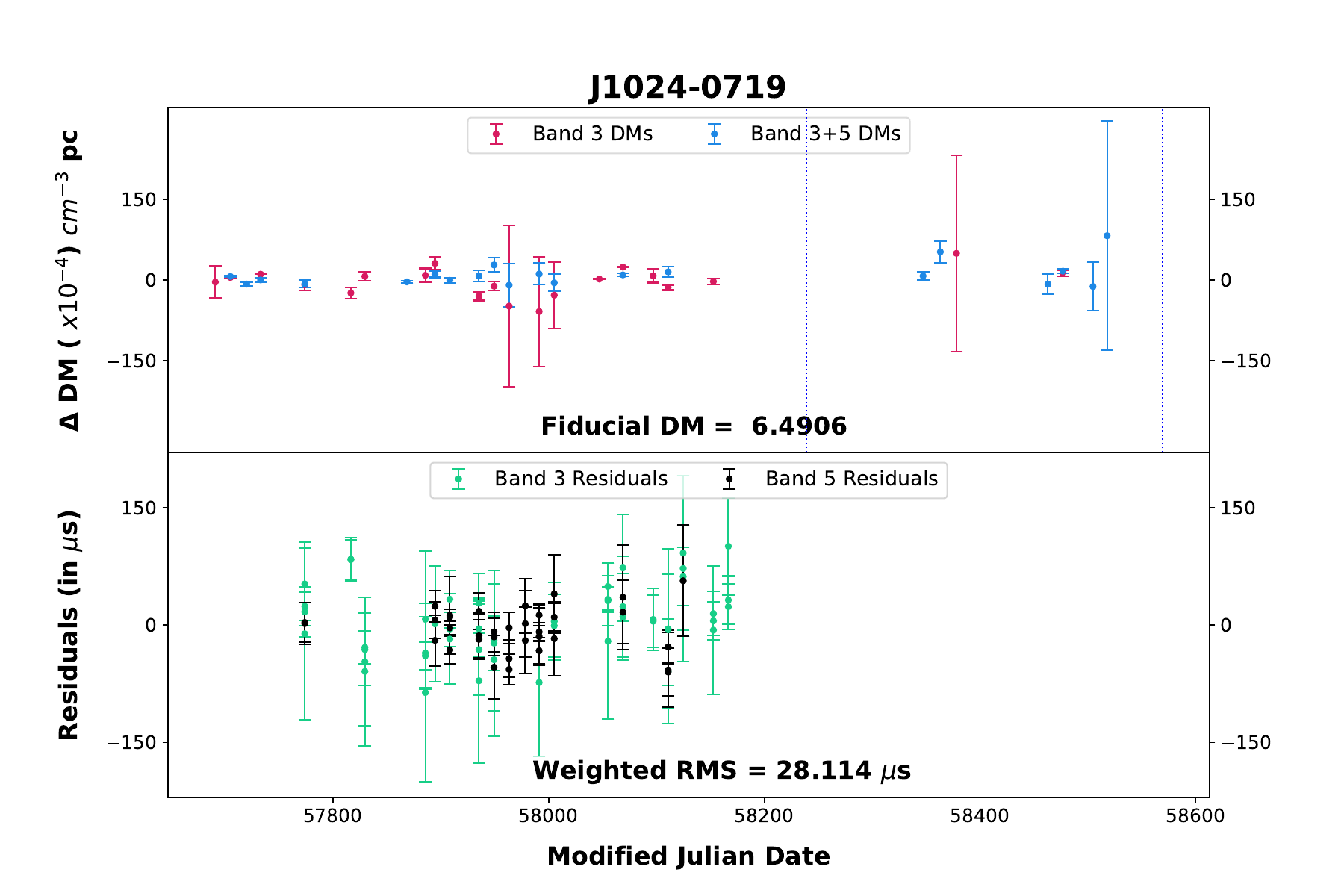}
\caption{Same as Figure \ref{fig:J0030}. Dispersion measure variations and timing residuals for J1024$-$0719 (B3 and B3+5). The two vertical lines at MJD 58239 and 58569 divide the DM time-series into two distinct sections as described in Figure \ref{fig: DMtimeseriesplot1}. } 
\label{fig:J1024}
\end{figure*}

\begin{figure*}[h!]
\includegraphics[width=0.9\linewidth]{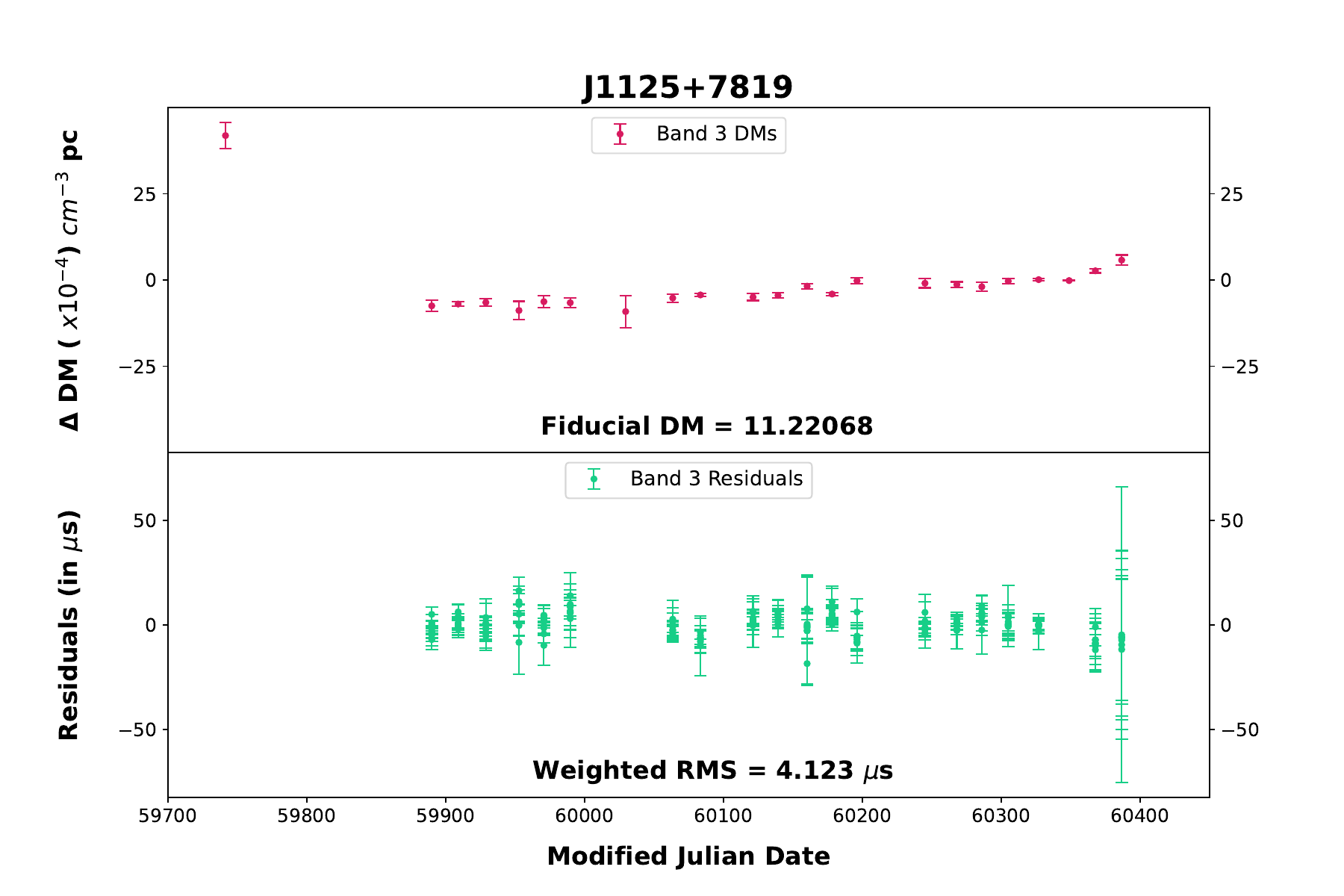}
\caption{Same as Figure \ref{fig:J0030}. Dispersion measure variations and timing residuals for J1125$+$7819 (B3). A possible DM jump event is seen at MJD 59741. }
\label{fig:J1125}
\end{figure*}

\begin{figure*}[h!]
\includegraphics[width=0.9\linewidth]{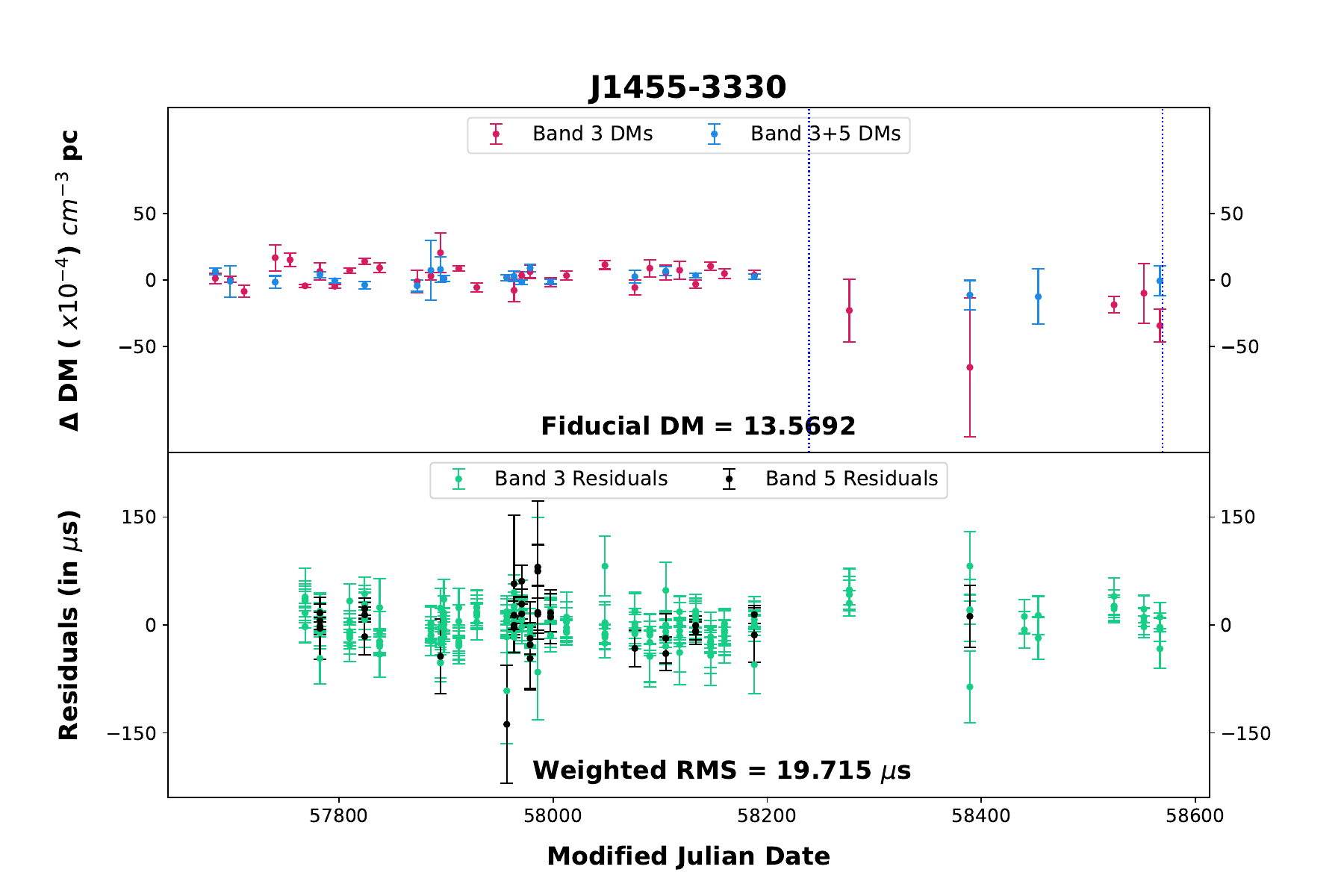}
\caption{Same as Figure \ref{fig:J0030}. Dispersion measure variations and timing residuals for J1455$-$3330 (B3 and B3+5). The two vertical lines at MJD 58239 and 58569 divide the DM time-series into two distinct sections as described in Figure \ref{fig: DMtimeseriesplot1}. } 
\label{fig:J1455}
\end{figure*}

\begin{figure*}[h!]
\includegraphics[width=0.88\linewidth]{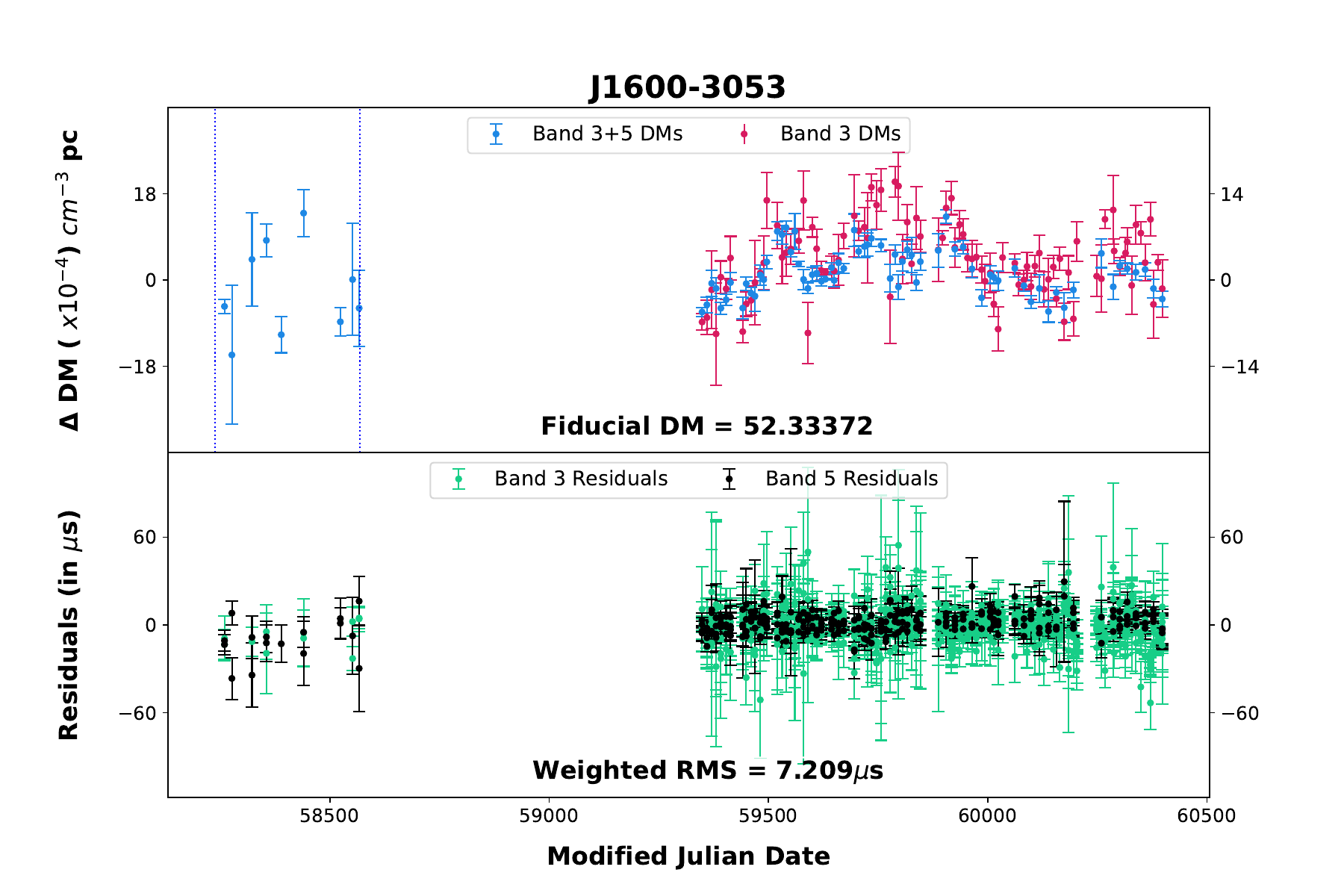}
\caption{Same as Figure \ref{fig:J0030}. Dispersion measure variations and timing residuals for  J1600$-$3053 (B3 and B3+5). The two vertical lines at MJD 58239 and 58569 divide the DM time-series into three distinct sections as described in Figure \ref{fig: DMtimeseriesplot1}. The vertical axes of DM time-series plot are scaled differently for epochs before and after MJD 58569 to reflect the improved DM precision achieved from cycle 37 onward. }
\label{fig:J1600}
\end{figure*}

\begin{figure*}[h!]
\includegraphics[width=0.88\linewidth]{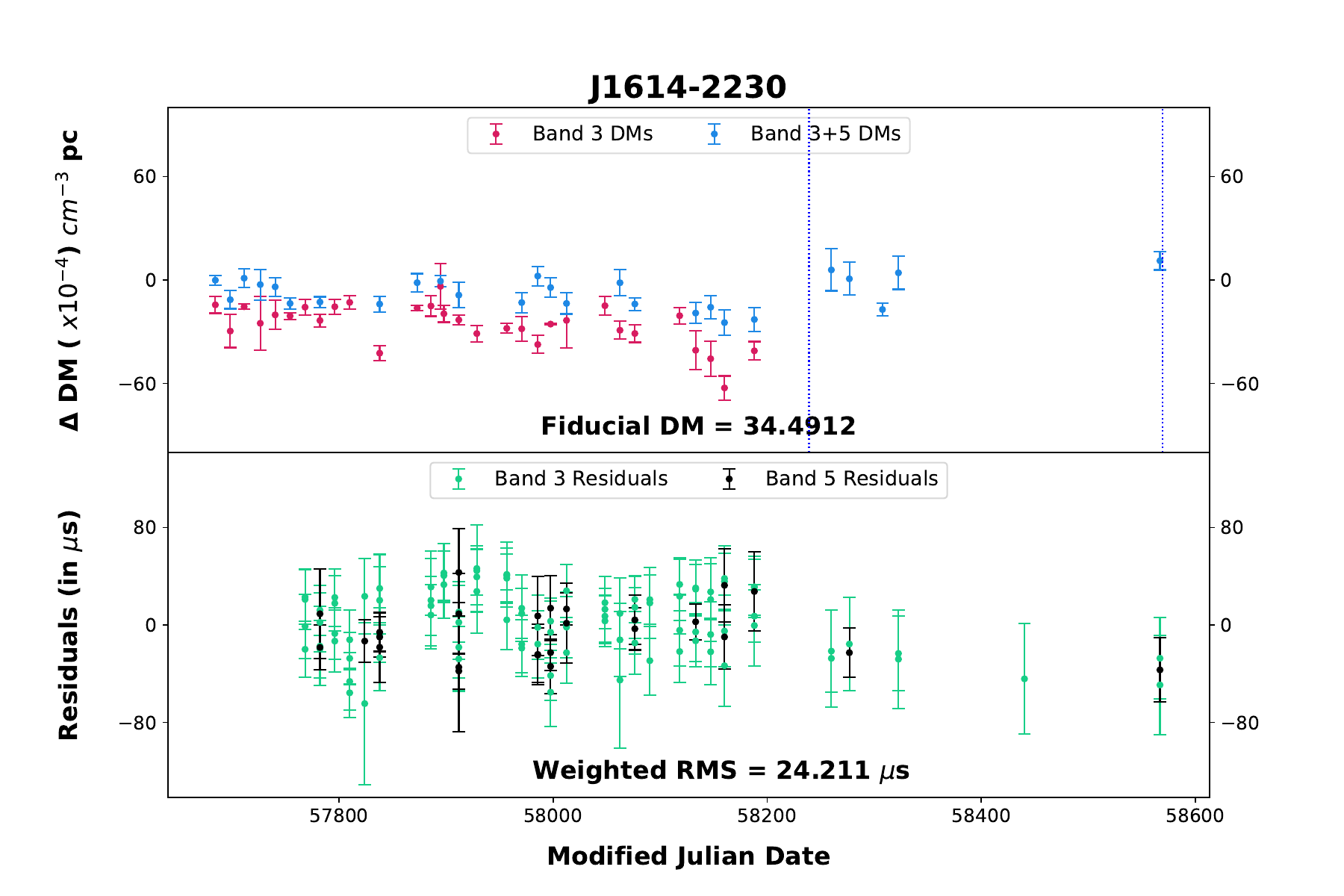}
\caption{Same as Figure \ref{fig:J0030}. Dispersion measure variations and timing residuals for  J1614$-$2230 (B3 and B3+5). The two vertical lines at MJD 58239 and 58569 divide the DM time-series into two distinct sections as described in Figure \ref{fig: DMtimeseriesplot1}.  }
\label{fig:J1614}
\end{figure*}

\begin{figure*}[h!]
\includegraphics[width=0.86\linewidth]{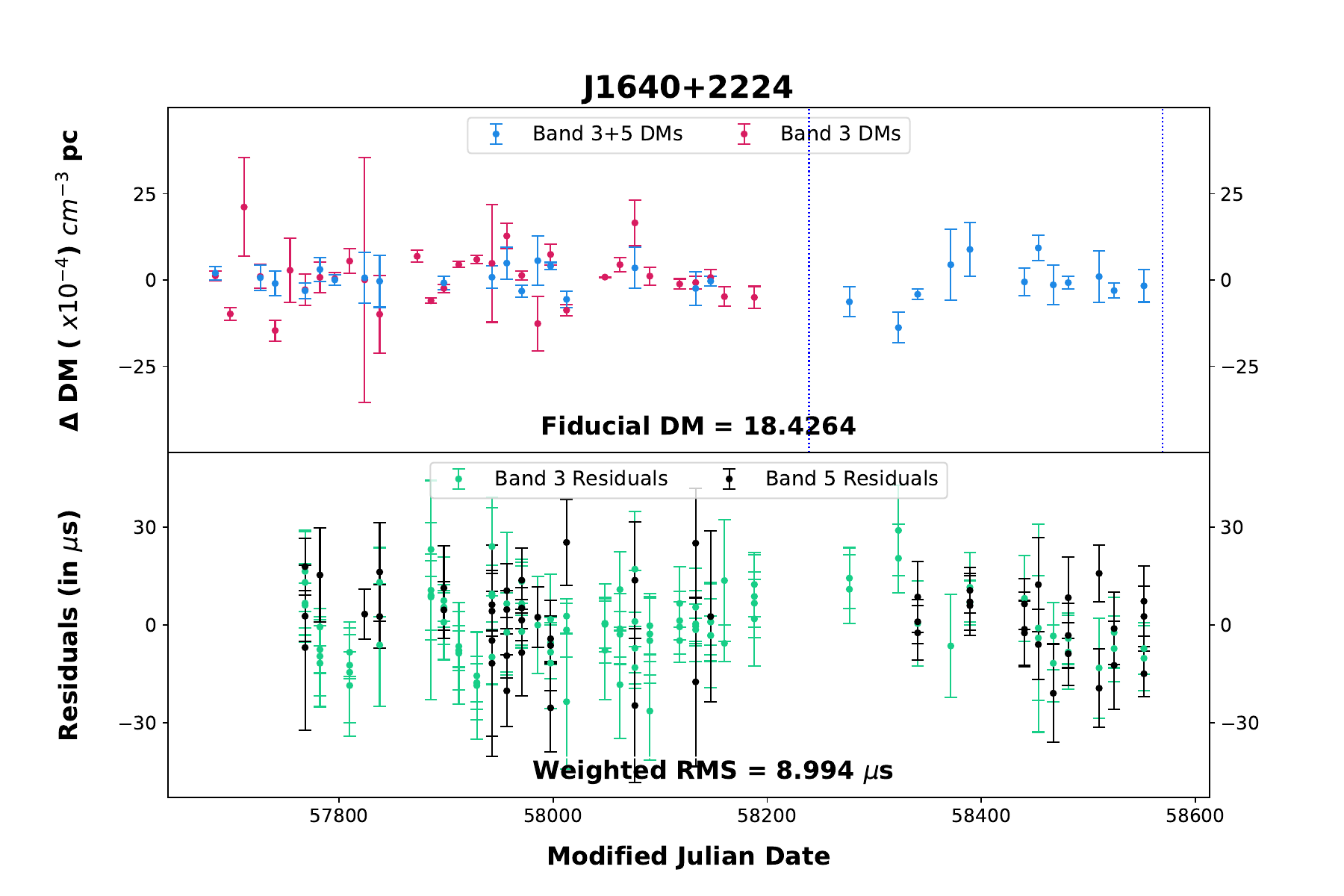}
\caption{Same as Figure \ref{fig:J0030}. Dispersion measure variations and timing residuals for J1640$+$2224 (B3 and B3+5). The two vertical lines at MJD 58239 and 58569 divide the DM time-series into two distinct sections as described in Figure \ref{fig: DMtimeseriesplot1}. }
\label{fig:J1640}
\end{figure*}

\begin{figure*}[h!]
\includegraphics[width=0.86\linewidth]{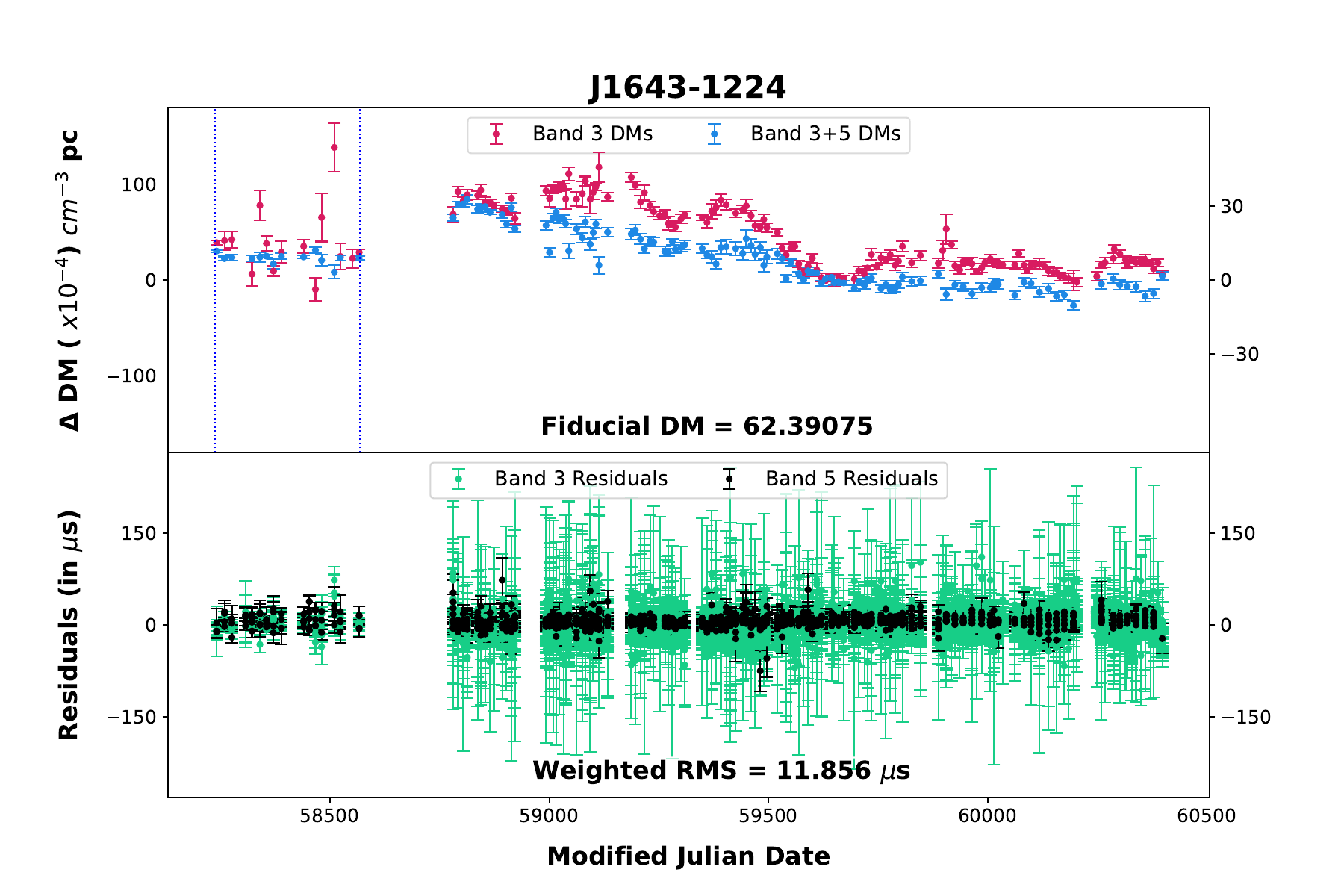}
\caption{Same as Figure \ref{fig:J0030}. Dispersion measure variations and timing residuals for J1643$-$1224 (B3 and B3+5). The two vertical lines at MJD 58239 and 58569 divide the DM time-series into three distinct sections as described in Figure \ref{fig: DMtimeseriesplot1}. The vertical axes of DM time-series plot are scaled differently for epochs before and after MJD 58569 to reflect the improved DM precision achieved from cycle 37 onward.  }
\label{fig:J1643}
\end{figure*}

\begin{figure*}[h!]
\includegraphics[width=0.87\linewidth]{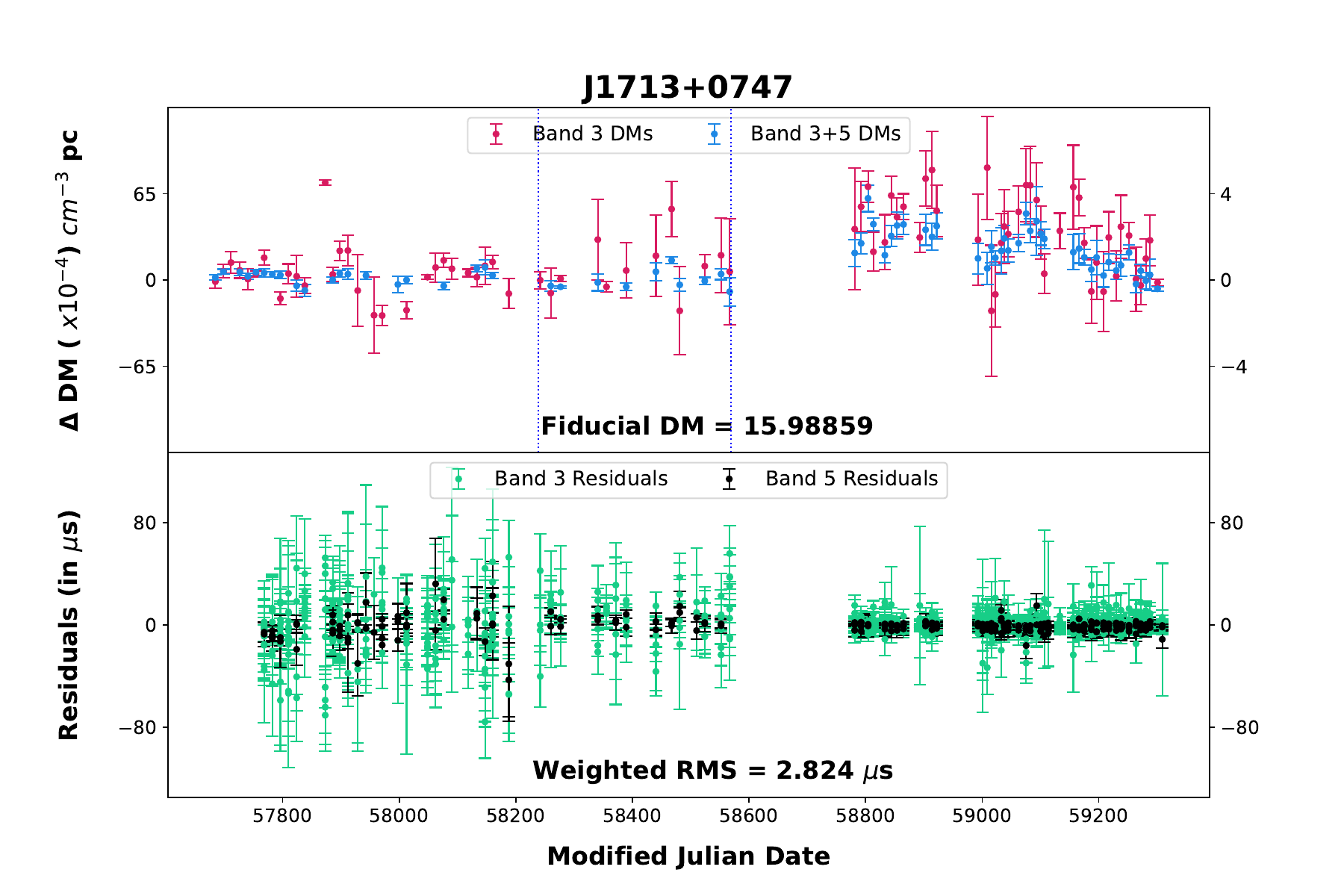}
\caption{Same as Figure \ref{fig:J0030}. Dispersion measure variations and timing residuals for J1713+0747 (B3 and B3+5) till the profile-change event in April 2021. The two vertical lines at MJD 58239 and 58569 divide the DM time-series into three distinct sections as described in Figure \ref{fig: DMtimeseriesplot1}. The vertical axes of DM time-series plot are scaled differently for epochs before and after MJD 58569 to reflect the improved DM precision achieved from cycle 37 onward. }
\label{fig:J1713}
\end{figure*}

\begin{figure*}[h!]
\includegraphics[width=0.87\linewidth]{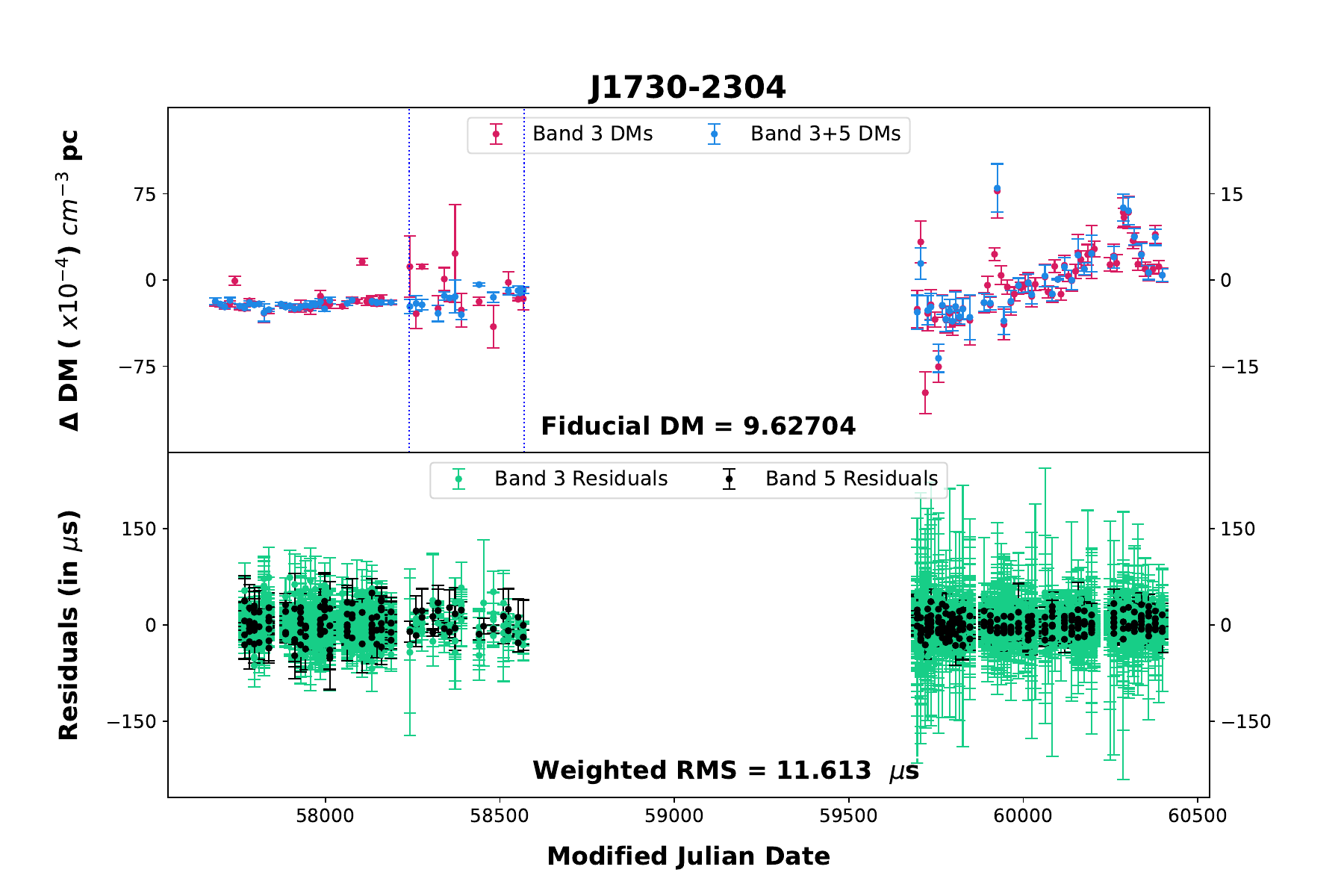}
\caption{Same as Figure \ref{fig:J0030}. Dispersion measure variations and timing residuals for J1730$-$2304 (B3 and B3+5). The two vertical lines at MJD 58239 and 58569 divide the DM time-series into three distinct sections as described in Figure \ref{fig: DMtimeseriesplot1}. The vertical axes of DM time-series plot are scaled differently for epochs before and after MJD 58569 to reflect the improved DM precision achieved from cycle 37 onward. }
\label{fig:J1730}
\end{figure*}

\begin{figure*}[h!]
\includegraphics[width=0.88\linewidth]{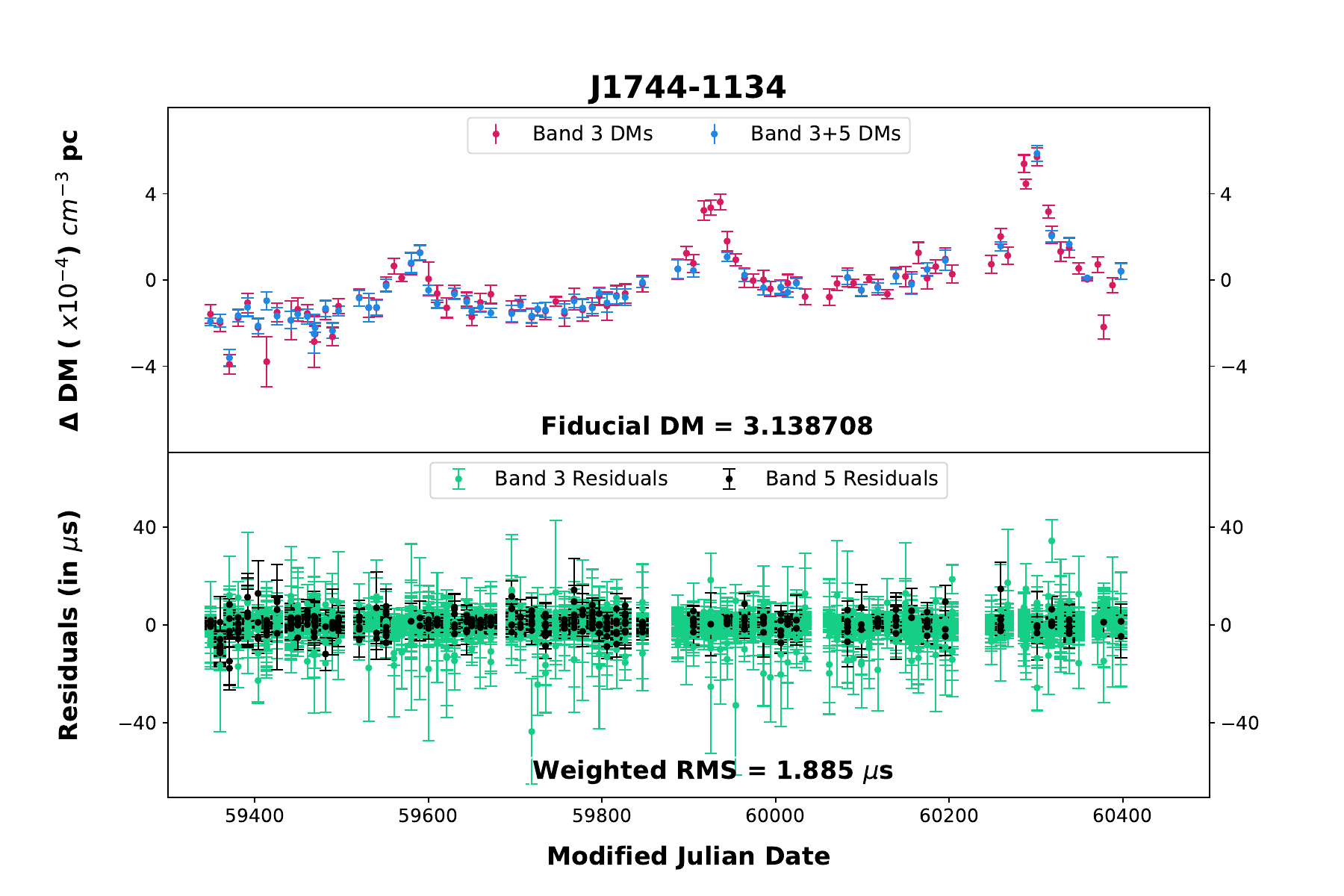}
\caption{Same as Figure \ref{fig:J0030}. Dispersion measure variations and timing residuals for J1744$-$1134 (B3 and B3+5). }
\label{fig:J1744}
\end{figure*}

\begin{figure*}[h!]
\includegraphics[width=0.88\linewidth]{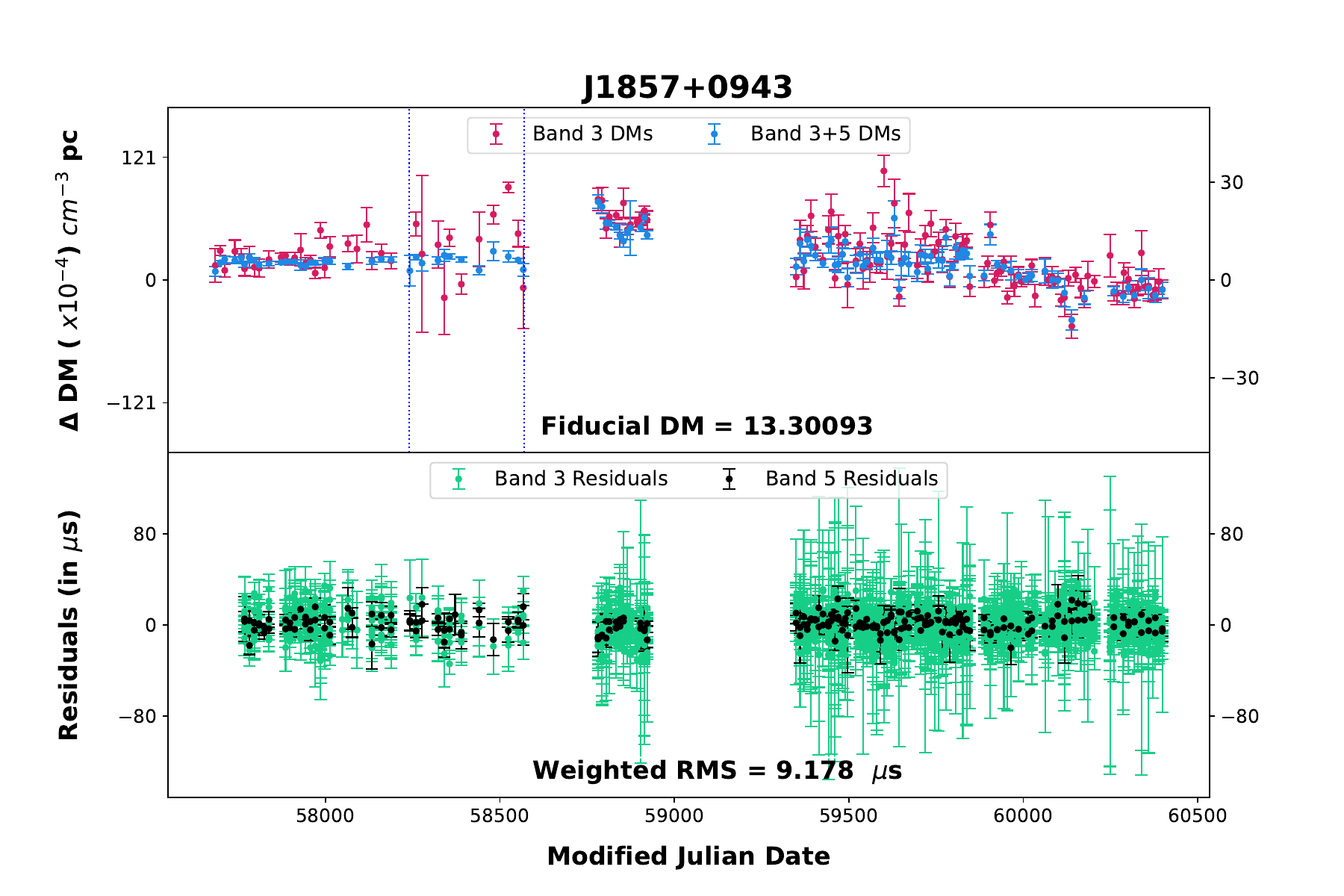}
\caption{Same as Figure \ref{fig:J0030}. Dispersion measure variations and timing residuals for J1857$+$0943 (B3 and B3+5). The two vertical lines at MJD 58239 and 58569 divide the DM time-series into three distinct sections as described in Figure \ref{fig: DMtimeseriesplot1}. The vertical axes of DM time-series plot are scaled differently for epochs before and after MJD 58569 to reflect the improved DM precision achieved from cycle 37 onward.  }
\label{fig:J1857}
\end{figure*}

\begin{figure*}[h!]
\includegraphics[width=0.87\linewidth]{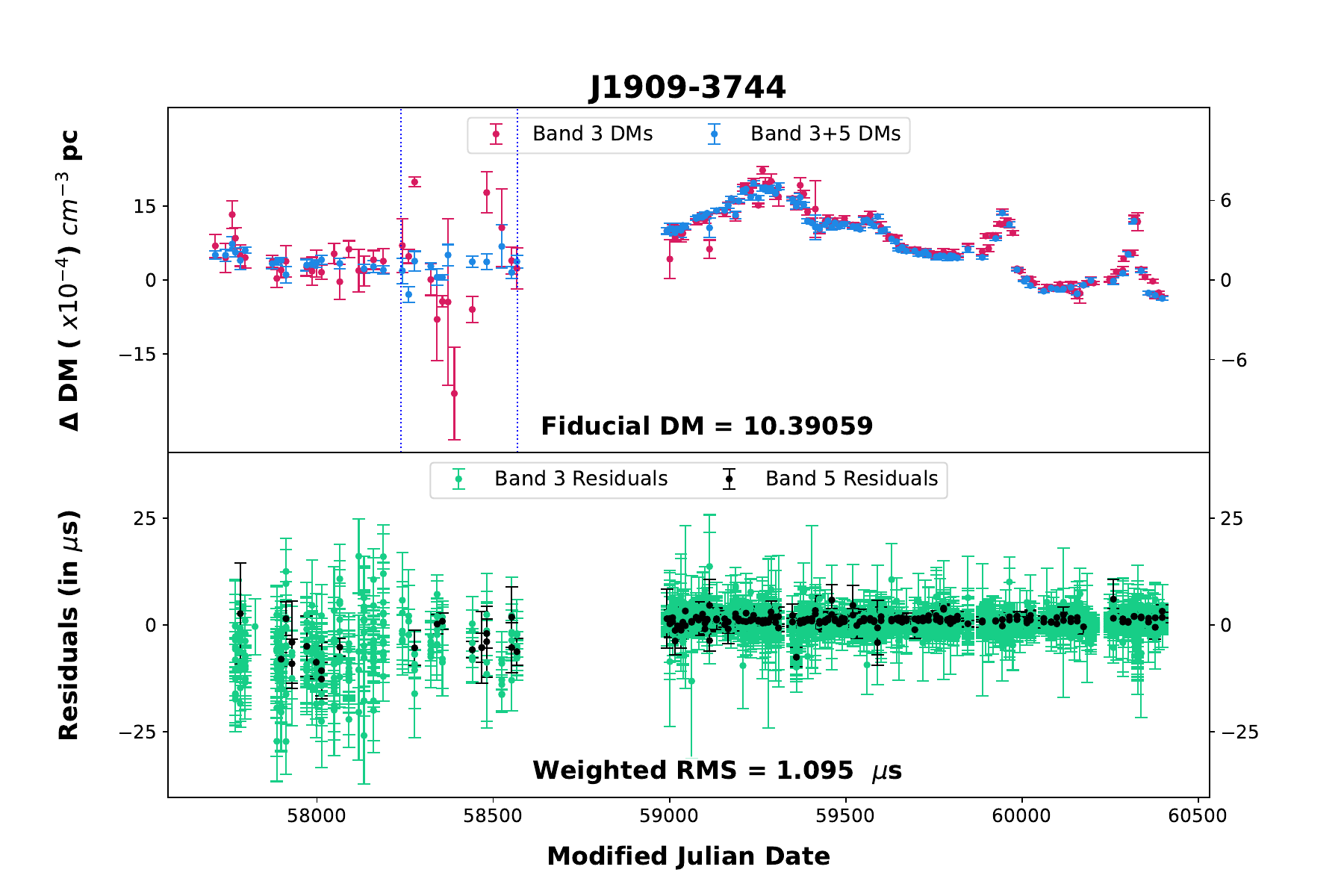}
\caption{Same as Figure \ref{fig:J0030}. Dispersion measure variations and timing residuals for  J1909$-$3744 (B3 and B3+5). The two vertical lines at MJD 58239 and 58569 divide the DM time-series into three distinct sections as described in Figure \ref{fig: DMtimeseriesplot1}. The vertical axes of DM time-series plot are scaled differently for epochs before and after MJD 58569 to reflect the improved DM precision achieved from cycle 37 onward. }
\label{fig:J1909}
\end{figure*}

\begin{figure*}[h!]
\includegraphics[width=0.87\linewidth]{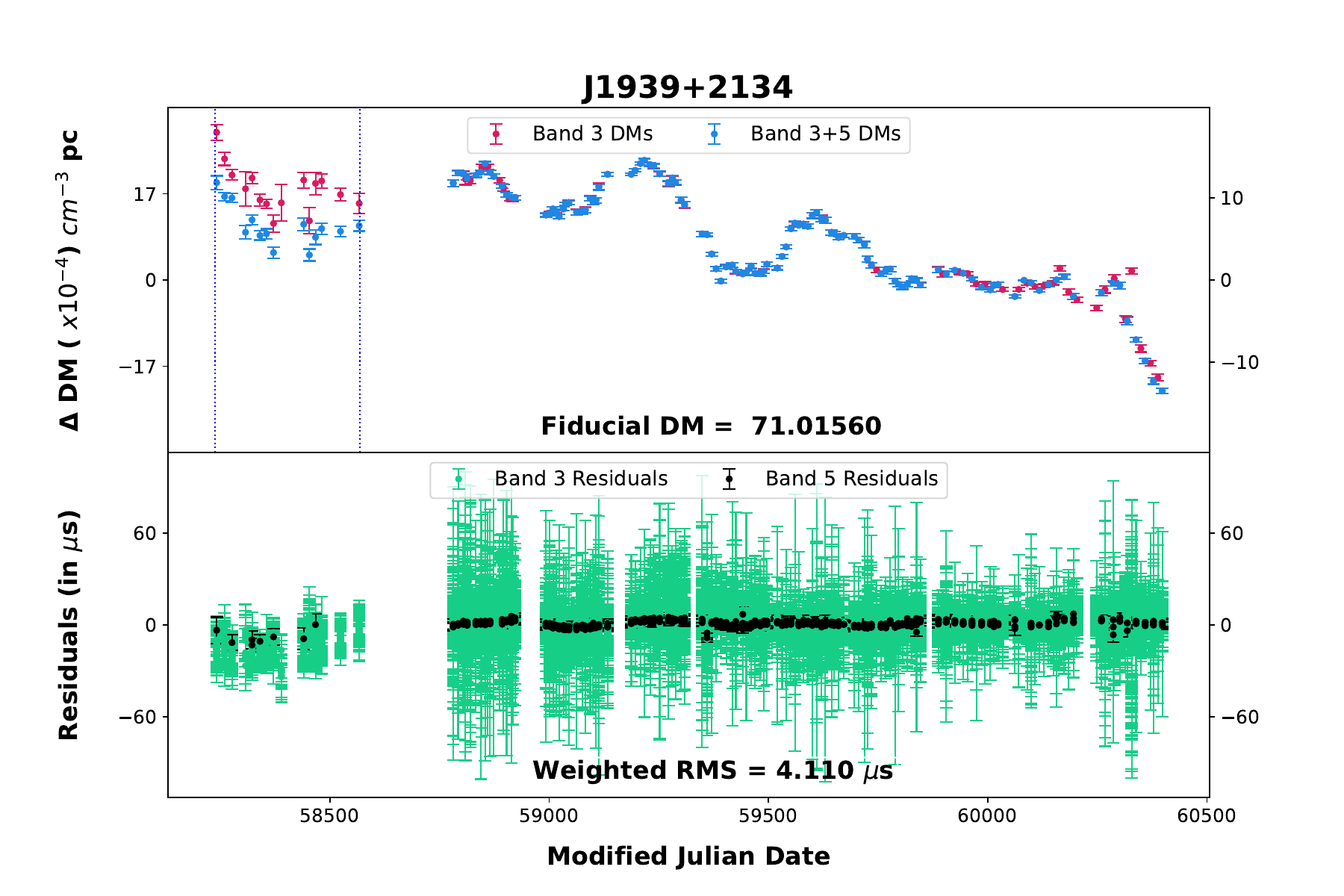}
\caption{Same as Figure \ref{fig:J0030}. Dispersion measure variations and timing residuals for J1939$+$2134 (B3 and B3+5). The two vertical lines at MJD 58239 and 58569 divide the DM time-series into three distinct sections as described in Figure \ref{fig: DMtimeseriesplot1}. The vertical axes of DM time-series plot are scaled differently for epochs before and after MJD 58569 to reflect the improved DM precision achieved from cycle 37 onward.  }
\label{fig:J1939}
\end{figure*}

\begin{figure*}[h!]
\includegraphics[width=0.88\linewidth]{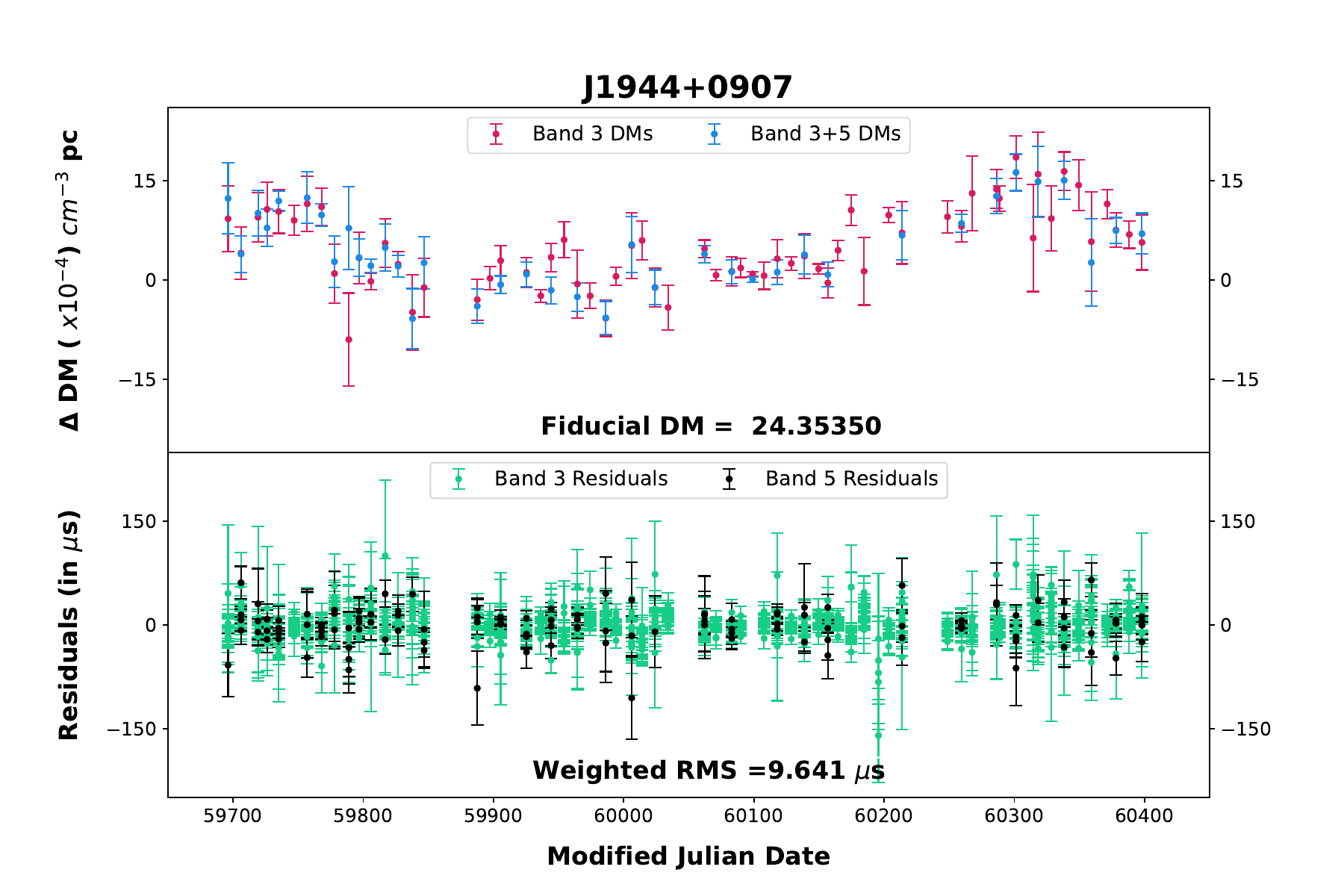}
\caption{Same as Figure \ref{fig:J0030}. Dispersion measure variations and timing residuals for J1944+0907 (B3 and B3+5). }
\label{fig:J1944}
\end{figure*}

\begin{figure*}[h!]
\includegraphics[width=0.88\linewidth]{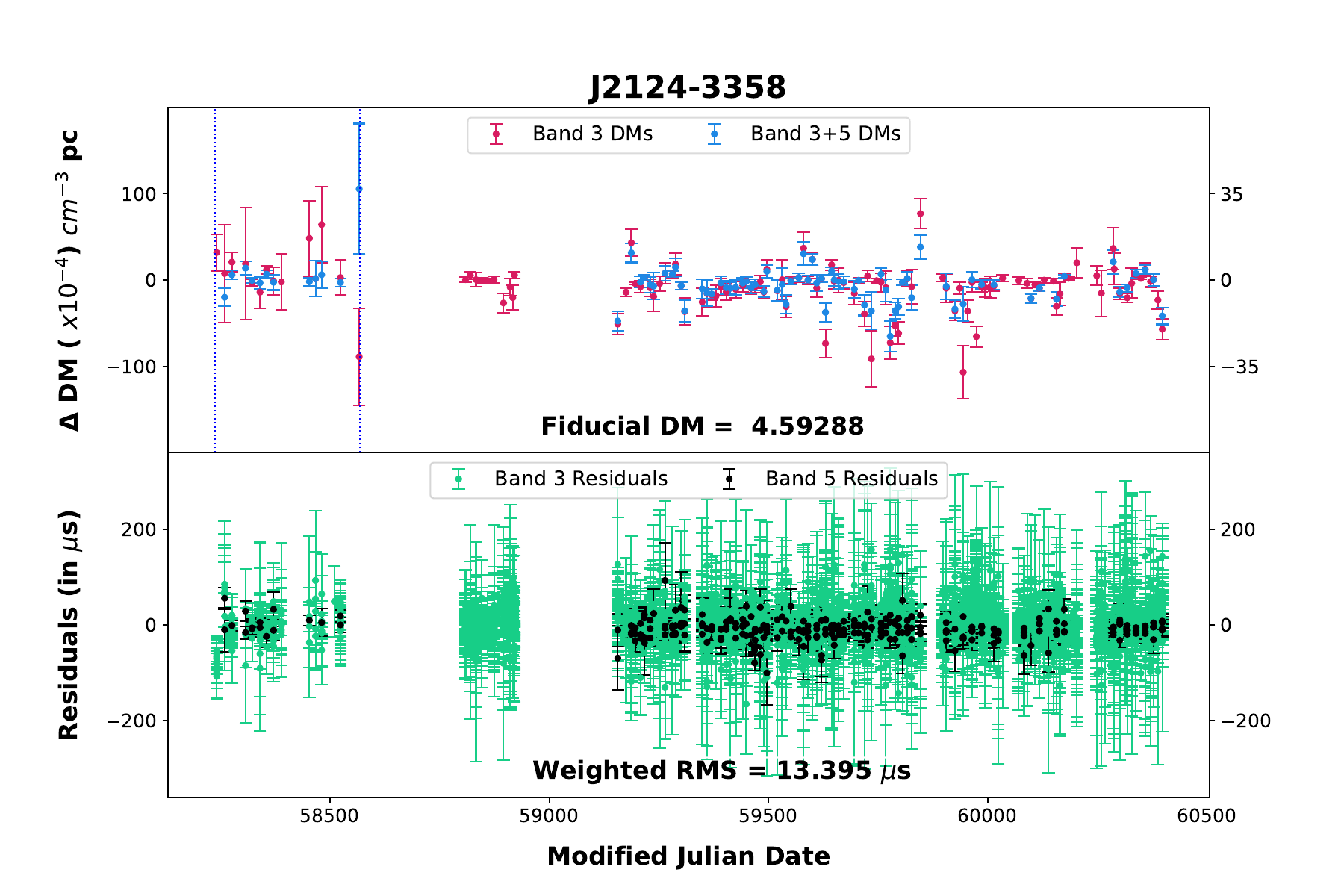}
\caption{Same as Figure \ref{fig:J0030}. Dispersion measure variations and timing residuals for J2124$-$3358 (B3 and B3+5). The two vertical lines at MJD 58239 and 58569 divide the DM time-series into three distinct sections as described in Figure \ref{fig: DMtimeseriesplot1}. The vertical axes of DM time-series plot are scaled differently for epochs before and after MJD 58569 to reflect the improved DM precision achieved from cycle 37 onward. }
\label{fig:J2124}
\end{figure*}

\begin{figure*}[h!]
\includegraphics[width=0.85\linewidth]{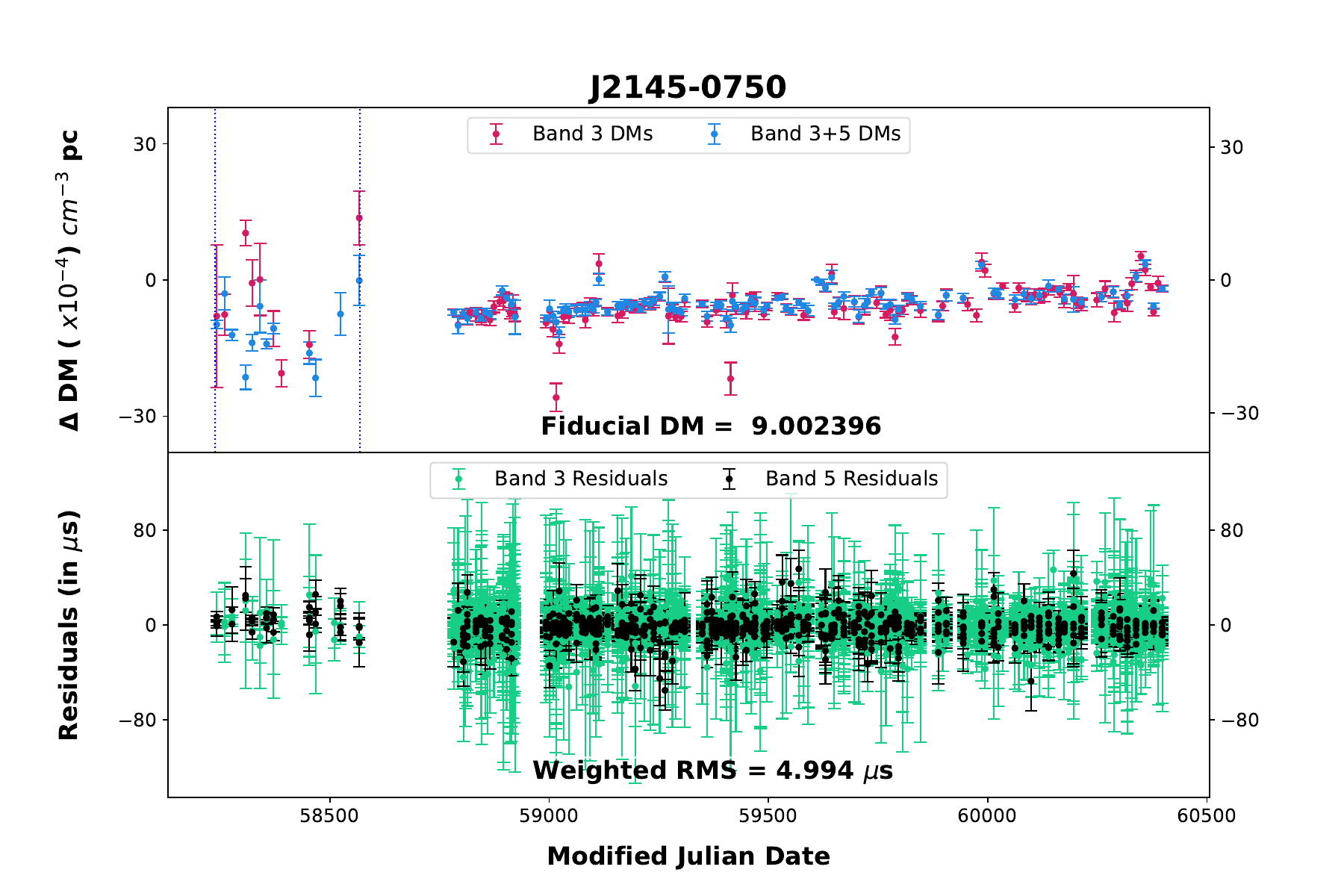}
\caption{Same as Figure \ref{fig:J0030}. Dispersion measure variations and timing residuals for J2145$-$0750 (B3 and B3+5). The two vertical lines at MJD 58239 and 58569 divide the DM time-series into three distinct sections as described in Figure \ref{fig: DMtimeseriesplot1}.  }
\label{fig:J2145}
\end{figure*}

\begin{figure*}[h!]
\includegraphics[width=0.85\linewidth]{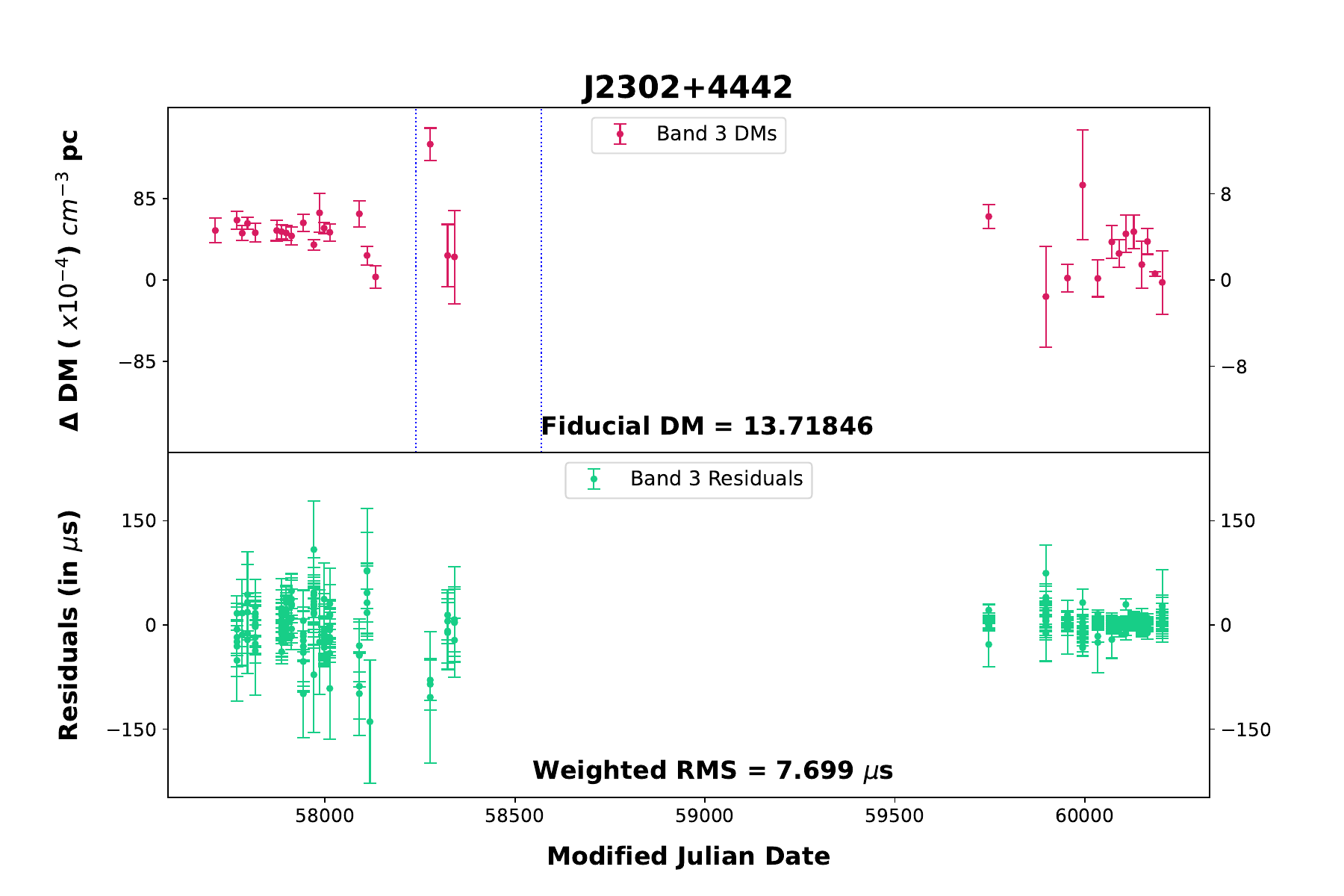}
\caption{Same as Figure \ref{fig:J0030}. Dispersion measure variations and timing residuals for J2302$+$4442 (B3). The two vertical lines at MJD 58239 and 58569 divide the DM time-series into three distinct sections as described in Figure \ref{fig: DMtimeseriesplot1}. The vertical axes of DM time-series plot are scaled differently for epochs before and after MJD 58569 to reflect the improved DM precision achieved from cycle 37 onward.}
\label{fig:J2302}
\end{figure*}

\end{document}